\documentclass[11pt,letterpaper]{article}
\usepackage[dvips]{epsfig}
\usepackage{amsmath,amssymb,amsfonts,enumerate,bbm}
\usepackage{caption}
\usepackage{url}
\usepackage{fullpage}
\usepackage{ifthen}
\usepackage{graphicx}

\usepackage{subfigure}

\usepackage{color,soul}
\def\dnsparagraph#1{\par\vspace{2pt}\noindent{\em #1}.}

\def\FTBFS{\mbox{\tt FT-BFS}}

\long\def\commabs #1\commabsend{#1}
\long\def\commful #1\commfulend{}
\long\def\comment #1\commentend{}

\def\ApproxSetCover{\mbox{\tt ApproxSetCover}}
\def\Set{\mathfrak{F}}
\newtheorem{theorem}{Theorem}[section]
\newtheorem{lemma}[theorem]{Lemma}
\newtheorem{observation}[theorem]{Observation}
\newtheorem{corollary}[theorem]{Corollary}

\newtheorem{claim}[theorem]{Claim}

\newtheorem{definition}[theorem]{Definition}
\newcommand{\REAL}{\mathbb R}
\def\deg{\mbox{\tt deg}}
\def\depth{\mbox{\tt depth}}

\def\Cost{\mbox{\tt Cost}}
\newcommand{\dist}{\mbox{\rm dist}}

\def\FTMBFS{\mbox{\tt FT-MBFS}}

\def\inline#1:{\par\vskip 7pt\noindent{\bf #1:}\hskip 10pt}
\def\Proof{\par\noindent{\bf Proof:~}}
\def\blackslug{\hbox{\hskip 1pt \vrule width 4pt height 8pt
    depth 1.5pt \hskip 1pt}}
\def\QED{\quad\blackslug\lower 8.5pt\null\par}
\newcommand{\New}[0]{\mbox{\tt New}}
\def\LAB{\mbox{\tt Label}}

\def\LastE{\mbox{\tt LastE}}

\def\NSource{\sigma}
\def\Root{\mbox{\tt r}}
\def\Leaf{\mbox{\tt Leaf}}
\def\NLeaf{\mbox{\tt nLeaf}}
\def\NodesIn{\mbox{\tt N}}

\def\ConstPath{\mbox{\sf Cons2FTBFS}}
\def\KernelGraph{\mathcal{K}}
\def\Breaker{\Psi}

\def\First{\mbox{\tt First}}
\def\Last{\mbox{\tt Last}}

\usepackage{geometry}
\title{Dual Failure Resilient BFS Structure}

\author{
Merav Parter
\thanks{Department of Computer Science and Applied Mathematics,
The Weizmann Institute of Science, Rehovot, Israel.
E-mail: {\tt merav.parter@ weizmann.ac.il}.
Supported in part by the Israel Science Foundation (grant 894/09),
the United States-Israel Binational Science Foundation (grant 2008348),
the Israel Ministry of Science and Technology (infrastructures grant),
and the Citi Foundation.
Recipient of the Google European Fellowship in distributed computing;
research is supported in part by this Fellowship.}
}

\begin{document}
\begin{titlepage}
\date{}
\maketitle

\begin{abstract}
We study {\em breadth-first search (BFS)} spanning trees,
and address the problem of designing a sparse {\em fault-tolerant} BFS structure,
or {\em FT-BFS } for short, resilient to the failure of up to two edges in the given undirected unweighted graph $G$, i.e., a sparse subgraph $H$ of $G$ such that subsequent to the failure of up to two edges,
the surviving part $H'$ of $H$ still contains a BFS spanning tree for (the surviving part of) $G$. FT-BFS structures, as well as the related notion of replacement paths, have been studied so far for the restricted case of a single failure.
It has been noted widely that when concerning shortest-paths in a variety of contexts, there is a sharp qualitative difference between a single failure and two or more failures \cite{DP09}.
%Indeed, a key barrier in handling dual failure events is the complex structure of shortest paths avoiding two edge failures.
%For a source node $s$, a target node $v$ and a pair $F=\{e_i,e_j\}\subseteq G$ of failed edges, the shortest $s-v$ path $P_{s,v,F}$ that does not go through the edge pair $F$ is the natural extension of the well studied single failure \emph{replacement path}. Thus, our dual failure \FTBFS\ structure must contain some replacement path $P_{s,v,F}$ for every $v \in V(G)$ and every edge pair $F \subseteq E(G)$.
Our main results are as follows. We present an algorithm that
for every $n$-vertex unweighted undirected graph $G$ and source node $s$ constructs
a (two edge failure) \FTBFS\ structure rooted at $s$ with
$O(n^{5/3})$ edges.  To provide a useful theory of shortest paths avoiding 2 edges failures, we take a principled approach to classifying the arrangement these paths. We believe that the structural analysis provided in this paper may decrease the barrier for understanding the general case of $f\geq 2$ faults and pave the way to the future design of $f$-fault resilient structures for $f \geq 2$.
We also provide a matching lower bound, which in fact holds for the general case of $f \geq 1$ and multiple sources $S \subseteq V$. It shows that for every $f\geq 1$, and integer $1 \leq \NSource \leq n$, there exist $n$-vertex graphs with a source set $S \subseteq V$ of cardinality $\NSource$ for which any \FTBFS\ structure rooted at each $s \in S$, resilient to up to $f$-edge faults has $\Omega(\NSource^{1/(f+1)} \cdot n^{2-1/(f+1)})$ edges. In particular, for $f=2$ and $\NSource=1$, a dual failure \FTBFS\ structure rooted at $s$ must have $\Omega(n^{5/3})$ edges in the worst case.
Finally, we also consider the optimization variant for this problem, and propose an $O(\log n)$ approximation algorithm for constructing \FTBFS\ structures resilient to up to $f$-faults for any constant $f \geq 1$ and any source set $S \subseteq V$.
\end{abstract}
\bigskip
{\bf Regular submission.}
\def\thepage{}
\end{titlepage}
%%%%%%%%%%%%%%%%%%%%%%
\section{Introduction}
%%%%%%%%%%%%%%%%%%%%%%
%%%%%%%%%%%%%%%%%%%%%%%%%
\dnsparagraph{\textbf{Background and motivation}}
%%%%%%%%%%%%%%%%%%%%%%%%%
Large network systems of electricity, telephony or communication are traditionally designed to withstand the possibility of a  \emph{single} failure in one of their components. This is partially justified by the optimistic view that a failure is a rare event. Yet, since in modern day huge communication networks
several components may fail or malfunction at any given time, the restriction to single failure events mainly stems from the unfortunate fact that supporting the capability of coping two failures or more is, in many cases, considerably more complex than having to overcome just a single failure. For example, when considering the setting of shortest path in some underlying graph, it has been widely noted that there is a sharp qualitative and quantitative difference between shortest paths avoiding just one failure and paths avoiding two or more failures.
%Fault-resilience can be introduced into the network in several different ways. Here we focus on a notion of fault-tolerance whereby the structure  is ``reinforced'' (by adding to it various components) so that subsequent to the failure of some of the network's edges, the surviving part of the structure is still operational. As this reinforcement carries certain costs, it is desirable to minimize the number of added components.
%
We consider the structure  of {\em breadth-first search (BFS)} spanning trees,
and address the problem of designing dual failure {\em fault-tolerant} BFS structure, or {\em \FTBFS} for short.
By this we mean a subgraph $H$ of the given network $G$,
such that subsequent to the failure up to two of the edges,
the surviving part $H'$ of $H$ still contains a BFS spanning tree for the surviving part of $G$.
\par Typical network design problems involve three types of objectives: (1) construction time (i.e., cost of the preprocessing phase) (2) quality of usage, i.e., efficiency of operations preformed in the constructed structure, and (3) the size of the constructed structure. The current work is motivated by settings in which objectives (2) and (3) play a dominant role. In particular, objective (2) is important in cases where using approximate shortest paths instead of exact ones (e.g., for routing), entails a high cost on the system and it is preferable to purchase a larger structure that will allow optimal operation (e.g., routing on shortest paths).
Subject to objective (2), it is still desirable to construct (or purchase) the minimum cost structure satisfying the usability requirements (e.g., optimum routing). A typical motivation for this is a setting where the graph edges represent the channels of a communication network, and the system designer would like
to purchase or lease a minimal collection of channels
(i.e., a subgraph $G' \subseteq G$) that maintains its functionality as a BFS tree with respect to the source $s$ upon failures in $G$. In such a context, the cost of computation at the preprocessing stage (i.e., objective (1)) may
be negligible compared to the purchasing/leasing cost of the resulting structure. Hence, our key cost measure in this paper is the \emph{size} of
the fault tolerant structure that provides the exact shortest paths distance from a given source vertex $s$, and our main goal is to achieve {\em sparse} (or {\em compact}) such structures (our construction time is still polynomial in $n$).
The notion of {\em \FTBFS}\ structure is closely related to the problem of constructing \emph{replacement paths} and in particular to its {\em single source} variant,
%the \emph{single-source replacement paths} problem,
studied in \cite{GW12} only for the single failure case.
For a source node $s$, a target node $v$ and an edge $e\in G$, the shortest $s-v$ path $P_{s,v,e}$ that does not go through $e$ is known as a \emph{replacement path}. The replacement path problem requires to compute the collection $\mathcal{P}_{s,v}$ of all $s-v$ replacement paths $P_{s,v,e}$ for every failed edge $e$ that appears on the $s-v$ shortest-path $\pi(s,v)$ in $G$.
Note that a replacement path is, by definition restricted to a single failure event.  Under this restricted setting, the replacement path $P_{s,v,e}$ admits a rather convenient form, consisting of three segments: a prefix of the shortest-path $\pi(s,v)$ up to some vertex $b \in \pi(s,v)$ occurring before the failing edge $e$, followed by a ``detour" avoiding the path $\pi(s,v)$ (and in particular the failing edge $e$), and terminating with a suffix of $\pi(s,v)$. This clean decomposition has led to the development of algorithms that compute the collection $\mathcal{P}_{s,v}$ efficiently (cf. \cite{Hershberger01,RTREP05,BK09,WY10,GW12}).
A replacement path $P_{s,v,e}$ is called \emph{new-ending} if its last edge is different from the last edge of the shortest path $\pi(s,v)$.
%, or in other words, its terminating suffix of $\pi(s,v)$ contains just the target vertex $v$.
Put another way, a new-ending replacement path $P_{s,v,e}$ has the property that once it diverges from the shortest-path $\pi(s,v)$ at the vertex $b$, it joins $\pi(s,v)$ again only at the final vertex $v$.
It is shown in \cite{PPFTBFS13} that for a given graph $G$ and source vertex $s$, a structure $H \subseteq G$ containing a BFS tree rooted at $s$ plus the last edge of each new-ending replacement path $P_{s,v,e}$ for every $e \in G$ and $v \in V$, is a single-failure \FTBFS\ structure. This means that
%to create a single-failure \FTBFS\ structure there is no need to include all replacement paths; rather,
it suffices to focus on the new-ending replacement paths and pick a single edge from each of them (specifically, the last).  Furthermore, by analyzing the special structure of the new-ending paths, it is shown therein that such a structure consists of $O(n^{3/2})$ edges where $n$ is the number of vertices in the graph.
This result is complemented by a matching lower bound
showing that for every sufficiently large integer $n$, there exist an $n$-vertex graph $G$ and a source $s \in V$, for which every single failure \FTBFS\
structure is of size $\Omega(n^{3/2})$.
Since {\em exact} \FTBFS\ structures may be rather expensive, \cite{PPFTBFSSODA14,BiloApBFS} exploit the structure of replacement-paths to construct \emph{approximate} \FTBFS\ structures with $O(n)$ edges for unweighted undirected graphs.
%
%This last observation motivated the approach of resorting to \emph{approximate}
%distances, in order to allow the design of a sparse subgraph with properties
%resembling those of an \FTBFS\ structure \cite{PPFTBFSSODA14,BiloApBFS}.
%Using the structure of the replacement paths, \cite{PPFTBFSSODA14} provides a construction of a subgraph $H \subseteq G$, with at most $4n$ edges, that yields {\em approximate} BFS structures (with multiplicative stretch $3$) that are resistant to a single edge failure.
\par Indeed, the convenient structure of the replacement paths has facilitated the development of solutions to many other related problems, such as dynamic algorithms for shortest paths and \emph{$f$-sensitivity distance oracles}, capable of efficiently answering proximity queries following a $f$-failures event \cite{DTCR08,BK09}.
%It has been shown in~\cite{DTCR08} that given a directed weighted $n$-vertex  graph $G$, it is possible to construct in time $\Ot(mn^2)$ a $1$-sensitivity fault tolerant distance oracle of size $O(n^2 \log n)$
%capable of answering distance queries in $O(1)$ time in the presence of a single failed edge or vertex.
%The preprocessing time was recently improved to $\Ot(mn)$,
%with unchanged size and query time ~\cite{BK09}.
%\par
Recently, distance sensitivity oracles have been considered for weighted
and directed graphs in the \emph{single source} setting \cite{GW12}.
%Specifically, Grandoni and Williams considered the problem of
%\emph{single-source replacement paths} where one aims to compute the collection of all replacement paths for a given source node $s$,
%and proposed an efficient randomized algorithm that does so in
%$\widetilde{O}(APSP(n,M))$ where $APSP(n,M)$ is the time required to compute
%all-pairs-shortest-paths in a weighted graph with integer weights $[-M,M]$.
An efficient construction of single source distance oracles for planar graphs is provided in \cite{BSPlanar}.
\par Yet, this long line of results, heavily exploits the structure of the single failure replacement path, and is consequently limited to handling no more than one fault in the network.
A natural goal is to generalize some of these results to settings with two or more failures.  It appears that the main barrier for such an extension is rooted in the fact that the structure of a replacement path $P_{s,v,F}$ avoiding an edge pair $F$ is rather involved and no longer admits a nice decomposition as its single failure counterparts.
Since understanding the structure of replacement paths and their interactions proved to be fundamental when designing fault resilient structures, understanding the structure $P_{s,v,F}$ is key essential step for making the desired jump from a single failure to at least two, for many network design tasks.
%\textbf{MP: below is a repetition, maybe it is better written..}
%Lacking this structural understanding for more than single failure, has been a barrier for extending a vast majority of fault tolerant structures, algorithms and distance oracle to support more than single failure.
A remarkable breakthrough in this direction is obtained in \cite{DP09}, presenting the first
$2$-sensitivity distance oracle of size $O(n^2 \log^3 n)$, capable of answering $2$-sensitivity queries in $O(\log n)$ time. Indeed, both the data structure and the query algorithm of \cite{DP09} are considerably more complex than the single failure case studied in \cite{DTCR08,BK09}.
An $f$-sensitivity distance oracle overcoming $f\geq 1$ failures is given in \cite{WY10}. By using fast matrix multiplication, \cite{WY10} yields the first sub-cubic
time (randomized) algorithm for the replacement paths problem when the edge-lengths are small integers.
Yet, despite the time efficient algorithm of \cite{WY10}
%for computing replacement paths that avoid multiple faults,
 the understanding of the underlying structural properties of these paths is still lacking.

We note that in certain cases the jump from one fault to multiple faults is quite natural and tractable.
For example, in the setting of \emph{fault tolerant spanners} for an arbitrary undirected weighted graph, it is shown in \cite{CLPR09-span} that there exists a (polynomially constructible) $f$-vertex fault tolerant $(2k-1)$-spanner of size $O(f^2 k^{f+1} \cdot n^{1+1/k}\log^{1-1/k}n)$ and an
$f$-edge fault tolerant $(2k-1)$-spanner of size $O(f\cdot n^{1+1/k})$ for a graph of size $n$.
A randomized construction attaining an improved tradeoff for vertex fault-tolerant spanners was shortly afterwards presented in \cite{DK11}.% yielding (with high probability) for every graph $G = (V,E)$, odd integer $s$ and integer $f$, an $f$-vertex fault-tolerant $s$-spanner
%with $O\left(f^{2-\frac{2}{s+1}}n^{1+\frac{2}{s+1}}\log{n}\right)$ edges. %Note that the ``cost'' of adding fault-tolerance in the setting of spanners is often low (e.g., merely polylogarithmic in $n$). As already shown in \cite{PPFTBFS13}, it turns out that our insistence on exact distances in $\FTBFS$ structures plays a dominant role and makes the problem significantly harder,
%outweighing our willingness to settle for a ``single source'' solution (compared to the approximate all-pairs solution of FT-spanners).\\
\par Finally, observe that the dual-failure FT-BFS structure studied in this paper is limited in three senses: (1) it is rather dense, although it matches the lower bound, (2) it deals with a single source, and (3) it supports up to two edge faults.
Given the density of the structure (i.e., (1)), one may claim that it may be better to use approximate structures as provided in \cite{PPFTBFSSODA14,PPFTBFDISC14} for example, instead of exact ones.
While this is true, we believe that it is still very important to understand the more fundamental exact problem first. The "theory" of paths avoiding two faults provided in this paper would surely be a key building block for designing \emph{approximate} structures that avoids two faults (e.g., in the same manner that the theory of single fault replacement paths of \cite{PPFTBFS13} laid the basis for approximate structures avoiding single fault in \cite{PPFTBFSSODA14,PPFTBFDISC14}).
%\textbf{MP: end of new lines}
In particular, we believe that understanding the single source case, beyond the single edge failure event, is an important milestone for designing fault tolerant structures under more generalized settings: One axis of generalization is increasing the number of supported sources, i.e., considering a setting where one is given a subset of sources $S \subseteq V$, and it is desired to provide a dual failure FT-BFS tree rooted at each source $s\in S$. Multi-source FT-BFS structures, referred to hereafter as \FTMBFS\, have been studied in \cite{PPFTBFS13} for the case of a single edge (or vertex) failure and have been later shown to provide an important building block in designing sparse \emph{fault tolerant additive spanners}, that provide a bounded additive stretch for \emph{all} pairs in the graph under the failing of a single edge (or vertex) \cite{PPFTBFDISC14}. An additional axis of generalization is increasing the number of supported faults. A natural generalized structure is an $f$-\FTBFS\, which contains the collection of all single source replacement paths avoiding up to $f$ edges in the graph. Combining these two axes results in $f$-\FTMBFS\ structure, that for a given source set $S \subseteq V$ provides an $f$-\FTBFS\ structure with respect to each source $s \in S$.
We believe that the structural theory of dual failure replacement paths developed in this paper paves the way to understanding these generalized structures. Towards this end, we provide two results for the generalized setting, namely, lower bound constructions and approximability results, as elaborated in next paragraph.
%%%%%%%%%%%%%%%%%%%%%%%%%
\dnsparagraph{\textbf{Contributions}}
%%%%%%%%%%%%%%%%%%%%%%%%%
We present an algorithm that for every
$n$-vertex unweighted graph $G$ and source node $s$, constructs a dual failure
\FTBFS\ structure rooted at $s$ with $O(n^{5/3})$ edges.
The size analysis of the output subgraph requires a deep understanding of the various configurations that may be assumed by a replacement path avoiding two faults.
An essential component in our analysis deals with the detour segment of the single failure replacement paths. %Specifically, we provide a structural classification (or glossary) of these detours and describe simplifying rules for each class.
%We believe that understanding the structure of dual failure replacement paths may decrease the current barrier of overcoming more than one fault and may ease the extension of many algorithms and data structures that currently can cope with only single failure events.
%
While a tight universal upper bound on the size of $f$-fault \FTBFS\ structures for general $f\geq 1$ is currently beyond our reach, we do have several results for the case of $f$ failures for any constant $f\geq 1$.
In Section \ref{sec:lb}, we present a lower bound stating that for every cardinality of sources $1\leq \NSource \leq n$, there exists an $n$-vertex graph and a source set $S \subseteq V$ where $|S|=\NSource$, for which any
$f$-fault \FTMBFS\ structure for each $s \in S$ requires $\Omega(\NSource^{1/(f+1)} \cdot n^{2-1/(f+1)})$ edges. Hence, for $f=2$ and $\NSource=1$ the lower bound translates into $\Omega(n^{5/3})$ edges, which matches our upper bound construction.
Finally, note that while our upper bound algorithm matches the worst-case lower bounds,
they might still be far from optimal for certain instances, see \cite{FTBFSArxiv}. Consequently, in Section \ref{sec:opt}, we complete the upper bound analysis by presenting an $O(\log n)$ approximation algorithm for the Minimum \FTMBFS\ problem in which one is given a graph $G=(V,E)$, constant integer $f \geq 1$, a source set $S \subseteq V$, and it is required to construct an $f$-failure \FTMBFS\ subgraph $H$ of minimum size (i.e., number of edges). This approximation algorithm is superior in instances where the graph enjoys
a sparse $f$-failure \FTMBFS\ tree (even linear in $O(n)$), hence paying $O(\NSource^{1/(f+1)} \cdot n^{2-1/(f+1)})$ edges is wasteful.
%In light of the hardness result for these problems (Theorem 1 of \cite{PPFTBFS13}), the approximability result is tight (up to constants).
\begin{theorem}[Upper Bound for dual failure FT-BFS]
\label{thm:upper}
For every unweighted undirected graph $G=(V,E)$ and source vertex $s \in V$, there is a (polynomially time constructible) dual failure FT-BFS structure $H \subseteq G$ with respect to $s$, with $O(n^{5/3})$ edges.
\end{theorem}

\begin{theorem}[Lower Bound for $f$-failure FT-MBFS]
\label{thm:lower}
For every constant $f \geq 1$, $n \geq o(1)$ and $1\leq \NSource \leq n$, there exist an $n$-vertex graph $G(V, E)$ and a source  set $S \subseteq V$ of cardinality $\NSource$ such that any $f$-\FTMBFS\ structure for the
source set $S$ has $\Omega(\NSource^{1-1/(f+1)} \cdot n^{2-1/(f+1)})$ edges. In particular, dual failure \FTBFS\ structures requires $\Omega(n^{5/3})$ edges.
\end{theorem}

%\textbf{MP: not sure if to give it such a dedicated space, since it is a natural extension from \cite{PPFTBFS13}}
\begin{theorem}[$\Theta(\log n)$-approximation for $f$-failure FT-BFS]
\label{thm:approx}
There exists a polynomial time algorithm that for every constant $f\geq 1$ and $n$-vertex graph $G$
and source set $S \subseteq V$ constructs an $f$-failure \FTMBFS\ structure $H$ whose size (i.e., number of edges) is larger by a factor of at most $\Theta(\log n)$ than the optimal structure $H^*$ (by Thm. 1 of \cite{PPFTBFS13}, this is tight up to constants, assuming $P\neq NP$).
\end{theorem}
\dnsparagraph{\textbf{Preliminaries and notation}}
Given an unweighted undirected graph $G=(V,E)$ and a source node $s$, let $T_0(s) \subseteq G$ be a shortest paths (or BFS) tree rooted at $s$. Throughout, the edges of these paths are considered to be directed away from the source node $s$.
For a path $P=[v_1, \ldots, v_k]$, let $\LastE(P)$ be the last edge of path $P$. Let $|P|$ denote the length of the path and $P[v_i, v_j]$ be the subpath of $P$ from $v_i$ to $v_j$. For paths $P_1$ and $P_2$, $P_1 \circ P_2$ denote the path obtained by concatenating $P_2$ to $P_1$.
A vertex $w$ is a \emph{divergence point} of the $s-v$ paths  $P_1$ and $P_2$ if $w \in P_1 \cap P_2$ but the next vertex $u$ after $w$ (i.e., such that $u$ is closer to $v$) in the path $P_1$ is not in $P_2$.
Given an $s-v$ path $P$ and an edge $e=(x,y) \in P$, let $\dist(s, e, P)$ be the distance (in edges) between $s$ and $e$ on $P$.
%%%%%%%%%%%%%%%%%%%%%%%%%%%%%%%%%%%%%%%%%%%%%%%%%%%%%%
\dnsparagraph{\textbf{Techniques and proof outline}}
%%%%%%%%%%%%%%%%%%%%%%%%%%%%%%%%%%%%%%%%%%%%%%%%%%%%%%
For a source node $s$, a target node $v$ and a pair $F=\{e_i,t_j\}\subseteq G$ of failed edges, the shortest $s-v$ path $P_{s,v,F}$ that does not go through the edge pair $F$ is the natural extension of the well studied single failure \emph{replacement path}. Thus, our dual failure \FTBFS\ structure must contain some replacement path $P_{s,v,F}$ for every $v \in V(G)$ and every edge pair $F \subseteq E(G)$.
%If no edge of $F$ lies on $\pi(s,v)$, then the replacement path $P_{s,v,F}$ is simply $\pi(s,v)$. If however at least one of the edges, say $e_i \in F$ lies on $\pi(s,v)$, a replacement path $P_{s,v,\{e_i\}}$
%
It is convenient to view the failing edges $F=\{e_i,t_j\}$ as corresponding to two subsequent independent failing events where first the edge $e_i$ fails and later on, the second edge $t_j$ fails.
If the first failing edge $e_i$ does not lie on the $s-v$ shortest-path $\pi(s,v)$, then the replacement path $P_{s,v,\{e_i\}}$ is simply $\pi(s,v)$. Otherwise, when $e_i \in \pi(s,v)$, the replacement path $P_{s,v,\{e_i\}}$ consists of a prefix of $\pi(s,v)$ followed by a detour $D_i$ avoiding $\pi(s,v)$ (and $e_i$), followed by a suffix of $\pi(s,v)$. Consider now the second failing edge $t_j$. Clearly, if $t_j$ is not on $P_{s,v,\{e_i\}}$ then the dual failure replacement path $P_{s,v,F}$ remains as is, i.e.,$P_{s,v,F}=P_{s,v,\{e_i\}}$. The interesting case is where $t_j \in P_{s,v,\{e_i\}}$. This case is further divided into two subcases. In the first subcase, $t_j$ appears on either the prefix or the suffix segments of $P_{s,v,\{e_i\}}$, i.e., $t_j$ appears on $\pi(s,v)$. A replacement path $P_{s,v,F}$ protecting against two faults on $\pi(s,v)$ is called hereafter a $(\pi,\pi)$\emph{-replacement path}. In the complementary subcase, the second failing edge $t_j$ appears on the detour segment $D_i$, i.e., $t_j \in P_{s,v,\{e_i\}}\setminus \pi(s,v)$. A replacement path $P_{s,v,F}$ for $F=\{e_i,t_j\}$ where $e_i$ lies on $\pi(s,v)$ and the $t_j$ lies on the detour $D_i$ is called hereafter a $(\pi,\sf D)$\emph{-replacement path}. Our algorithm for constructing the dual failure \FTBFS\ structure, Alg. $\ConstPath$, carefully selects a replacement path $P_{s,v,F}$ for every $v \in V$ and for every edge pair $F \subset E$.
Essentially, for each vertex $v$, the algorithm constructs a subgraph $H(v)$ consisting of the \emph{last edges} of the replacement paths $P_{s,v,F}$, i.e., $H(v)=\bigcup_{F \subseteq E, |F|\leq 2}\LastE(P_{s,v,F})$ where $\LastE(P_{s,v,F})$ is the last edge of the replacement path $P_{s,v,F}$. The final structure $H$ is then given by taking the union, i.e., $H=\bigcup_{v \in V} H(v)$. In the analysis section, we show that (a) taking the last edge of each replacement path is \emph{sufficient} and (b) the size (number of edges) of each $H(v)$ is bounded by $O(n^{2/3})$. A replacement path $P_{s,v,F}$ is called a \emph{new-ending} path if its last edge was not present in the structure at the time that the path was selected by the algorithm. \footnote{Note that in \cite{PPFTBFS13}, a path is new-ending if its last edge is not in the initial BFS tree $T_0(s)$. Here the definition is more strict and depends on the time step in which the path was considered by the algorithm. Yet, since the initial graph $H_0$ used by the algorithm at step $0$ contains the BFS tree $T_0(s)$, a new-ending path in the current definition, is also new-ending according to the definition of \cite{PPFTBFS13} (but not vice-versa).}
Since only the last edges of the replacement paths are taken into the structure, it is required to bound the number of new-ending paths $P_{s,v,F}$. Indeed, the lion share of this paper is dedicated to bounding the size of $H(v)$, which turns out to be significantly more involved compared to the single failure case of \cite{PPFTBFS13}.
We first consider the simplified case where the two failing edges lie on $\pi(s,v)$ and bound the number
of new-ending $(\pi,\pi)$-replacement paths by $O(\sqrt{n})$. This is shown by using a very similar argument to that of the single failure case.
The most technically involved task is bounding the number of new-ending $(\pi,\sf D)$-replacement paths $P_{s,v,F}$. We classify these paths into two main classes. The first class consists of the paths $P_{s,v,F}$ that do not intersect the edges of the detour of the replacement path protecting their \emph{first} failing edge. A new-ending path in this class has the following structure: it diverges from the shortest-path $\pi(s,v)$ at some vertex $b$ (above the failing edge $e_i$) and joins $\pi(s,v)$ again only at the final vertex $v$, without intersecting the detour $D_i$ at all (see Fig. \ref{fig:apathtype}(d)). The second class consists of new-ending paths $P_{s,v,F}$ that intersect their detour $D_i$ in at least one edge. Any  path in this class has the following structure: it diverges from the shortest-path $\pi(s,v)$ at the first vertex of the detour $D_i$, it then follows the detour $D_i$ up to some vertex $c$ above the failing edge $t_j$, and joins $\pi(s,v)$ and $D_i$ again only at the final vertex $v$. In other words, such a path has two divergence points: a unique $\pi$-divergence point $b$ where it departs from $\pi(s,v)$ and a $\sf D$-divergence point $c$ where it departs from $D_i$ (see Fig. \ref{fig:apathtype}(c))

\begin{figure}[htbp]
\begin{center}
\includegraphics[width=5in]{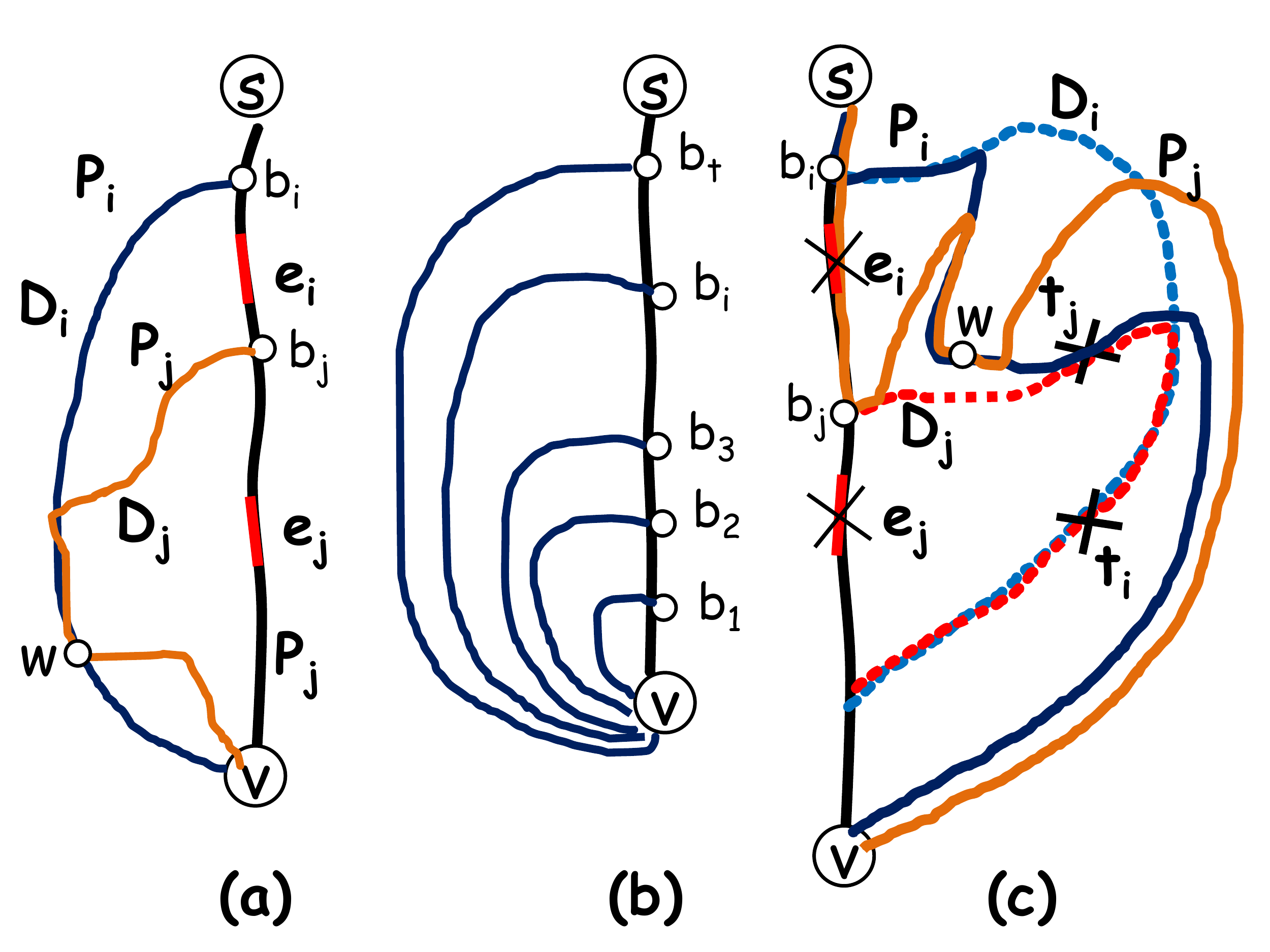}
\end{center}
\caption{(a) The single failure case.
The suffixes $\widetilde{P}_i=P_i[b_i,v]$ are disjoint. The existence of a vertex $w$ in the intersection implies that there are two safe routes from $w$ to $v$ (of the same lengths). Note that these routes are safe since they do not intersect with the edges of $\pi(s,v)$. (b) The disjointness of the suffixes implies that the divergence points are distinct and hence can be ordered on the $\pi(s,v)$ path in increasing distance from $v$. (c) The dual failure case. Shown are two $(\pi, \sf D)$ replacement paths $P_i=P_{s,v,\{e_i,t_i\}}$ and $P_j=P_{s,v,\{e_j,t_j\}}$ where $t_i \in D_i$ and $t_j \in D_j$. The suffixes $\widetilde{P}_i$ and  $\widetilde{P}_j$ intersect at the common vertex $w$, however the subpath $P_i[w,v]$ contains the failing edge $t_j$ and hence cannot be used by the path $P_j$.
\label{fig:singlef}}
\end{figure}

We proceed by briefly outlining the proof for the single failure case, i.e., bounding the number of $s-v$ new-ending $P_{s,v,\{e_i\}}$ paths by $O(\sqrt{n})$. We then consider the simplifying case where all replacement paths in $G \setminus F$ are unique (there are no two equally shortest replacement-paths). Finally, we highlight the technicalities that arise in the general case (whose detailed treatment is deferred to Section \ref{sec:analysis}).
\dnsparagraph{\textbf{Recap for the single failure case and first attempt}}
Assume that all shortest-paths are computed according to a weight assignment $W$ that guarantees the uniqueness of the shortest-paths (i.e, breaking ties in a consistent manner).
Consider the collection $P_1, \ldots, P_t$ of $s-v$ new-ending replacement paths where $P_i=P_{s,v,\{e_i\}}$ for $e_i \in \pi(s,v)$ and every path $P_i$ ends with a distinct edge of $v$, i.e., $\LastE(P_i)\neq \LastE(P_j)\neq \LastE(\pi(s,v))$ for every $i, j \in \{1, \ldots, t\}$.\footnote{The replacement-paths are computed according to the weight assignment $W$ that breaks the shortest-path ties. Since only the last edges of each replacement-path are taken into the structure, in our analysis, we consider one representative replacement-path for each new edge incident to $v$.} We now bound $t$ by $O(\sqrt{n})$ and as there are $n$ vertices, overall there are total of $O(n^{3/2})$ edges in a \FTBFS\ structure that contains the last edges of all replacement paths. For every path $P_i$, let $b_i$ be the unique divergence point from $\pi(s,v)$.
The following observation is crucial in this context. 
\begin{observation}
\label{obs:singlef}
The suffixes $\widetilde{P}_i=P_{i}[b_i,v]\setminus \{v\}$ are vertex-disjoint, i.e., $\widetilde{P}_i \cap \widetilde{P}_j=\emptyset$ for every $i, j \in \{1,\ldots,t\}$.
\end{observation}
\Proof
Since $b_i$ is the unique divergence point of $P_i$ from $\pi(s,v)$, it holds that $ \widetilde{P}_i \cap E(\pi(s,v))=\emptyset$, for every $i \in \{1,\ldots,t\}$.
Assume towards contradiction that there exists a common vertex $w \in \widetilde{P}_i \cap \widetilde{P}_j$ in the intersection. For an illustration see Fig. \ref{fig:singlef}(a).
This implies that there two distinct $w-v$ paths, namely, $P_i[w,v]$ and $P_j[w,v]$ in $G \setminus E(\pi(s,v))$, leading to contradiction by the uniqueness of $W$. (Informally, since in this case the failing edge $e_i$ protected by $P_i$ is not on $P_j[w,v]$ and vice-versa, it implies that one of the last edges, namely, $\LastE(P_i)$ or $\LastE(P_j)$ can be avoided in the structure.)
The observation follows.
\QED
In particular, by Obs. \ref{obs:singlef}, we have that the collection of divergence points $b_1, \ldots, b_t$ are \emph{distinct}. For an illustration see Fig. \ref{fig:singlef}(b).
This allows us to order the paths $P_{1}, \ldots, P_{t}$ in increasing distance between $b_{j}$ and $v$ where $\dist(b_{1},v,G)<\ldots<\dist(b_{t},v,G)$.
For every $j \in \{1, \ldots, t\}$, we then have that $|\widetilde{P}_{j}|\geq \dist(b_{j},v,G)\geq j-2$.
Finally, by exploiting the disjointness of the suffixes, we can bound the total number of vertices occupied by these suffixes, by $n \geq |\bigcup_{j=1}^t \widetilde{P}_{j}|=\sum_{j=1}^t|\widetilde{P}_{j}|\geq \sum_{j=1}^t j-2=\Omega(t^2),$ hence $t=O(\sqrt{n})$. In addition, by Obs. \ref{obs:singlef}, it also holds that $t \leq \dist(s,v,G)$ and hence the \FTBFS\ structure contains $O(\min\{\sqrt{n}, D\}\cdot n)$ edges where $D$ is the depth of the BFS tree.

Unfortunately, when considering the dual failure case, the key observation, Obs. \ref{obs:singlef}, fails to hold. Consider two dual failure new-ending replacement paths $P_i=P_{s,v,\{e_i,t_i\}}$ and $P_j=P_{s,v,\{e_j,t_j\}}$ where $t_k$ is on the detour segment $D_k$ of $P_{s,v,\{e_k\}}$ for $k \in \{i,j\}$. In addition, since we only care for bounding the number of edges incident to $v$, these paths are selected so that each ends with a new and distinct edge, i.e., $\LastE(P_i)\neq \LastE(P_j)\neq \LastE(\pi(s,v))$.
Let $b_i$ (resp., $b_j$) be the unique divergence point of $P_i$ (resp., $P_j$) from $\pi(s,v)$. By definition it holds that the suffix $\widetilde{P}_k=P_k[b_k,v]\setminus \{v\}$ is disjoint with $\pi(s,v)$ for both $k \in \{i,j\}$.
Yet, in contrast to the single failure case, we can no longer show that these suffixes are disjoint.
To see this, assume there exists a common vertex $w$ in the intersection where $w \in \widetilde{P}_i \cap \widetilde{P}_j$. In the single failure case, since both failing edges $e_i$ and $e_j$ lie on $\pi(s,v)$, we had the guarantee that they do not appear on either of the segments $P_i[w,v]$ and $P_j[w,v]$. Hence, in such a case, the two $w-v$ subpaths $P_i[w,v]$ and $P_j[w,v]$ are interchangeable and safe to be used by both of the paths $P_i$ and $P_j$ (i.e., safe in the sense that they do not contain the failing edges of these paths). Unfortunately, in our case, since the second failing edge of $P_j$, namely, $t_j$, is not on $\pi(s,v)$ (but rather on the detour $D_j$), we no longer have such guarantee. Specifically, it might be the case that $t_j$ appears on the suffix $P_{i}[w,v]$ and hence the subpath $P_{i}[w,v]$ is no longer \emph{safe} for $P_j$, which justifies the introduction of the two new edges, $\LastE(P_i)$ and $\LastE(P_j)$. For an illustration see Fig. \ref{fig:singlef}(c).
This toy example illustrates that dual failure replacement paths may share many vertices, which makes the mission of bounding their number much less tractable.

%%%%%%%%%%%%%%%%%%%
%%%%%%%%%%%%%%%%%%%

%%%%%%%%%%%%%%%%%%
\dnsparagraph{Easy case (1) : $f$-faults on $\pi(s,v)$}
To warm up, we proceed by claiming that the collection of last edges of the $s-v$ replacement-paths protecting against at most $f$ faults on the shortest path $\pi(s,v)$ is bounded by $O(\sqrt{n})$.
Consider the collection of $s-v$ replacement paths  $\mathcal{P}_v=\{P_{s,v,F} ~\mid~ F\subseteq \pi(s,v) , |F| \leq f\}$ and let $\widehat{E}_v=\{\LastE(P) ~\mid~ P \in \mathcal{P}_v\}$.
\begin{lemma}
$|\widehat{E}_v|=O(\sqrt{n})$ for every $v \in V$.
\end{lemma}
\Proof
Fix $v$ and consider $\mathcal{P}'_v$, the collection of representative paths from $\mathcal{P}_v$, each ending with a distinct last edge,
i.e., $\LastE(P)\neq \LastE(P')$ for every $P,P' \in \mathcal{P}'_v$. For every $P \in \mathcal{P}'_v$, let $d(P)$ be the last divergence point from $\pi(s,v)$. We first claim that the suffix segments $P[d(P),v]$ are vertex disjoint besides the common endpoint $v$.
To see this, assume towards contradiction that there exists a mutual vertex $w$ in the intersection of $P[d(P),v]$ and $P'[d(P'),v]$ for two distinct paths $P,P' \in \mathcal{P}'_v$. Since $P$ and $P'$ ends with a distinct last edge, and as $d(P),d(P')$ are the last divergence points from $\pi(s,v)$, we get that there are two distinct $w-v$ paths
in $G \setminus E(\pi(s,v))$, namely, $P_1=P[w,v]$ and $P_2=P'[w,v]$, contradiction to the uniqueness of the shortest-paths.

We can then sort the paths $P$ in $\mathcal{P}'_v$ in increasing distance of $d(P)$ and $v$, and the argumentation follows the exact same line as for the single edge fault case (i.e., the $i$'th segment $P_i[d(P_i),v]$ is of length at least $i$ for every $i \in \{1, \ldots, |\widehat{E}_v|\}$ and these segments are vertex disjoint).
\QED

\dnsparagraph{Easy case (2) : small FT-diameter graphs}
Let $D_f(G)=\max\{\dist(s,v,G \setminus F) ~\mid~ F \subseteq E, |F|\leq f-1\}$ be the $f$-\emph{FT-diameter} of the graph $G$.
We proceed by claiming that graphs of small $f$-FT-diameter have relatively sparse $f$-\FTBFS\ structures (i.e., BFS structures that are resilient against the failing of at most $f$ edges.)
Since it is sufficient to collect the last edge from each replacement-path (by the same argument as for the single fault case), we have the following.
\begin{observation}
For every $n$-vertex graph $G$ and source vertex $s \in V$, there exists an $f$-\FTBFS\ structure $H\subseteq G$ with $O((D_f(G))^f \cdot n)$ edges.
\end{observation}

\section{Notation}
\label{sec:notation}
Given a graph $G=(V,E)$ and a source node $s$, let $T_0(s) \subseteq G$ be
a shortest paths (or BFS) tree rooted at $s$.
Let $\pi(s, v,T_0)$ be the $s-v$ shortest-path in tree $T_0$, when the tree $T_0=T_0(s)$, we may omit it and simply write $\pi(s,v)$.
%\\ {\bf Later, this notation sometimes appear as $\pi_{s, v}$. Change.} \\
Let $\Gamma(v, G)$ be the set of $v$ neighbors in $G$. Let $E(v,G)=\{(u,v) \in E(G)\}$ be the set of edges incident to $v$ in the graph $G$ and let $\deg(v,G)=|E(v,G)|$ denote the degree of node $v$ in $G$. When the graph $G$ is clear from the context, we may omit it and simply write $\deg(v)$.
Let $\depth(s, v) = \dist(s,v,G)$ denote the {\em depth} of $v$
in the BFS tree $T_0(s)$. When the source node $s$ is clear from the context,
we may omit it and simply write $\depth(v)$.
For a subgraph $G'=(V', E') \subseteq G$
(where $V' \subseteq V$ and $E' \subseteq E$)
and a pair of nodes $u,v \in V$, let $\dist(u,v, G')$ denote the
shortest-path distance in edges between $u$ and $v$ in $G'$.
Assuming an edge weight function $W: E(G)\to \REAL^{+}$, let $SP(s, v_i, G, W)$ be the set of $s-v_i$ shortest-paths in $G$ according to the edge weights of $W$.  If $W$ is a weight assignment that guarantees the uniqueness of the shortest paths, then we override the definition and let $P=SP(u,v, G, W)$ be the unique $u-v$ shortest path in $G$ according to $W$. Throughout, the edges of these paths are considered to be directed away from the source node $s$. The edges on any $s-v$ path $P$ are considered from the top $s$ to the bottom $v$, hence an edge $e_i \in P$ is a \emph{above} $e_j \in P$ if $e_i$ is closer to $s$ then $e_j$.
For an edge $e=(x,y)\in T_0(s)$, define $\dist(s,e)=i$ if $\depth(x)=i-1$ and $\depth(y)=i$.
A vertex $w$ is a \emph{divergence point} of the $s-v$ paths  $P_1$ and $P_2$ if $w \in P_1 \cap P_2$ but the next vertex $u$ after $w$ (i.e., such that $u$ is closer to $v$) in the path $P_1$ is not in $P_2$. We view the $\pi(s,v)$ path from top (i.e., $s$) to bottom $v$. An edge $e_i$ is said to be \emph{above} $e_j$, if it is closer to $s$ on the path $\pi(s,v)$.
%\begin{definition}
A subgraph $H$ is an $f$-\FTMBFS\ structure (multi-source FT-BFS) for $G$ with respect to a source set $S \subseteq V$, iff
$\dist(s, v, H\setminus F)=\dist(s, v, G \setminus F)$ for every vertex pair $(s,v) \in S \times V$ and every failing sequence $F \subseteq E$, $|F| \leq f$.
Single source $f$-\FTMBFS\ structures (with $|S|=1$)
are referred to here as $f$-\FTBFS\ structures. In addition, $f$-\FTMBFS\ structures with $|S|=1$ and $f=2$ are referred to here as dual failure \FTBFS\ structures.

%%%%%%%%%%%%%%%%%%%%%%%
\dnsparagraph{\textbf{A bit harder: the dual failure case with simplifying assumptions}}
We next sketch the size analysis for dual failure FT-structures, for a very degenerate case.
We focus on vertex $v \in V$ and show that it has at most $O(n^{2/3})$ edges in the final structure $H$.
The following notation is useful in our analysis.
For every $(\pi,\sf D)$ replacement-path $P=P_{s,v,\{e_i,t_i\}}$, let $D(P)=P_{s,v,\{e_i\}}\setminus \pi(s,v)$ be the detour segment of $P_{s,v,\{e_i\}}$ such that $t_i \in D(P)$ (including the endpoints on $\pi(s,v)$),
let $F(P)=\{e_i,t_i\}$ be the failing edges protected by $P$, $F_1(P) =e_i \in \pi(s,v)$ and $F_2(P)=t_i \in D(P)$.
For two $s-v$ $(\pi,\sf D)$-replacement paths $P_i,P_j$, we say that $P_i$ \emph{interferes} with $P_j$, if $F_2(P_j) \in P_i \setminus D(P_i)$. The $(\pi,\sf D)$ paths $P_i,P_j$ are \emph{independent} if $P_i$ does not interfere with $P_j$ and vice-versa. For a fixed $v \in V$, define $H(v)=\{\LastE(P_{s,v,F}) ~\mid~ F \subset E, |F| \leq 2\}$ as the collection of last edges of all $s-v$ replacement-paths where $P_{s,v,F}$ is the (unique) $s-v$ shortest-path in $G \setminus F$.
%In general, this is \emph{not} the case and a crucial component of our scheme is to carefully select the replacement-paths so that the upper bound would hole. For the sake of conversation, we defer these details to Appendix Sec. \ref{append:analysis}, where we describe Alg. $\ConstPath$ which carefully select the replacement paths $P_{s,v,F}$.
%Let $T_0$ be a BFS tree rooted in $s$ and for path $P$ define $\LastE(P)$ to be the the last edge of $P$. Let $\pi(s,v)$ be the $s-v$ shortest-path in $G$.
%For a pair of edges $F=\{e_i,e_j\} \in E$, let $P_{s,v,F}$ be the (unique) $s-v$ shortest-path in $G \setminus F$. The dual failure FT-BFS structure $H$ is given by
%$$H=\bigcup_{v \in V} \bigcup_{F \subset E, |F|\leq 2} \LastE(P_{s,v,F}) \cup T_0.$$
%In the analysis section, we show that it is sufficient to include in the structure only the last edges of the replacement paths. The proof is similar to the single edge case, and hence it is deferred to appendix.
%
%To bound the number of edges in $H$, we show that the every vertex $v$ contributes $O(n^{2/3})$ edges to the structures.
%
%Let $\mathcal{P}^1_v=\{P_{s,v,e}, e \in \pi(s,v)\}$ be the collection of $s-v$ replacement paths protecting against single fault and let  $\mathcal{P}^2_v=\{P_{s,v,\{e_i,e_j\}}\}$
It is sufficient to consider one representative replacement-path for each new edge of $v$ in $H(v)$.  Hence, assume throughout, that the last edge of each path in the collection of new-ending $s-v$ replacement paths is \emph{distinct}.
Since bounding the collection of $(\pi,\pi)$ replacement-paths (protecting against two edges faults on $\pi(s,v)$) is very similar to the single fault case, we restrict attention to the more technically challenging part of bounding $(\pi,\sf D)$-paths.
%To get the flavor of the analysis, we now restrict attention to the collection of paths $P_{s,v,e_i,e_j}$ whose detour $D(P_{s,v,e_i,e_j})$ are \emph{edge disjoint}.
%Since only the last edge of each replacement path is taken into the structure, it is sufficient to take one representative replacement-path for each such edge and we assume the each replacement path in our collection is ended with a distinct edge. Towards the end of this section, we show that there are at most $O(n^{2/3})$
%such paths.
We now bound $H(v)$ in the special case obtained by making the following simplifying assumptions:
(S1) all $s-v$ replacement paths in $G \setminus F$ are \emph{unique} for every $F \subseteq E$, $|F|\leq 2$,
(S2) the detour segments $D_i$ of the $s-v$ single edge replacement paths $P_{s,v,\{e_i\}}$, $e_i \in \pi(s,v)$, are edge disjoint, and
(S3) all replacement-paths are \emph{independent}.
We then classify the $(\pi,\sf D)$ replacement paths into two classes depending on whether or not they intersect their detour (i.e, the detour that protects their first failing edge and contains their second failing edge). Let $\mathcal{P}_{nodet}$  be the subset of replacement paths $P$ that do not intersect the edges of their detours and let $\mathcal{P}_{inter}$ be the remaining paths.
%Each class is bounded separately.
\dnsparagraph{\textbf{$(\pi,\sf D)$-paths that do not intersect their detour}}
%We now consider the set of replacement-path $P_{s,v,\{e_i,e_j\}}$ such that $e_i \in \pi(s,v)$ and $e_j$ occurs on the detour segment $D$ of $P_{s,v,\{e_i\}}$ and in addition $P_{s,v,\{e_i,e_j\}}$ do not intersect the edges of $D$.
%Let $P_1, \ldots, P_N \in \mathcal{P}_{nodet}$ be the collection of $(\pi,\sf D)$ paths, such that $P_i$ and $D(P_i)$ are edge-disjoint and $\LastE(P_i) \neq \LastE(P_j)$ for $i \neq j \in \{1, \ldots,N\}$.
To bound this  class, it is sufficient to use assumptions (S1) and (S2).
We begin by noting that each path $P_i \in \mathcal{P}_{nodet}$ protects a distinct edge on $\pi(s,v)$.
Order these paths $\mathcal{P}_{nodet}=\{P_1, \ldots, P_{N}\}$ in increasing distance between $s$ and $F_1(P_j)$, i.e.,\\ $\dist(s, F_1(P_1))<\ldots<\dist(s, F_1(P_{N}))$. Let $e_i=F_1(P_i)$ and $M=\lfloor N/2 \rfloor$. We now restrict attention to the set of first $M$ paths $\mathcal{P}_M=\{P_{j} ~\mid~ 1 \leq j \leq M\}$.  Let $\mathcal{D}_M=\{D(P) ~\mid~ P \in \mathcal{P}_M\}$ be the collection of their corresponding detours and let $V(\mathcal{D}_M)=\bigcup_{D \in \mathcal{D}_M} V(D)$.
The paths of $\mathcal{P}_{nodet}$ are classified into two classes depending on whether or not they intersect the edges of
$\mathcal{D}_M$. In a way similar to the proof of the single failure case, one can show that there are $O(\sqrt{n})$ paths in $\mathcal{P}_M$ that do \emph{not} intersect the edges of $\mathcal{D}_M$. Hence, it remains to bound the remaining paths in $\mathcal{P}_M$. For every such  path $P_i$, let $a_i$ be their last mutual vertex in $V(\mathcal{D}_{M})\setminus \{v\}$. Again, by the uniqueness of the shortest-path, we can show the following.
\begin{lemma}
\label{obs:a_dist}
(a) $F_1(P_i) \neq F_1(P_j)$ for every $P_i, P_j \in \mathcal{P}_{nodet}$ and hence by (S2) $D(P_i)$ and $D(P_j)$ are edge disjoint;
(b) $V(P_i[a_i,v]) \cap V(P_j[a_j,v])=\emptyset$.
\end{lemma}
We next classify the detours $D_i \in \mathcal{D}_M$ according to their lengths. A detour $D_i$ is \emph{expensive} if $|D_i|\geq M/2$, otherwise it is \emph{cheap}. Next, the new-ending paths $P_i \in \mathcal{P}_M$ that intersect $\mathcal{D}_M$ are classified according to the detour $D(a_i)$ on which $a_i$ (the last common vertex of $P_i\setminus \{v\}$ and $V(\mathcal{D}_M)$) appears. Then $P_i$ is expensive (resp., cheap) if $D(a_i)$ is expensive (resp., cheap). Let $\mathcal{P}_{cheap}=\{P_i \in \mathcal{P}_M~\mid~ D(a_i) \mbox{~is cheap~}\}$ and $\mathcal{P}_{expen}=\{P_i \in \mathcal{P}_M~\mid~ D(a_i) \mbox{~is expensive~}\}$. We next separately bound $|\mathcal{P}_{cheap}|$ and $|\mathcal{P}_{expen}|$.
\begin{claim}
\label{cl:cheap_main}
$|\mathcal{P}_{cheap}|=O(\sqrt{n})$.
\end{claim}
\Proof
Let $V_{cheap}=\bigcup_{P_i \in \mathcal{P}_{cheap}}V(P_i[a_i, v])\setminus \{v\}$.
By Obs. \ref{obs:a_dist},
$|V_{cheap}|=\sum_{P_i \in \mathcal{P}_{cheap}} (|P_i[a_i, v]|-1)$. We now focus on some $P_i$ and show that $|P_i[a_i, v]|\geq M/2$. First note that $P_i[a_i, v]$ and $\pi(s,v)$ are vertex disjoint (except for the common endpoint $v$), as $a_i$ occurs after the unique $\pi$-divergence point of $P_i$ from $\pi(s,v)$. Hence,
\begin{equation}
\label{eq:nodet1_main}
|P_i[a_i, v]|\geq \dist(a_i, v, (G \setminus V(\pi(s,v))) \cup \{v\})~.
\end{equation}
Let $D_j=D(a_i) \in \mathcal{D}_M$ be the detour protecting against the failing of the edge $e_j$. Then,
\begin{eqnarray}
\label{eqn:nondet2_main}
\dist(x_j, v, G \setminus \{e_j\})&\geq& \dist(x_j, v, G)
\geq \dist(e_j, v) \geq M~,
\end{eqnarray}
where the penultimate inequality follows as $x_j$ appears above the failing edge on $\pi(s,v)$ and last inequality follows by the fact that $D_j \in \mathcal{D}_M$.
Since $a_i$ appears on a cheap detour $D_j$, we get that
%\begin{eqnarray}
%\label{eqn:nondet33_main}
$\dist(x_j,a_i, G \setminus \{e_j\})\leq |D_j| < M/2~,$
%\end{eqnarray}
and combining this with Eq. (\ref{eqn:nondet2_main}), we get that $\dist(a_i, v, G \setminus \{e_j\})\geq M/2$.
By combining with Eq. (\ref{eq:nodet1_main}), we get that overall $|P_i[a_i, v]|\geq M/2$. We therefore have that $M/2 \cdot |\mathcal{P}_{cheap}| \leq |V_{cheap}|\leq n$. It follows that $|\mathcal{P}_{cheap}|\leq 2n/M$. Since clearly also $|\mathcal{P}_{cheap}|\leq M$, we have $|\mathcal{P}_{cheap}|\leq \min\{M, 2n/M\}\leq \sqrt{2n}$. The claim follows.
\QED
\begin{claim}
\label{cl:expens_detours_main}
$|\mathcal{P}_{expen}|=O(n^{2/3})$.
\end{claim}
\Proof
Let $\mathcal{D}_{expen}=\{D_j \in \mathcal{D}_M ~\mid~ |D_j|\geq M/2\}$ be the collection of
expensive detours, $z=|\mathcal{D}_{expen}|$.
We now classify the expensive paths of $\mathcal{P}_{expen}$ into $z$ classes where each path $P_i$ is mapped to the class of the detour $D_j \in \mathcal{D}_{expen}$ on which $a_i$ appears.
For every $D_j \in \mathcal{D}_{expen}$,
let $\mathcal{P}_j=\{P_i \in \mathcal{P}_{expen} ~\mid~ D(a_i)=D_j\}$, and let $N_j=|\mathcal{P}_j|$ be the cardinality of this set.
\par We begin by bounding the number of vertices appearing in the expensive detours, let $V_{\sf D}=\bigcup_{D_j \in \mathcal{D}_M ~\mid~ |D_j| \geq M/2} V(D_j)$ be the vertices appearing on the expensive detours. By edge-disjointness of the detours (assumption (S2)), we get that $|V_{\sf D}| \geq z \cdot (M/2-2)$.
We now proceed by bounding the number of vertices appearing on the expensive replacement paths,
$V_P=\bigcup_{P_i \in \mathcal{P}_{expen}}V(P_i[a_i,v])\setminus \{a_i, v\}.$
Note that for every expensive path $P_i$, its segment $P_i[a_i,v]$ is vertex disjoint (expect for its endpoints $a_i$ and $v$) with the vertex set $V_{\sf D}$. Fix some $j \in \{1, \ldots, z\}$, with $N_j$ expensive paths $\mathcal{P}_j$. We now claim that $V_j=\bigcup_{P_i \in \mathcal{P}_j} P_i[a_i,v]$ contains $\Omega(N_j^2)$ vertices. By Cl. \ref{obs:a_dist}, the $P_i[a_i,v]$ segments are disjoint. Order the paths of $\mathcal{P}_j$ in increasing distance of $a_i$ from $v$. Since $a_i \in D_j$ for every $P_i \in \mathcal{P}_j$ and the $a_i$'s are distinct it holds that $|V_j|\geq (N_j-1)^2/2$ and summing over all $j$ (as the suffixes $P_i[a_i,v] \setminus \{v\}$ are disjoint) and using the Cauchy-Schwarz inequality, we get that $|V_P|\geq \sum_{j=1}^z (N_j-1)^2/2~$.
Recall that the sets $V_P$ and $V_{\sf D}$ are disjoint, and thus, we get that
$n \geq |V_P \cup V_{\sf D}|=|V_P|+|V_{\sf D}|\geq\sum_{j=1}^z (N_j-1)^2/2+ z \cdot M/2=\Omega(M^{3/2})~.$
We get $M = O(n^{2/3})$, as required.
\QED
\dnsparagraph{\textbf{$(\pi,D)$-paths that intersect their detour}}
%For every new-ending path $P_i=P_{s,v,F} \in \mathcal{P}_v$, recall that $D(P_i)=D_i$ is the detour segment protecting the first failing edge $F_1(P_i) \in \pi(s,v)$ of $v$.
We now consider the replacement-paths in $\mathcal{P}_{inter}$ that intersect the edges of their detour , under assumptions (S1-S3). For every $P_i \in \mathcal{P}_{inter}$, let $D_i=D(P_i)$, and $x_i,y_i \in \pi(s,v)$ be the first (resp., last) vertices of the detour $D_i$. Let $b_i$ (resp., $c_i$) be the first divergence point of $P_i$ and $\pi(s,v)$ (resp., $D(P_i)$). Let $e_i=F_1(P_i)$ be the first failing edge protected by $P_i$. It is easy to see that by the uniqueness of the shortest-paths, $x_i=b_i$ and $P_i=\pi(s, x_i) \circ D_i(x_i,c_i) \circ P_i[c_i, v]$. That is, $b_i$ and $c_i$ are unique divergence points from $\pi(s,v)$ and $D(P_i)$ respectively and thus the suffix $P_i[c_i,v]$ is edge disjoint with $\pi(s,v)$ and $D_i$. In addition, since the detour segments are disjoint (by assumption (S2)), we have the following.
\begin{observation}
\label{obs:dist_indep_easy}
For every two paths $P_i,P_j \in \mathcal{P}_{inter}$,
(1) $P'_i=P_i[c_i, v]\setminus \{v\}$ and $P'_j=P_j[c_j,v]\setminus \{v\}$ are vertex disjoint.
(2) If $e_i=e_j$ then $D(P_i)=D(P_j)$ and $c_i \neq c_j$.
\end{observation}
We now induce an $(e,c)$-ordering on the paths of $\mathcal{P}_{inter}$, which can be viewed as based on treating $e_i$ and $c_i$ lexicographically:
For $e_i\neq e_j$, we say that $(e_i,c_i)<(e_j,c_j)$ if $\dist(s,e_i,\pi(s,v))< \dist(s,e_j,\pi(s,v))$. For $e_i=e_j$, let $(e_i,c_i)<(e_j,c_j)$ if $\dist(x_i, c_i, D(P_i))<\dist(x_i, c_j, D(P_i))$. By Obs. \ref{obs:dist_indep_easy}, this is well defined.
We next order the paths of $\mathcal{P}_{inter}$ in increasing $(e,c)$ order. Let $\overrightarrow{\mathcal{P}}_{inter}=\{P_1, \ldots, P_{\ell}\}$ where $(e_1,c_1)< \ldots<(e_\ell,c_\ell)$. By showing that $F(P_j) \nsubseteq P_i$ for every $i <j$, we have that the lengths of the paths in the ordered set $\overrightarrow{\mathcal{P}}_{inter}$ are strictly monotone decreasing.
\begin{lemma}
\label{lem:ordering_simple}
$|P_1| > \ldots >|P_\ell|$
(or alternatively, if $(e_i,c_i)<(e_j,c_j)$ then $|P_i|>|P_j|$).
\end{lemma}
\Proof
Let $i<j \in \{1, \ldots, \ell\}$. We begin by showing that $F(P_j) \notin P_i$. Recall that $P_k=\pi(s,x_k) \circ D_k(x_k, c_k) \circ P_k[c_k,v]$ where $D_k=D(P_k)$ for $k \in \{i,j\}$. Let $x_k,y_k$ be the first (resp., last) vertices of the detour $D_k$ for $k \in \{i,j\}$.

Since $P_i$ diverges from $\pi(s,v)$ above $e_i$ which is not below $e_j$ , it holds that $e_j \notin P_i$. So, it remains to show that $F_2(P_j) \notin P_i$. Assume towards contradiction that $F_2(P_j)=(q_1,q_2)$ occurs on $P_i$. First, assume that $D_i=D_j$. Since $F_2(P_j) \in D_i$ and $P_i[c_i,v]$ is edge disjoint with $D_i$ (i.e. $c_i$ is a unique divergence point from $D_i$), it holds that $F_2(P_j) \in D_i[x_i, c_i]$. By the ordering $\dist(x_i, c_i)>\dist(x_i, c_j)$ and since $D_i[x_i, c_i] \subset D_i[x_i,c_j]\subset P_j$, we end with contradiction. Next, assume that $D_i \neq D_j$.
We show that in such a case there are two $q_2-v$ shortest paths in $G \setminus \{F(P_i), e_j\}$, namely, $Z_1=P_i[q_2,v]$ and $Z_2=D_j[q_2, y_j] \circ \pi(y_j, v)$, hence leading to a contradiction by the uniqueness of the shortest-paths. First, note that since $P_i$ is new-ending, indeed $Z_1 \neq Z_2$. Since $e_i$ is above $e_j$ on $\pi(s,v)$, $e_i \notin Z_2$ and since $F_2(P_i)\in D_i$ and $E(D_i)\cap E(D_j)$, it holds that $F(P_i) \notin P_j$. By the optimality of $P_i$ and $P_{s,v,\{e_j\}}$ it holds that $|Z_1|=|Z_2|$, leading to a contradiction by the uniqueness of the shortest-paths
in $G \setminus \{F(P_i), e_j\}$.

Assume towards contradiction that $|P_i|\leq |P_j|$. Since $P_i \neq P_j \in G \setminus F(P_j)$, we end with contradiction to the uniqueness of the $s-v$ shortest paths in $G \setminus F(P_j)$.
\QED

We now group the ordered paths $P_i$ of $\mathcal{P}_{inter}$ into classes depending on their $e_i$-value (i.e., the first failing edge they protect in $\pi(s,v)$). For every vertex $e_k \in \pi(s,v)$, let $N_k$ be the number of replacement paths in $\mathcal{P}_{inter}$ whose first failing edge is $e_k$. Let $z=|\dist(s,v)|$. By assumptions (S1-S3) and the ordering of Lemma \ref{lem:ordering_simple}, we get:
\begin{lemma}
\label{lem:multi}
For every $P_i \neq P_j \in \mathcal{P}_{inter}$:
(a) $V(P_i[c_i,v]) \cap V(P_j[c_j,v])=\{v\}$.
(b) If $F_1(P_i) \neq F_1(P_j)$, then $P_i[b_i, v]\setminus \{b_i,v\}$ and $P_{j}[b_j,v]\setminus\{b_j,v\}$ are vertex disjoint.
(c) The total number of vertices occupied by these paths is $\Omega(\sum_{i=1}^z \sum_{k=1}^i N_k+\sum_{i=1}^z N^2_i)$.
\end{lemma}
Note that $|\mathcal{P}_{inter}|=\sum_{i=1}^z N_i$.
Hence, by combining this with Lemma \ref{lem:multi}(c), we get that there are $O(n^{2/3})$ such paths. This completes the analysis for the simplified case.
\paragraph{Road map.}
We now provide a high level road map of the general proof of the upper bound \emph{without} assuming (S1-S3).
First, the algorithm needs to support the case where the replacement-paths are not unique and hence have to be carefully chosen. The guiding principle for selecting the desired replacement paths is to favor replacement-paths that diverge from $\pi(s,v)$ as close to $s$ as possible. Among these, the algorithm favors replacement-paths that diverge from their detour segment as early as possible. Second, when removing assumption (S2), one has to incorporate into the analysis the optional \emph{complex interactions} between detour segments. The main structural theory developed in this paper is meant to deal this complication.
Hence, a crucial step for understanding the structure of dual failure replacement paths is the understanding of the structure of the detours. To do that, we focus on pairs of detours $D_i$ and $D_j$ and classify their structural dependency into six classes (see Fig. \ref{fig:detours}).
We then provide some simplifying rules for each class that are frequently used in our argumentation. Quite interestingly, understanding the pairwise relation between the detours was sufficient in order to obtain an understanding of the \emph{global} picture (i.e., which might contain complex interaction between many detours).
Finally, removing assumption (S3) entails another major complication in our analysis. In particular, when two paths $P_i$ and $P_j$ interfere, Lemma \ref{lem:multi}(a) is no-longer guaranteed to hold. In our analysis, the set of interfering paths is further classified into two subsets by distinguishing between two types of interference, namely, $\pi$-\emph{interference} and $\sf D$-\emph{interference}. We show that each of these two classes imposes different structural constraints which allow us to bound their cardinality.
Our tool kit consists of two main components: (a) complete mapping of the pairwise interactions between detours and (b) a subgraph $\KernelGraph$ denoted hereafter as a \emph{kernel subgraph} that contains the entire required information from $G$ but has some convenient properties that facilitate the analysis. This structure is heavily based on the detour configuration machinery established in (a) (see Section \ref{sec:deturs}). For every vertex $v$, the kernel subgraph $\KernelGraph_v(\mathcal{D})$ is imposed on a given collection of detours $\mathcal{D}=\{D_1, \ldots, D_t\}$. Clearly the set of relevant faulty edges of the $(\pi, \sf D)$ replacement paths is given by the subgraph $G_v(\mathcal{D})=\pi(s,v) \cup \{D_i \mid D_i \in \mathcal{D}\}$. Quite surprisingly, we show that in order to analyze the structure of the new-ending $(\pi,\sf D)$ replacement paths, it is sufficient to consider the subgraph $\KernelGraph_v(\mathcal{D})$ which contains all the relevant faulty edges. The kernel graph is used, for example, to bound the number of replacement-paths that do not intersect their detours. For example, it is essential for establishing Lemma \ref{cl:cheap_main} and \ref{cl:expens_detours_main} without assuming (S2).
%Finally, our third tool (c) is a delicate classification of structural interference between replacement paths.
%This kernel subgraph is shown to have an elegant structure, which makes it a rather convenient tool in our analysis.
%
We note that these tools might be used in further contexts to pave the way to the future design of $f$-fault resilient structures for $f \geq 2$. Equipped with these tools, to bound the number of new-ending $(\pi, \sf D)$ paths, we employ the same high level strategy as taken for the single failure case: new-ending paths consume many vertices, and since the number of vertices is limited by $n$, the number of new-ending paths is bounded as well (as a function of $n$). To do that, we would like to show that every new-ending path has an nonnegligible number of distinct vertices, not appearing on any other path. The main technical question is to identify a subpath of the new-ending path that is guaranteed to be sufficiently long and disjoint from all others. Since our replacement paths may overlap and share many common vertices, towards achieving this goal we classify the new-ending paths into five classes and bound that size of each class separately.
For schematic illustrations of this classification, see Fig. \ref{fig:rpclass}. The size analysis of each class exploits the tools described above and provides a deeper understanding of the complex behavior of dual failure replacement paths.
\dnsparagraph{\textbf{Beyond two faults}}
In the current analysis, a crucial step for understanding the structure of dual-failure replacement paths is the understanding of the detour structure of single failure replacement paths. The understanding of $f$-failure replacement paths becomes much less tractable as the number of faults $f$ increases.
Consider for example the case of $f=3$.
In this case, there are two types of detours: (1) $\sf D_1$ detours, the detours of the
single failure replacement paths, e.g., $P_{s,v,\{e_i\}}\setminus \pi(s,v)$ for $e_i \in \pi(s,v)$; and (2) $\sf D_2$ detours, the detours of the dual failure replacement paths, e.g., $P_{s,v,\{e_i,t_j\}}\setminus P_{s,v,\{e_i\}}$ for $e_i \in \pi(s,v)$ and $t_j \in P_{s,v,\{e_i\}}\setminus \pi(s,v)$. It is then required to understand the interactions between two detours of type $\sf D_2$ as well as the interaction between a detour of type $\sf D_1$ and of type $\sf D_2$. The generalization of $(\pi,\pi)$ and $(\pi,\sf D)$ replacement path classification in the case of $3$ faults gives raise to the following classes:
(a): $(\pi,\pi,\pi)$ replacement paths protecting against three faults on $\pi(s,v)$; (b) $(\pi,\pi,\sf D_1)$ replacement paths protecting against two faults on $\pi(s,v)$ and one fault on a detour of type $\sf D_1$; (c) $(\pi,\sf D_1, \sf D_1)$ replacement paths protecting against single fault on $\pi(s,v)$ and two faults on $\sf D_1$ and (d) $(\pi,\sf D_1, \sf D_2)$ replacement paths protecting against single fault on $\pi(s,v)$, single fault on $\sf D_1$ detour and single fault on $\sf D_2$ detour.
By using similar arguments to the single failure case, the edges added due to type (a) replacement paths can be bounded by $O(n^{3/2})$ (and this can be generalized to any $f\geq 1$ faults). The main difficulty arises when considering the other types, as this calls for a deep understanding of the interactions between detours of type $\sf D_1$ and $\sf D_2$. For a general integer $f\geq 1$,  a detour $D'$ is said to be of type $\sf D_j$ for $j\in \{1,\ldots, f-1\}$, if there exists an $j$-failure replacement path $P_{s,v,F}$, $F=\{e_1,\ldots,e_j\}$ such that $D'=P_{s,v,F} \setminus P_{s,v,F'}$ for $F'=\{e_1,\ldots,e_{j-1}\}$.
It is then required to understand the interactions between detours of type $\sf D_1, \ldots, \sf D_{f-1}$.
An additional source of difficulty arises when attempting to generalize the notion of \emph{interference}. In the dual-failure case, we considered two types of interference, namely, $\pi$-interference and $\sf D$-interference and each such class called for different tools. In the case of general $f\geq 1$, one needs to consider many more options, e.g., interference of types $(\pi,\pi,\sf D_1)$, $(\pi,\sf D_{i_1},\sf \ldots, D_{i_k})$ for $i_1,...,i_k \in \{1, \ldots, f-1\}$ etc. Each such class may impose different structural constraints
which would eventually provide the basis for bounding its cardinality. We note that the lower-bound construction of $\Omega(n^{2-1/(f+1)})$ should give us some useful hints for attaining (hopefully a matching) upper bound.
%Theorem \ref{thm:upper} is established in Sec. \ref{sec:upper}

\section{Description and Analysis of Algorithm $\ConstPath$}
\label{sec:analysis}
%%%%%%%%%%%%%%%%%%%%%%%
In this section, we establish Thm. \ref{thm:upper}.
For useful notation, see Sec. \ref{sec:notation}.
We present an algorithm that given an unweighted undirected  $n$-vertex graph $G=(V,E)$ and a source $s \in V$, constructs a dual failure \FTBFS\ subgraph $H \subseteq G$. We then analyze the correctness of the algorithm and bound the size of the output structure. The size analysis of the subgraph $H$ constitutes the main technical contribution of this paper.
%We begin now describe the Algorithm for constructing the collection of replacement paths $P_{s,v,F}$ for every pair $e_i,e_j \in E$ and every target $v$.
\paragraph{Algorithm $\ConstPath$.}
%\label{sec:const_paths}
Let $W$ be a weight assignment that guarantees the uniqueness of the shortest-paths.\footnote{Note that the given graph is unweighted and the fractional weights of $W$ only break the unweighted shortest-path ties in a consistent manner.} Let $T_0=\bigcup_{v \in V} \pi(s,v)$ be the BFS tree rooted at $s$ where $\pi(s,v)$ is the shortest path from $s$ to $v$ in $G$, namely, $\pi(s,v)=[s=u_0, u_1, \ldots, u_\ell=v]=SP(s,v, G, W)$. \par
\begin{figure}[htbp]
\begin{center}
\includegraphics[width=4in]{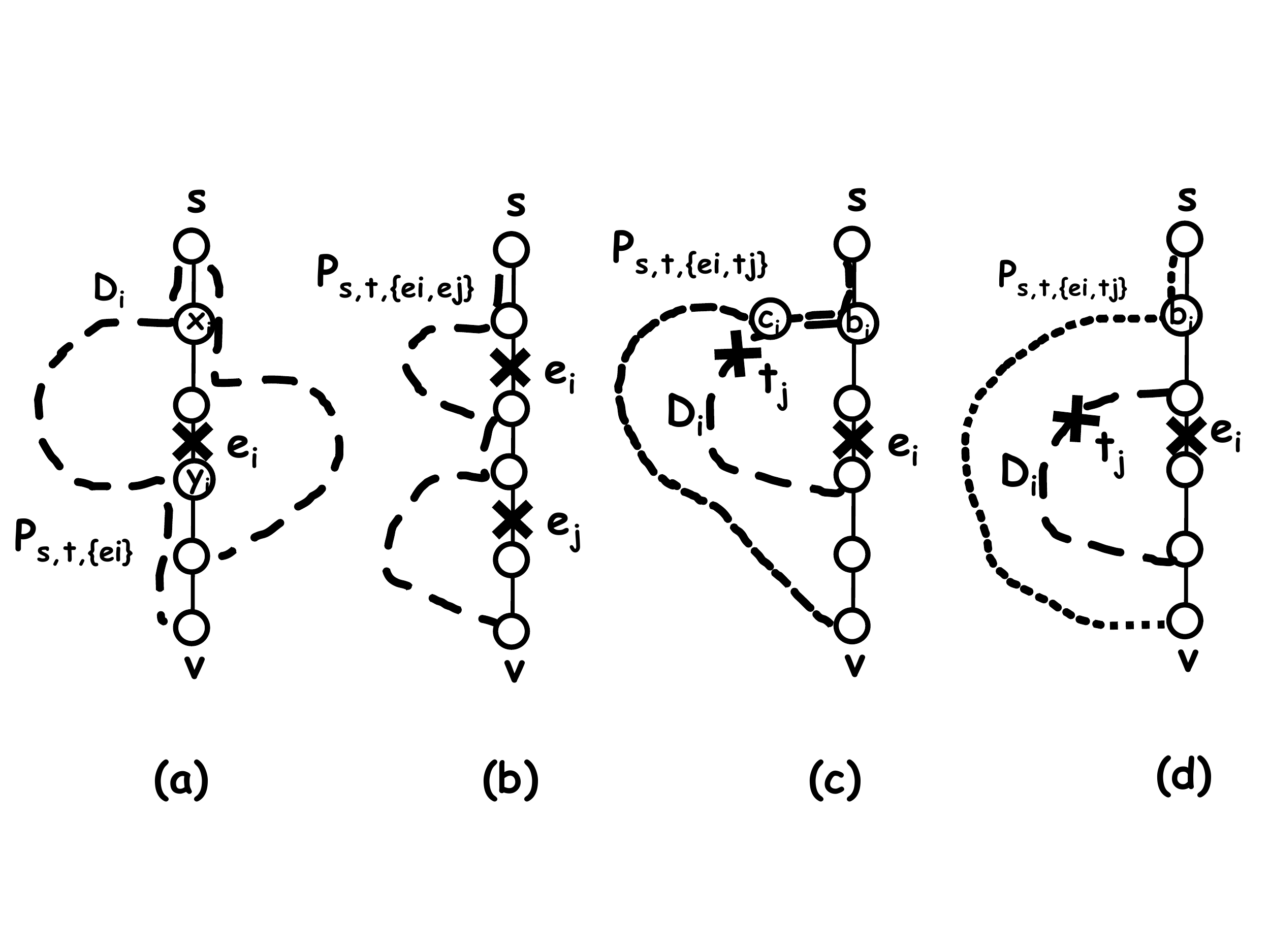}
\includegraphics[width=4in]{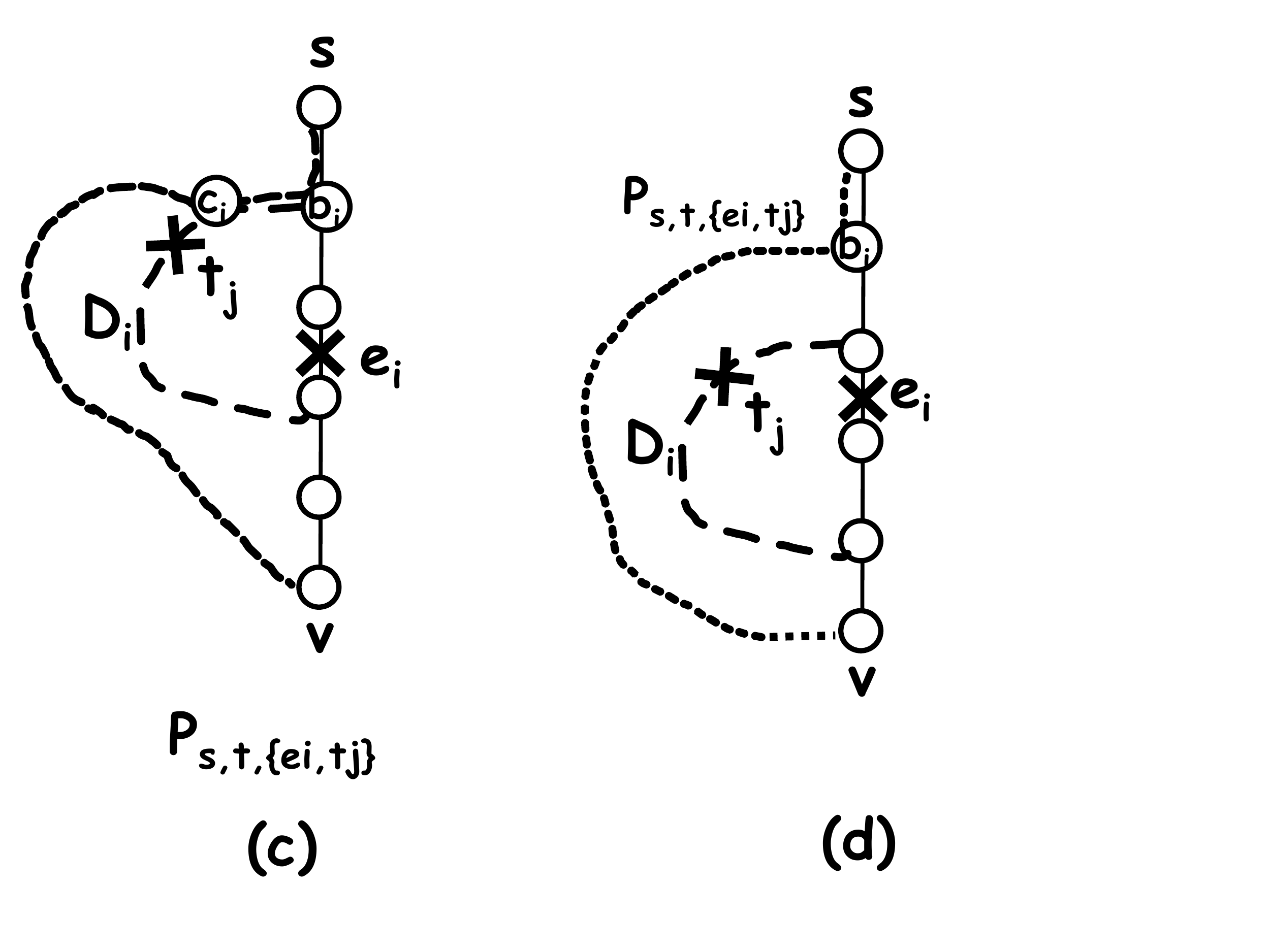}
\end{center}
\caption{Three types of replacement paths each constructed by a distinct steps of Alg. $\ConstPath$. (a) Single edge failure $e_i \in \pi(s,v)$. The algorithm selects the replacement path $P_{s,v,\{e_i\}}$ whose divergence point from $\pi(s,v)$ is as close to $s$ as possible. (Shown on the right hand side is another candidate path of the same length with a lower distinct divergence point, the was not chosen.) (b) Two edge faults both occurring on $\pi(s,v)$. The corresponding replacement path $P_{s,v, \{e_i,e_j\}}$ may have two divergence points from $\pi(s,v)$. (c) One edge fault $e_i$ is on $\pi(s,v)$ and one $t_j$ is on of the detour segment $D_i$. The algorithm selects the replacement path $P_{s,v,\{e_i,t_j\}}$ with the ``highest" (closest to $s$) divergence points $b_i$ from $\pi(s,v)$ and $c_i$ from the detour segment $D_i$.
(d) As in (c), only that the replacement path $P_{s,v,\{e_i,t_j\}}$ does not intersect with its detour $D_i$.
\label{fig:apathtype}}
\end{figure}

  For a source node $s$, a target node $v$ and an edge pair $F=\{e_i,e_j\}\in G$, the shortest $s-v$ path $P_{s,v,F}$ that does not go through $F$ is known as a \emph{replacement path}. Thus, a dual \FTBFS\ structure contains the collection of all replacement paths $P_{s,v,F}$ for every $v \in V(G)$ and every failed pair of edges $F=\{e_i,e_j\} \subseteq E(G)$.
Hereafter, we fix one vertex $v$ and concentrate on constructing $s-v$ replacement paths protecting against at most two failures in $E(G)$.
Algorithm $\ConstPath$ consists of three steps depending on the type of the faulty edges. See Fig. \ref{fig:apathtype} for a schematic illustration. First, it constructs a collection of paths $P_{s,v,F} \in SP(s,v, G\setminus F)$ where only one edge failure occurs, i.e., $F=\{e_i\}$ for every $e_i \in \pi(s,v)$.
The selection prefers the replacement path that diverges from $\pi(s,v)$ as early as possible. Then, the algorithm considers the case where the two failing edges occur on $\pi(s,v)$. Finally, letting $D_i=P_i \setminus \pi(s,v)$ be the detour segment of $P_i=P_{s,v,\{e_i\}}$ from $\pi(s,v)$, the last step considers the case where the second failing edge occurs on $D_i$. In this case, the procedure would attempt to construct a replacement path whose divergence point from $\pi(s,v)$ is as close to $s$ as possible and under certain conditions it imposes also the requirement that the divergence point from $D_i$ is as closest to $s$ as well. Eventually, only the \emph{last} edge of each replacement path is added to the construction.
The following definition is useful.
For every $e_i=(u_{i},u_{i+1}) \in \pi(s,v)$, and $k \in \{0, \ldots, i\}$, we would like to consider the possibility that $u_k$ is the point where the replacement path protecting against a failure in $e_i$ diverges from $\pi(s,v)$. To enforce that possibility, for every two vertices, $u_{k}, u_{\ell} \in \pi(s,v)$, we define the graph
\begin{equation}
\label{eq:gk_def}
G(u_k,u_\ell)=\left(G \setminus V(\pi(u_k,u_\ell)\right)  \cup \{u_k,v\}~.
\end{equation}
that contains $u_k$ and $v$ but does not contain the other vertices on the segment $\pi(u_k,u_\ell)$.
Intuitively, for an edge $e_i=(u_{i},u_{i+1}) \in \pi(s,v)$ and a vertex $u_k \in \pi(s,u_{i})$, the first divergence point of the $s-v$ replacement path $P \in G(u_k,u_{i})\setminus \{e_i,e_j\}$ from $\pi(s,v)$ is $u_k$.
Since the divergence point from $\pi(s,v)$ of any replacement path protecting against the failing of $e_i$ must occur above the failing edge $e_i$, it holds that $u_k \in \pi(s, u_{i})$. Analogously, an $s-v$ replacement path $P$ in $G(u_k,v)\setminus \{e_i,e_j\}$ diverges from $\pi(s,v)$ at the point $u_k$ and its last edge is not in $\pi(s,v)$. In such a case, $P[u_k, v]$ and $\pi(s,v)$ are edge disjoint.
The algorithm would attempt to find the  upmost divergence point $u_k \in \pi(s,u_i)$ such that $G(u_k,u_i)$ contains a replacement path for the failures $e_i,e_j$.
\dnsparagraph{(1) Single edge fault replacement paths}
The first step considers single edge failure scenarios. Denote the collection of possible edge failures by
$\mathcal{F}^1_{v}(\pi)=\{\{e_i\} ~\mid~ e_i \in \pi(s,v)\}$.
Let $k_0 \in \{0, \ldots, i\}$ be the minimal index $k$ satisfying that $\dist(s,v, G(u_k,u_i) \setminus \{e_i\})=\dist(s,v, G \setminus \{e_i\})$.
Define $P_{s,v,\{e_i\}} \in SP(s,v, G(u_{k_0},u_i) \setminus \{e_i\},W)$.
For each $e_i \in \mathcal{F}^1_{v}(\pi)$, let $D_i=P_{s,v,\{e_i\}} \setminus \pi(s,v)$ be the detour segment of the replacement path chosen for $e_i$.
In Cl. \ref{cl:onefstru}, we show that $P_{s,v,\{e_i\}}=\pi(s,x_i) \circ D_i \circ \pi(y_i,v)$ where $x_i$ (resp., $y_i$) is the first (resp., last) vertex of $D_i$ and $e_i \in \pi(x_i,y_i)$.
As mentioned earlier, we do not have to add the entire replacement paths to the constructed structure; we later prove that it suffices to add the last edge of each replacement paths. Let $E_1(\pi)=\{\LastE(P_{s,v,\{e_i\}}), e_i \in \pi(s,v)\}$ be the last edges of replacement paths protecting against faults in $E(\pi(s,v))$. These edges will be added to the constructed structure.
%Let $E'_v=E(v,G) \setminus E(v,T_0)$ be the set of edge adjacent to $v$ that are not in $T_0$.
%For every sequence $F \subseteq V$, the algorithm first check if $\dist(s,v, G \setminus (F \cup E'_v))>\dist(s,v, G \setminus F)$. If so, then any replacement path $P_{s,v,F} \in SP(s,v, G \setminus F)$ must end with an edge $e \notin T_0$, we refer to such path a new-ending replacement paths. If the
%$\dist(s,v, G \setminus (F \cup E'_v))=\dist(s,v, G \setminus F)$, let $P_{s,v,F} \in SP(s,v, G \setminus (F \cup E'_v),W)$ (i.e., we choose $P_{s,v,F}$ that ends with an edge in $T_0$).
%From now on we describe the construction of new-ending replacement paths.
For every edge $e_i \in \pi(s,v)$, let
$D_i= \in E(P_{s,v, \{e_i\}})\setminus E(\pi(s,v))$
be the detour segment of $P_{s,v, \{e_i\}}$. In Cl. \ref{cl:onefstru}, we show that
$P_{s,v,\{e_i\}}$ can be decomposed into three segments such that $P_{s,v,\{e_i\}}=\pi(s,x_i) \circ D_i \circ \pi(y_i,v)$ where $D_i=P_{s,v,\{e_i\}}[x_i,y_i]$ is the detour segment.
\dnsparagraph{(2) Two faults on $\pi(s,v)$}
The second step considers pairs of failures occurring both on $\pi(s,v)$. The collection of failure events considered is thus $\mathcal{F}^2_{v}(\pi)=\{F=\{e_i,e_j\} \mid F \subseteq \pi(s,v)\}$.\\
Without loss of generality, assume throughout that $e_i$ appears above $e_j$ on the path $\pi(s,v)$. Recall that $D_i$ (respectively, $D_j$) is the detour segment of $P_{s,v,\{e_i\}}$ (resp., $P_{s,v,\{e_j\}}$). The procedure constructs the shortest path $P_{s,v,F} \in SP(s,v, G \setminus F)$ in the following manner.  First the algorithm prefers a replacement path that is composed of the detours $D_i$ and $D_j$ constructed at step (1).
Specifically, if the intersection $D_i \cap D_j \neq \emptyset$, then let $w \in D_i \cap D_j$ be the last point on $D_j$ that is common to $D_i$.
Define the path $P=\pi(s,x_i) \circ D_i[x_i,w] \circ D_j[w,y_j] \circ \pi(y_j,v)$. If $|P|=\dist(s,v, G \setminus F)$, then let $P_{s,v,F}=P$. Otherwise, define $P_{s,v,\{e_i,e_j\}}= SP(s,v, G \setminus \{e_i,e_j\},W)$.
%Else, let $k \in \{1, \ldots, i-1\}$ be the minimal index satisfying that $\dist(s,v, \widetilde{G}_{k} \setminus \{e_i,e_j\})=\dist(s,v, G \setminus \{e_i,e_j\})$.
%Define $P_{s,v,\{e_i,e_j\}} \in SP(s,v, \widetilde{G}_k \setminus \{e_i,e_j\},W)$.
%Define $P_{s,v,F} \in SP(s, v, G \setminus F,W)$ for every $F \in \mathcal{F}^2_{v}(\pi)$.
The set of edges to be added in this step is
$E_2(\pi)=\{\LastE(P_{s,v,F}), F \in \mathcal{F}^2_{v}(\pi)\}$, the collection of last edges of replacement paths protecting against two edges faults on $\pi(s,v)$.
%Consider a pair $F=(e_i,e_j) \in F_{v}(\pi^2)$ and without loss of generality assume that $e_i=(u_i,u_{i+1})$ appears above $e_j$ on $\pi(s,v)$ and define $P_{s,v,F} \in SP(s,v, G_{k_{i,j}} \setminus F,W)$ where $k_{i,j}$ is the minimal $k \in \{1, \ldots, i-1\}$ satisfying that $\dist(s,v, G_{k_i} \setminus \{e_i,e_j\})=\dist(s,v, G \setminus \{e_i,e_j\})$.
%\textbf{MP: Edit this section, so that you had a new-ending path, only if there exists no path that uses the dual failure faults paths.}
\dnsparagraph{(3) One fault on $\pi(s,v)$ and one on the detour}
The third step considers the remaining (relevant) case where one of the failing edge $e_i$ occurs on the path $\pi(s,v)$ and the second failing edge occurs on the detour segment $D_i$.
Hence, the collection of failure scenarios considered in this step is $\mathcal{F}_v(\sf D)=\{\{e_i, t_j\} ~\mid~ e_i \in \pi(s,v), t_j \in D_i\}$.
We now order the pairs $F=\{e_i,t_j\} \in \mathcal{F}_v(\sf D)$ in the following manner. Let $F_{i_1}=\{e_{i_1},t_{j_1}\}$ and $F_{i_2}=\{e_{i_2},t_{j_2}\}$.
If $e_{i_1}\neq e_{i_2}$, then let $F_{i_1}>F_{i_2}$ iff $\dist(s,e_{i_1},G)>\dist(s,e_{i_2},G)$. Else, if $e_{i_1}=e_{i_2}$, then we use the second coordinate $t_i$ to break the tie where $F_{i_1}>F_{i_2}$ iff $\dist(x_{i_1},t_{j_1}, D_{i_1})>\dist(x_{i_1},t_{j_2},D_{i_1})$ where $D_{i_1}=P_{s,v,\{e_{i_1}\}}[x_{i_1},y_{i_1}]$ is the detour segment of $P_{s,v,\{e_{i_1}\}}$.
Let $\overrightarrow{\mathcal{F}}_v(D)=\{ F_1, F_2, \ldots, F_k\}$ be the ordering of the faulty pairs  $\mathcal{F}_v(\sf D)$ in decreasing order where $F_1>F_2>\ldots >F_k$.
\par Let $E_0(v)=E(v,T_0) \cup E_1(\pi) \cup E_2(\pi)$ be an initial collection of edges incident to $v$ be added to the structure by steps (1) and (2). The algorithm considers the faulty pairs $F$ according to the ordering of $\overrightarrow{\mathcal{F}}_v(D)$, where at step $\tau\geq 1$, given $E_{\tau-1}(v)$, it considers the pair $F_\tau=(e_\tau,t_\tau)\in \overrightarrow{\mathcal{F}}_v(D)$ and computes the replacement path $P_{s,v, F_\tau} \in SP(s,v, G \setminus F_\tau)$ in the following manner. Let $G_{\tau-1}(v)=(G \setminus E(v,G)) \cup E_{\tau-1}(v)$ be a subgraph of $G$ in which the edges incident to $v$ are only the edges of $E_{\tau-1}(v)$.  First, if there exists a shortest replacement path for $F_{\tau}$ in $G_{\tau-1}(v)$, namely, one that uses the edges of $E_{\tau-1}(v)$, then no new edge of $v$ should be introduced.
I.e., If $\dist(s,v, G_{\tau-1}(v) \setminus F_{\tau})=\dist(s,v, G \setminus F_{\tau})$, then let $P_{s,v,F_\tau}=SP(s,v, G_{\tau-1}(v) \setminus F_{\tau},W)$.
Otherwise, a new edge of $v$ that is not in $E_{\tau-1}(v)$ is essential for satisfying the pair $F_{\tau}$. The algorithm then aims to select a new-ending replacement path whose first divergence point $b_{\tau}$ from $\pi(s,v)$ is as close to $s$ as possible. Let $x_\tau$ be the first vertex of the detour $D_\tau$.
The point $b_\tau$ is found as follows. Let $e_\tau=(u_{i_{\tau}}, u_{i_{\tau}+1})$, define $u_k \in \pi(s, u_{i_{\tau}})$ as the closest vertex to $s$ satisfying that $\dist(s,v, G(u_k,v) \setminus F_{\tau})=\dist(s,v, G\setminus F_{\tau})$. Then $b_{\tau}=u_k$ and let $P=SP(s,v, G(u_k,v) \setminus F_{\tau},W)$.
If the first divergence point $b_\tau$ of $P$ from $\pi(s,v)$ is not $x_\tau$ (i.e., the divergence point is not as that of $P_{s,v,\{e_\tau\}}$ and $\pi(s,v)$) then let $P_{s,v,F_\tau}=P$.
Else, if $b_{\tau}=x_\tau$, the replacement path $P_{s,v,F_\tau}$ is selected so that its unique divergence point from the detour $D_{\tau}$ is as close to $x_\tau$ as possible. To enforce that,
let $D_\tau=[x_\tau=w_0, \ldots, w_q=y_\tau]$ where the second failing edge is $t_\tau=(w_j,w_{j+1})$, then for every $j\geq 1$ and every $\ell \in \{0, \ldots, j\}$, define
\begin{equation}
\label{eq:gelld}
G_{\sf D}(w_\ell)=\left(G(x_\tau,v) \setminus V(D_\tau[w_\ell, y_\tau])\right) \cup \{w_\ell\}~.
\end{equation}
That is, an $s-v$ replacement path in the subgraph $G_{\sf D}(w_\ell) \setminus F_\tau$ diverges from $\pi(s,v)$ at the unique point $b_{\tau}$ and diverges from $D_\tau$ at the point $w_\ell$. Since
the divergence point from $D_\tau$ must occur above the second failing edge $t_\tau$ it holds that $w_\ell \in D[w_0,w_j]$. The algorithm computes a $P_{s,v,F_{\tau}}$ path whose divergence point from $D_\tau$ is as close to $w_0$ on the detour $D_{\tau}$ as possible:
let $\ell \in \{0, \ldots, j\}$ be the minimum index satisfying that $\dist(s,v,G_{\sf D}(w_\ell) \setminus F_\tau)=\dist(s,v,G \setminus F_\tau)$. Let
$P_{s,v,F_\tau} =\pi(s, x_\tau) \circ D_{\tau}[x_\tau, w_{\ell}] \circ SP(w_\ell,v, G_{\sf D}(w_\ell) \setminus F_\tau,W).$
Finally, let $E_\tau(v)=E_{\tau-1}(v) \cup \{\LastE(P_{s,v,F_\tau})\}$. This completes the description of the algorithm.
\par Let $\mathcal{F}_v=\mathcal{F}^1_v(\pi) \cup \mathcal{F}^2_v(\pi) \cup \mathcal{F}_v(\sf D)$ be the collection of single edge and edge pair failure events for which an $s-v$ replacement path $P_{s,v,F}$ was constructed.
Let
$H(v)=\bigcup_{F \in \mathcal{F}_v} \{\LastE(P_{s,v,F})\}$
be the collection of last edges of all replacement paths in $P_{s,v,F}$. Finally, the algorithm outputs
$H=\bigcup_{v \in V} H(v) \cup T_0$
as the resulting dual failure $\FTBFS$ structure.
%\def\APPENDFIGREPALL{
%%%%%%%%%%%%%%%%%%%
%%%%%%%%%%%%%%%%%%%
%}%\APPENDFIGREPALL

%\section{Analysis}
In this section, we show that the subgraph $H$, the output of Alg. $\ConstPath$, is a dual failure $\FTBFS$ structure and then bound its size. Recall that a path $P_{\tau}=P_{s,v,F_{\tau}}$ is a $(\pi,\pi)$-replacement path if its two failing edges appear on the $\pi(s,v)$ path, i.e.,  $|F_{\tau}|=2$ and $F_{\tau} \subseteq \pi(s,v)$. Otherwise, if the first failing edge $e_i$ appears on the $\pi(s,v)$ path and the second failing edge $t_j$ appears on the detour segment $D_i$ of $P_{s,v,\{e_i\}}$, it is a $(\pi,\sf D)$-replacement path.
Hence, step (2) constructs the collection of $(\pi,\pi)$-replacement paths and step (3) constructs the collection of $(\pi,\sf D)$-replacement paths.
\par An edge $e \in H$ is \emph{new} if $e \in H \setminus T_0$, i.e, it is not part of the original fault free BFS tree $T_0$ computed in $G$. A $(\pi,\sf D)$ replacement path $P_{\tau}=P_{s,v,F_{\tau}}$ is \emph{new-ending} if $G_{\tau-1}(v)$ did not satisfy the faults of $F_{\tau}$, i.e., $\LastE(P_{\tau})$ was first added to the constructed $H(v)$ by $P_{\tau}$. In particular, for a new-ending replacement path $P_{\tau}$, we  have $\LastE(P_{\tau})\notin T_0$.
Note that Alg. $\ConstPath$ adds only the last edge of new-ending paths to the structure.
Hence, our goal is to bound the number of new edges in $H$. Let $\New(v)=H(v) \setminus E(v,T_0)$ be the collection of new edges incident to $v$.
Throughout, we focus on a single vertex $v \in V\setminus \{s\}$ and show that $|\New(v)|=O(n^{2/3})$.
For every $e \in \New(v)$, let $P(e)=P_{\tau}$ be the new ending replacement path that first introduced $\LastE(P)=e$ to $H(v)$.
To bound the size of $\New(v)$, we study the structure of new-ending paths.
Let $\mathcal{P}_v=\{P(e) ~\mid~ e \in \New(v)\}$ be the collection of new-ending $s-v$ replacement paths, each representing one distinct new edge from $\New(v)$.
\par The following notation is useful in our setting.
We view the $\pi(s,v)$ path from top (i.e., $s$) to bottom $v$. An edge $e_i$ is said to be \emph{above} $e_j$, if it is closer to $s$ on the path $\pi(s,v)$.
For vertices $u_i, u_j \in \pi(s,v)$, we denote $u_i< u_j$ if $\dist(s, u_i, G) < \dist(s, u_j, G)$ (i.e., $u_i$ appears on $\pi(s,v)$ \emph{before} $u_j$).
For a given edge pair $F \in \mathcal{F}_v$ and a replacement path $P=P_{s,v,F}$, let $F_1(P)=e_i$ be the first failing edge in $F$ (note that this edge, by convention, is always on the shortest path $\pi(s,v)$) and let $F_2(P)$ be the second failing edge in $F$ (if exists), where $F_2(P)$ might be either on $\pi(s,v)$ or on the detour segment $D_i$ of $P_{s,v,\{e_i\}}$. Let $F(P)=\{F_1(P),F_2(P)\}$ be the two failing edges (i.e., $P \in SP(s, v, G \setminus F(P))$). Let $D(P)=D_i$, be the detour segment protecting against the failing of the edge $F_1(P)=e_i \in \pi(s,v)$.
Throughout, we assume $D_i=P_{s,v,\{e_i\}}[x_i,y_i]$. We denote the first (resp., last) vertex of the detour $D_i$ by $x(D_i)$ (resp. $y(D_i)$) , i.e., $x(D_i)=x_i$ and $y(D_i)=y_i$.

%An $s-v$ path $P$ is a $(\pi,\pi)$-replacement path if $F(P) \subseteq \pi(s,v)$, otherwise it is a $(\pi,\sf D)$-path, i.e., $F_1(P) \in \pi(s,v)$ and $F_2(P) \in D(P)$.
%
Note that a replacement path $P_i$ does not necessarily intersect with the detour $D(P_i)$ (e.g., see Fig. \ref{fig:apathtype}(d)). Let $b(P_i)$ (or $b_i$ for short) be the first divergence point of the path $P_i$ from $\pi(s,v)$. We denote this point as the $\pi$-\emph{divergence point} of $P_i$. If $P_i$ intersects with its detour $D(P_i)$, then let $c(P_i)$ (or $c_i$) be the first divergence point of $P_i$ from $D(P_i)$. We denote this point by the $D$-\emph{divergence point} of $P_i$.
See Fig. \ref{fig:apathtype}(c).
\subsection{Correctness}
The correctness analysis consists of two steps.
First, we show the correctness of the construction of the replacement paths $P_{s,v,F}$ by Alg. $\ConstPath$. Then, we show that taking the last edge of every replacement path $P_{s,v,F}$ for every $v \in V$ and $F \in \mathcal{F}_v$ is sufficient for making $H$ a dual failure $\FTBFS$ structure.
\begin{lemma}
\label{lem:correct_conspaths}
For every $v \in V$ and $F \in \mathcal{F}_v$,
$P_{s,v,F} \in SP(s,v, G \setminus F)$.
\end{lemma}
\Proof
Note that $P_{s,v,F}$ is not necessarily in $SP(s,v, G \setminus F,W)$. In particular, $SP(s,v, G \setminus F,W)$ correspond to a unique replacement-path which may not be the one the we want. To establish correctness, we thus show that the replacement path chosen is indeed a shortest-path in $G \setminus F$.
First, consider the case where $P_{s,v,F}$ was constructed in step (1), hence $F=\{e_i=(u_i,u_{i+1})\}$ where $e_i \in \pi(s,v)$. It is sufficient to show that there exists $u_k \in \pi(s,u_i)$, satisfying that $\dist(s,v,G(u_k,u_i)\setminus F_\tau)=\dist(s,v,G\setminus F_\tau)$. This holds as by Eq. (\ref{eq:gk_def}), $G(u_{i},u_{i})=G$.
Next, consider the case where $P_{s,v,F}$ was constructed in step (2). Hence, $P_{s,v,F}$ is a $(\pi,\pi)$-replacement path. This case is immediate.
\par Finally, consider the case where $P_{s,v,F}$ is a $(\pi,\sf D)$-replacement path. Let $\tau$ be the iteration in which the pair $F=F_{\tau}=(e_i,t_j)$ was considered by the algorithm in step (3).
It is sufficient to consider the case where $F_\tau$ is not satisfied by the current graph $G_{\tau-1}$, i.e., the path $P_{s,v,F}$ is a new-ending path. We first claim that there always exists an $s-v$ new-ending path with a unique divergence point $b$ from $\pi(s,v)$ that appears above the failing edge $e_i \in \pi(s,v)$.
Let $P =SP(s,v, G \setminus F_\tau,W)$ and let $b$ be the first divergence point of $P$ and $\pi(s,v)$.
Assume towards contradiction that $b$ is not unique and let $b' \in P[b,v]\cap \pi[b,v]$ be another divergence point.
There are two cases. If $e_i \in \pi(b,b')$, then $e_i \notin \pi(b',v)$ and hence by the uniqueness of $W$, $\pi(b',v)=P[b',v]$, contradiction to the fact that $b'$ is a divergence point.
Else, if $e_i \in \pi(b',v)$ (i.e., $e_i \notin \pi(b,b')$) then by the uniqueness of $W$,
$\pi(b,b')=P[b,b']$, contradiction to the fact that $b$ is a divergence point. Hence, the divergence point $b$ is unique and therefore it also holds that $P=SP(s,v, G(b,v) \setminus F_\tau,W)$.
If $b \neq x_\tau$ where $x_\tau$ is the first vertex of the detour $D_\tau=P_{s,v,\{e_i\}}[x_\tau,y_\tau]$ (i.e., the detour segment of $P_{s,v,\{e_i\}}$) then the correctness follows, since in this case the algorithm let $P_{s,v,F_\tau}=P$.
It remains to consider the case where $b=x_\tau$. We claim that in such a case the path $P$ has unique divergence point $c$ from the detour $D_\tau$. Assume towards contradiction that there exists an additional common point in the intersection $w \in \left(P[c,v] \cap D_\tau[c,y_\tau] \right) \setminus \{c\}$. Observe that $t_j \in D_\tau[w,y_\tau]$, as otherwise the path $P'=P[s,w]\circ D_\tau[w,y_\tau] \circ \pi(y_\tau,v)$ is in $SP(s,v, G\setminus F_\tau)$ and ends with an edge in $T_0$, contradiction to the fact that $P_{\tau}$ was not satisfied by $G_{\tau-1}(v)$. Therefore, $t_j \in D_\tau[w,y_\tau]$ and by the uniqueness of the weight assignment $W$ it holds that $P[c,w]=D_\tau[c,w]$, contradiction to the fact that $c$ is a divergence point. Hence, $c$ is a unique divergence point from $D_{\tau}$ and thus $P \subseteq G_{\sf D}(c)$ (see Eq. (\ref{eq:gelld})).

Letting $t_j=(q_1,q_2)$, the algorithm then selects the
closest vertex $u_{\ell} \in D_{\tau}[x_{\tau},q_1]$ to $x_{\tau}$  satisfying that $\dist(s,v, G_{\sf D}(u_{\ell})\setminus F_{\tau})=\dist(s,v, G\setminus F_{\tau})$. Since by the above, this holds for at least one vertex $c \in D_{\tau}$, correctness is established.
\QED

We now turn to show that taking the last edges of the constructed replacement path into the structure $H$ is sufficient.
\begin{lemma}
\label{lem:correct_conspaths}
For every $e_i,e_j \in E$ and every vertex $v \in V$, $\dist(s,v, H \setminus \{e_i,e_j\})=\dist(s,v, G \setminus \{e_i,e_j\})$.
\end{lemma}
\Proof
Assume, towards contradiction, that the claim does not hold.
Let
$$BP=\{(v,F) \mid v \in V, F \subset E, |F|\leq 2 \mbox{~and~}
\dist(s,v, H \setminus F) >  \dist(s,v,G \setminus F)\}$$
be the set of ``bad pairs," namely, pairs $(v,F)$ for which the $s-v$ shortest path distance in $H \setminus F$ is greater than that in $G\setminus F$.
(By the assumption, it holds that $BP\ne \emptyset$.)

First, note that for every bad pair $(v,F) \in BP$, it holds that $F \in \mathcal{F}_v$ and hence a replacement path $P_{s,v,F}$ was constructed for it by Algorithm $\ConstPath$.
For each bad pair $(v,F) \in BP$, define
$BE(v,F)=P_{s,v,F} \setminus E(H)$ to be the set of ``bad edges,''
namely, the set of $P_{s,v,F}$ edges that are missing in $H$
(due to the sparsification phase that maintains only ``last" new edges).
By definition, $BE(v,F) \neq \emptyset$ for every bad pair $(v,F) \in BP$.
Let $d(v,F)=\max_{e \in BE(v,F)}\{\dist(s,e,P_{s,v,F})\}$ be the maximal depth
of a missing edge in $BE(v,F)$, and let $DM(v,F)$ denote that ``deepest
missing edge'', i.e., the edge $e$ on $P_{s,v,F}$ satisfying
$d(v,F) = \dist(s,e,P_{s,v,F})$.
Finally, let $(v',F') \in BP$ be the pair that minimizes $d(v,F)$,
and let $e_1=(u_1,v_1) \in BE(v',F')$ be the deepest missing edge on
$P_{s,v',F'}$, namely, $e_1=DM(v',F')$. Note that $e_1$ is the {\em shallowest}
``deepest missing edge'' over all bad pairs $(v,F) \in BP$.
%$e_1$ is the most distant edge from $s$ on the path $P^{*}_{i',j'}$ such that
%$d(i',j')=\dist(s,e_1,P^{*}_{i',j'})$.
%

%{\bf The next sentence seems like an inaccurate description of the actual contradiction process:}
%We now show that $\dist(s, v_{i'}, T^{*} \setminus \{e_{j'}\})=|P^{*}_{i',j'}|$.
%Since $|P^{*}_{i',j'}|=\dist(s, v_{i'}, G \setminus \{e_{j'}\})$, this results in
%contradiction, hence establishing the correctness of the algorithm.
%
\begin{claim}
$(v_{1},F') \in BP$.
\end{claim}
\Proof
Assume towards contradiction otherwise and let $P_1 \in SP(s, v_{i_1}, H \setminus F')$. By the contradictory assumption, $|P_1|=|P_{s,v,F'}(s, v_{1})|=\dist(s, v_{1}, G \setminus F')$. Then, the path $P_2=P_1 \circ P_{s,v,F}(v_{1},v)$ is in $H \setminus F$ and in addition, $|P_2|=|P_{s,v,F'}|$, contradiction to the fact that $(v,F') \in BP$. The claim holds.
\QED
If $F'=(e_i,e_j) \in \mathcal{F}_{v_1}$, let $P'=P_{s,v_1,F'}$. Else, since $(v_1,F')$ is a bad pair there must be an edge $e_1 \in F' \cap \pi(s,v)$ as otherwise the path $\pi(s,v_1) \subseteq H \setminus F'$. Since $F' \notin \mathcal{F}_v$, it implies that $e_2 \in F' \setminus \{e_1\}$ does not appear on the detour of $P_{s,v_1,\widetilde{F}}$ where $\widetilde{F}=\{e_1\}$.
Therefore, $P_{s,v_1, \widetilde{F}} \in SP(s, v_1, G \setminus F')$ and let $\widetilde{P}=P_{s,v_1,\widetilde{F}}$.
By the construction of $H$, $\LastE(\widetilde{P}) \in H(v_1) \subseteq H$, and therefore the deepest missing edge of $(v_1,\widetilde{F})$ must be shallower, i.e.,
$d(v_1,\widetilde{F})<d(v',F')$. However, this is in contradiction to our choice of the pair $(v',F')$. The lemma follows.
\QED
%To do that, we begin by selecting a representative path $P(e)$ from $\mathcal{P}(e)$
%by taking the path $P$ who first introduced $e$ to the subgraph $H$.
%I.e., if there exists a path $P \in \mathcal{P}(e)$, such that $F_1(P),F_2(P) \in \pi(s,v)$, let $P(e)=P$ (if there several such paths, pick one arbitrary). Else, for every $P \in \mathcal{P}(e)$, define $\depth(P)=\dist(s, F_1(P))$ to be the the depth of the first failing edge on $\pi(s,v)$. Then,   $P_e$ is chosen to be the path in $\mathcal{P}(e)$ that maximizes $\depth(P)$, i.e., $\depth(P)=\max_{P' \in \mathcal{P}(e)} \depth(P')$ (if there are several such $P \in \mathcal{P}(e)$ one is chosen arbitrary).
We now provide two useful claims on the structure of the $s-v$ replacement paths and begin by considering the replacement path $P_{s,v,\{e_i\}}$ protecting against single edge fault $e_i \in \pi(s,v)$.
\begin{claim}
\label{cl:onefstru}
(1) Every $P_{s,v,\{e_i\}}$ can be decomposed into three segments such that $P_{s,v,\{e_i\}}=\pi(s,x_i) \circ D_i \circ \pi(y_i,v)$ where $D_i=P_{s,v,\{e_i\}}[x_i,y_i]$ is the detour segment, which is edge disjoint with $\pi(s,v)$.\\
(2) There is no alternative replacement path whose unique divergence point is closer to $s$ than $x_i$.
\end{claim}
\Proof
Begin with part (1).
Let $e_i=(u_{i},u_{i+1}) \in \pi(s,v)$.
Let $x_i \in \pi(s,u_i)$ be the closest vertex to $s$ satisfying that $\dist(s,v, G(x_i, u_i) \setminus \{e_i\})=\dist(s,v, G \setminus \{e_i\})$.
Then, Alg. $\ConstPath$ define $P_{s,v, \{e_i\}}=SP(s,v, G(x_i, u_i) \setminus \{e_i\},W)$.
We first claim that $P_{s,v, \{e_i\}}[s,x_i]=\pi(s,x_i)$. Since $\pi(s,x_i) \subseteq G(x_i, u_i)$, and $\pi(s,x_i)=SP(s,v, G,W)$, it also holds that $\pi(s,x_i)=SP(s,v, G(x_i, u_i),W)$. The claim holds as $P_{s,v, \{e_i\}}=SP(s,v, G(x_i, u_i),W)$. Let $y_i \in \left(P_{s,v, \{e_i\}}[x_i,v]\cap \pi(s,v)\right) \setminus \{x_i\}$ be the first vertex on $P_{s,v, \{e_i\}}$ appearing \emph{after} $x_i$ that is in $\pi(s,v)$. Note that by the definition of $G(x_i,u_i)$ and the fact that the failing edge is $e_i$, it holds that $y_i \in \pi(u_{i+1},v)$. Since $\pi(y_i,v) \subseteq  G(x_i, u_i)$, we have that
$$\pi(y_i,v)=SP(y_i,v, G,W)=SP(y_i,v, G(x_i, u_i),W)=P_{s,v, \{e_i\}}[y_i,v].$$
Note that $x_i$ is the unique divergence point as $\pi(x_i,u_i)$ is not in $P_{s,v,\{e_i\}}$ and in addition for the first vertex $y_i$ in $\pi(s,v)$ appearing after $x_i$, it holds that the paths collide. Hence, $D_i\cap \pi(s,v)=\{x_i,y_i\}$ where $e_i \in \pi(x_i,y_i)$. Part (1) follows. Part (2) follows immediately by the construction of the algorithm.
\QED
For every replacement path $P=P_{s,v,F}$, let $b(P)$ be the first divergence point of $P$ from $\pi(s,v)$. We call this point the $\pi$-divergence point of $P$. For a $(\pi,\sf D)$-replacement path $P=P_{s,v,\{e_i,t_j\}}$ that intersects its detour $D_i$, let $c(P)$ be the first divergence point of $P$ from $D_i$. We call this point the $\sf D$-divergence point of $P$. Note that while the $\pi$-divergence point is defined for every $s-v$ replacement path, the $\sf D$-divergence point is defined only for $(\pi,\sf D)$-replacement paths that intersect their detours. We conclude this section by showing that the $\pi$-divergence point of every replacement path is \emph{unique}.
\begin{claim}
\label{cl:unique_nnew}
(1) Every $(\pi,\sf D)$-replacement path $P=P_{s,v, F}$ has a unique $\pi$-divergence point $b(P)$ from $\pi(s,v)$.\\
(2) If  $P=P_{s,v, F}$ is also new-ending, then $P[b(P),v]$ and $\pi(s,v)$ are edge-disjoint.
\end{claim}
\Proof
Let $F=F_{\tau}=\{e_\tau,t_{\tau}\}$ be considered at time $\tau$ in step (3) of Alg. $\ConstPath$. If $P_{s,v,F_{\tau}}$ is new-ending, i.e., was not in $G_{\tau-1}(v)\setminus F_{\tau}$, then the claim follows immediately by construction, since $P_{s,v,F_{\tau}}$ is computed in $G(u_{k},v)$ for some $u_k \in \pi(s, u_{\tau})$ where $e_{\tau}=(u_{\tau},u_{\tau+1})$ and hence $u_k$ is the unique divergence point and $P_{s,v,F_{\tau}}[u_k,v]$ and $\pi(s,v)$ are edge disjoint.

It remains to consider claim (1) for the case where $P_{s,v,F_{\tau}}$  is not new-ending, i.e., exists in $G_{\tau-1}(v)\setminus F_{\tau}$. In such a case, $P_{s,v,F_{\tau}}=SP(s,v, G_{\tau-1}(v)\setminus F_{\tau},W)$.
Let $b_1$ be the first divergence point of $P_\tau=P_{s,v,F_{\tau}}$ and $\pi(s,v)$.
Assume towards construction that there exists an additional divergence point $b_2 \in P_\tau[b_1,v] \cap \pi(b_1,v) \setminus \{b_1,v\}$.
Since $P_{\tau}$ is not a $(\pi,\pi)$-replacement path, its second failing edge $t_{\tau}$ is not in $\pi(s,v)$. There are two cases. If $e_\tau \in \pi(b_1, b_2)$,
there are two $b_2-v$ paths in $G_{\tau-1}\setminus F_{\tau}$, namely, $\pi(b_2,v)$ and $P_{\tau}[b_2,v]$,
contradiction to the uniqueness of $W$.
Else, if $e_\tau \in \pi(b_2,v)$, there are two $b_1-b_2$ paths in $G_{\tau-1}(v)\setminus F_{\tau}$, namely, $\pi(b_1,b_2) \neq P_{\tau}[b_1,b_2]$,
leading to contradiction to the uniqueness of $W$ again.
The claim follows.
\QED

\subsection{Structural properties of detours}
\label{sec:deturs}
A crucial step for understanding the structure of the replacement path $P_{s,v,F}$ protecting against two edge failure in $G$, is the understanding of the structure of the replacement path $P_{s,v, \{e_i\}}$ protecting against single failure $e_i$ on $\pi(s,v)$. In particular, it is important to understand the detour segments $D_i=P_{s,v, \{e_i\}} \setminus \pi(s,v)$ of these paths.
\par In this section, we present some basic structural properties of detours, that will provide the tools for bounding the size of the final structure later on.
For detour $D_i$, recall that $x(D_i)$ (resp., $y(D_i)$) is the first (resp., last) common vertex with $\pi$.
Throughout, we consider two detours $D_1,D_2$. For $i \in \{1,2\}$, define $x_i=x(D_i)$ and $y_i=y(D_i)$. Let $e_i$ be the single edge on $\pi(s,v)$ that the detour $D_i$ protects, i.e., such that $P_{s,v, \{e_i\}}=\pi(s, x_i) \circ D_i \circ \pi(y_i,v)$.
Hence, $e_i \in \pi(x_i,y_i)$ for $i \in \{1,2\}$.
Two detours $D_1, D_2$ are \emph{independent}, $V(D_1) \cap V(D_2)=\emptyset$, otherwise they are \emph{dependent}.
We now provide a useful claim which follows by the fact that we use the weight assignment $W$ that guarantees the uniqueness of the shortest-paths.
\begin{claim}
\label{cl:jointdet}
Let $w_1, w_2 \in D_1 \cap D_2$ then $D_1[w_1, w_2]=D_2[w_1,w_2]$.
\end{claim}
\Proof
For $i \in \{1,2\}$, let $e_i=(u_i,u_{i+1})\in \pi(s,v)$ such that $P_{s,v, \{e_i\}}=\pi(s,x_i) \circ D_i \circ \pi(y_i,v)$.
By construction, $P_{s,v, \{e_i\}}[x_i,v] =SP(x_i,v, G\setminus \pi(x_i,u_{i+1}),W)$ for $i \in \{1,2\}$.
Assume, towards contradiction otherwise, then it implies that there are two distinct $w_1-w_2$ shortest paths in $G \setminus \pi(s,v)$, given by $D_i[w_1,w_2]=SP(w_1,w_2, G \setminus \pi(s,v),W)$ for $i \in \{1,2\}$, contradiction to the uniqueness of $W$.
\QED

Throughout we consider the detour segment $D_1[x_1,y_1]$ to be directed away from $x_1$, i.e., going from the starting vertex $x_1$ to the ending vertex $y_1$.

Note that by Cl. \ref{cl:jointdet}, every detour $D_1$
can be decomposed into three segments according to some dependent detour $D_2$: the noncommon prefix $D_1[x_1,w_1]$, the common segment $D_1[w_1,w_2]=D_2[w_1,w_2]$, and the noncommon suffix $D_1[w_2,y_1]$. It is important to note that this does not necessarily imply that this common segment is used by the two detours in the same direction. In particular, it might be the case that the detours visit the common segment in \emph{opposite} directions where $D_1=D_1[x_1,w_1] \circ D_1[w_1,w_2] \circ D_1[w_2,y_1]$ while
$D_2=D_2[x_2,w_2] \circ D_2[w_2,w_1] \circ D_1[w_1,y_2]$, (i.e. $D_1$ visits $w_1$ before $w_2$ and $D_2$ visits $w_2$ before $w_1$).

\subsubsection{Detour configurations and ordering}
\begin{figure}[htbp]
\begin{center}
~~
\includegraphics[width=3in]{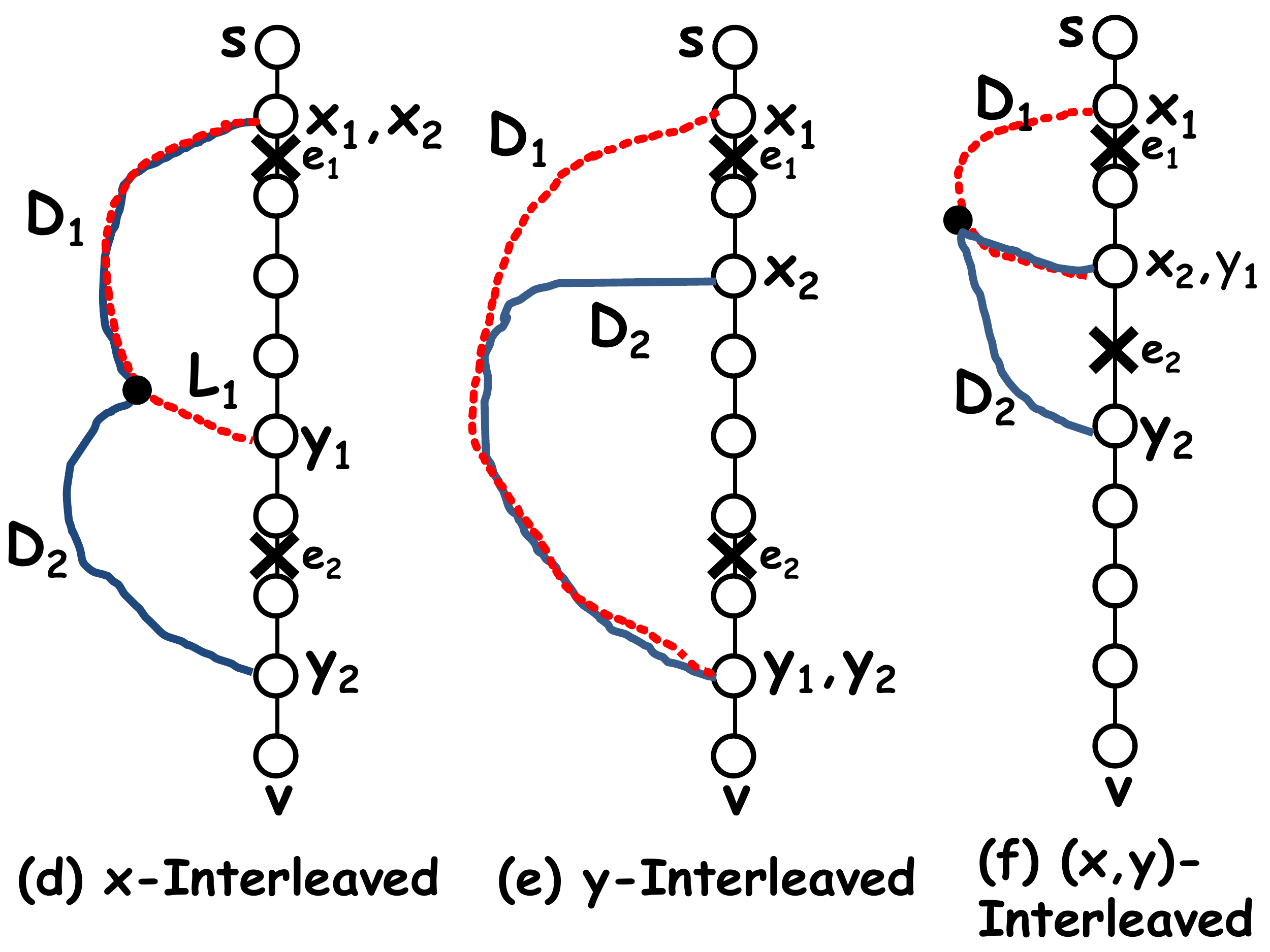}
\end{center}
\caption{Schematic illustration of the detours configurations.
\label{fig:detours}}
\end{figure}

In this subsection, we consider the possible detours configurations of two detours $D_1$ and $D_2$ where $x_1\leq x_2$. These configurations depend upon the lexicographic order of $x_1,y_1$ and $x_2,y_2$.
\begin{definition}[Detours Configurations]
\label{def:detour_conf}
\item{(Non-nested):}
$y_1 < x_2$.
\item{(Nested):}
$x_1<x_2<y_2<y_1$.
\item{(Interleaved):}
$x_1<x_2<y_1<y_2$.
\item{($x$-Interleaved):}
$x_1=x_2<y_1<y_2$.
\item{($y$-Interleaved):}
$x_1<x_2<y_1=y_2$.
\item{($(x,y)$-Interleaved):}
$x_1<y_1=x_2<y_2$.
\end{definition}
For a schematic illustration of these configurations, see Fig. \ref{fig:detours}.
%%%%%%%%%%%%%%%%%%%
%%%%%%%%%%%%%%%%%%%

\paragraph{The $(x,y)$-ordering the detours.}
The $(x,y)$-ordering of a collection of detours $\mathcal{D}$, namely, $\overrightarrow{\mathcal{D}}=\{D_1, \ldots, D_t\}$
is an ordering according to the lexicographic ordering of $(x(D_i), y(D_i))$ pairs. For ease of notation, let $x_i=x(D_i)$ and $y_i=y(D_i)$.
We say that $(x_i,y_i)>(x_j,y_j)$, if $x_i \geq x_j$ and if $x_i=x_j$ then $y_i >y_j$ (i.e., deeper on $\pi(s,v)$). Then, in an $(x,y)$ ordering the detours are ordered in \emph{decreasing} order of their $(x_i,y_i)$ pairs. I.e.,
$x_1 \geq x_2 \geq \ldots \geq x_t$ and if $x_i=x_j$ then $D_i$ precedes $D_j$ (denoted by $D_i \prec D_j$) in the ordering iff $y_i>y_j$.

\begin{claim}
\label{cl:disjindep}
If $D_1$ and $D_2$ are non-nested, then they are independent. Formally, if $y_1 < x_2$ then $D_1 \cap D_2=\emptyset$.
\end{claim}
\Proof
Assume towards contradiction there exists a common vertex $w \in D_1 \cap D_2$. See Fig. \ref{fig:detours}(a).
There are now two $y_1-w$ paths in $G\setminus \{e_1,e_2\}$, namely, $Q_1=D_1[y_1,w]$ and $Q_2=\pi(y_1,x_2) \circ D_2[x_2,w]$.
By the optimality of $P_{s,v, \{e_i\}}$ for both $i \in \{1,2\}$, it holds that $|Q_1|=|Q_2|$. Hence, the path $Q_3=\pi(s, y_1) \circ Q_1 \circ D_2[w,y_2] \circ \pi(y_2,v)$ is also in $SP(s,v, G\setminus \{e_2\})$, but its unique divergence point from $\pi(s,v)$, namely $y_1$, is strictly above $x_2$, in contradiction to the selection of $P_{s,v, \{e_2\}}$ by  Algorithm $\ConstPath$ (which was supposed to prefer the divergence point that is closest to $s$). The claim follows.
\QED

\begin{claim}
\label{cl:depen_detdisjoint}
If $D_2$ is nested in $D_1$, then  they are independent. Or, formally, if $x_1<x_2<y_2<y_1$, then $D_1 \cap D_2=\emptyset$.
\end{claim}
\Proof
Assume, towards contradiction, that there is a common vertex $w \in D_1 \cap D_2$. Let $e_1=(u_1,u_2)$ and $e_2=(u_3,u_4)$. Clearly, $e_2 \in \pi[x_2, y_2]$.
We consider two cases depending on where $e_1$ resides.
Case (1): $e_1 \notin \pi[x_1, x_2]$. In this case there are two $x_1-w$ paths in $G\setminus \{e_1,e_2\}$ given by $Q_1=\pi(x_1,x_2) \circ D_2[x_2,w]$ and $Q_2=D_1[x_1,w]$. By the optimality of $D_1$ and $D_2$, we get that $|Q_1|=|Q_2|$.
Since $x_1$ appears strictly above $x_2$, we end with contradiction the selection of $P_{s,v, \{e_2\}}$ by  Algorithm $\ConstPath$ (which was supposed to prefer the divergence point that is closest to $s$).

Case (2): $e_1 \in \pi[x_1, x_2]$.
In this case, there are two  $w-y_1$ shortest paths in $G \setminus \{e_1,e_2\})$, namely, $Q_1=D_1[w, y_1] \subseteq P_1$ and $Q_2=D_2[w, y_2] \circ \pi[y_2,y_1] \subseteq P_2$.
By the optimality of $P_{s,v, \{e_i\}}$ for both $i \in \{1,2\}$, it holds that $|Q_1|=|Q_2|$.
Note that $P_1 =SP(s,v, G(x_1,u_1) \setminus \{e_1\},W)$ and $P_2 =SP(s,v, G(x_2,u_3) \setminus \{e_2\},W)$.
Since both $Q_1$ and $Q_2$ exist in $G(x_1,u_4)$, it holds that $Q_1,Q_2=SP(w,y_1,G(x_1,u_4),W)$ leading to contradiction.
The claim follows.
\QED

\begin{claim}
\label{cl:det_nested}
If $D_1$ and $D_2$ are \emph{dependent} such that $D_2 \prec D_1$, i.e., $x_1 \leq x_2 \leq y_1 \leq y_2$,
then \\
(a) $e_1 \in \pi[x_1, x_2]$, if $x_1 \neq x_2$, and \\
(b) $e_2 \in \pi[y_1, y_2]$, if $y_1 \neq y_2$.
\end{claim}
\Proof
Let $w \in D_1 \cap D_2$ be a common vertex (see Fig. \ref{fig:detours}(c)).
Assume that $x_1<x_2$. Clearly, $e_1=(u_1,u_2) \in \pi[x_1, y_1]$ and $e_2=(u_3,u_4) \in \pi[x_2, y_2]$. Assume towards contradiction that $e_1 \notin \pi[x_1, x_2]$.
In this case there are two $x_1-w$ paths in $G \setminus \{e_1,e_2\}$, namely, $Q_1=\pi(x_1,x_2) \circ D_2[x_2,w]$ and $Q_2=D_1[x_1,w]$. By the optimality of $P_{s,v,\{e_i\}}$, for both $i \in \{1,2\}$, it holds that $|Q_1|=|Q_2|$. Since $x_1$ is strictly above $x_2$, we end with contradiction to the selection of $P_{s,v, \{e_2\}}$ by  Algorithm $\ConstPath$ (which was supposed to prefer the divergence point that is closest to $s$).
Consider (b) in case where $D_1$ and $D_2$ are not $y$-interleaved. Assume towards contradiction that $e_2=(u_3,u_4) \notin \pi[y_1, y_2]$, hence $e_2 \in \pi[x_2,y_1]$. Note that in such a case, $x_2 \neq y_1$ (i.e., $D_1$ and $D_2$ are not $(x,y)$-interleaved).
There are now two $w-y_2$ paths in $G \setminus \{e_1,e_2\}$, namely,  $Q_1=D_1[w,y_1] \circ \pi[y_1, y_2]$ and $Q_2=D_2[w, y_2]$. By optimality, $|Q_1|=|Q_2|$.
Since $Q_1, Q_2 \subseteq G(x_1,u_4)$, it holds that $Q_1,Q_2=SP(w, y_2,G(x_1,u_4),W)$, leading to contradiction. The claim follows.
\QED

\begin{figure}[h!]
\begin{center}
\includegraphics[width=4in]{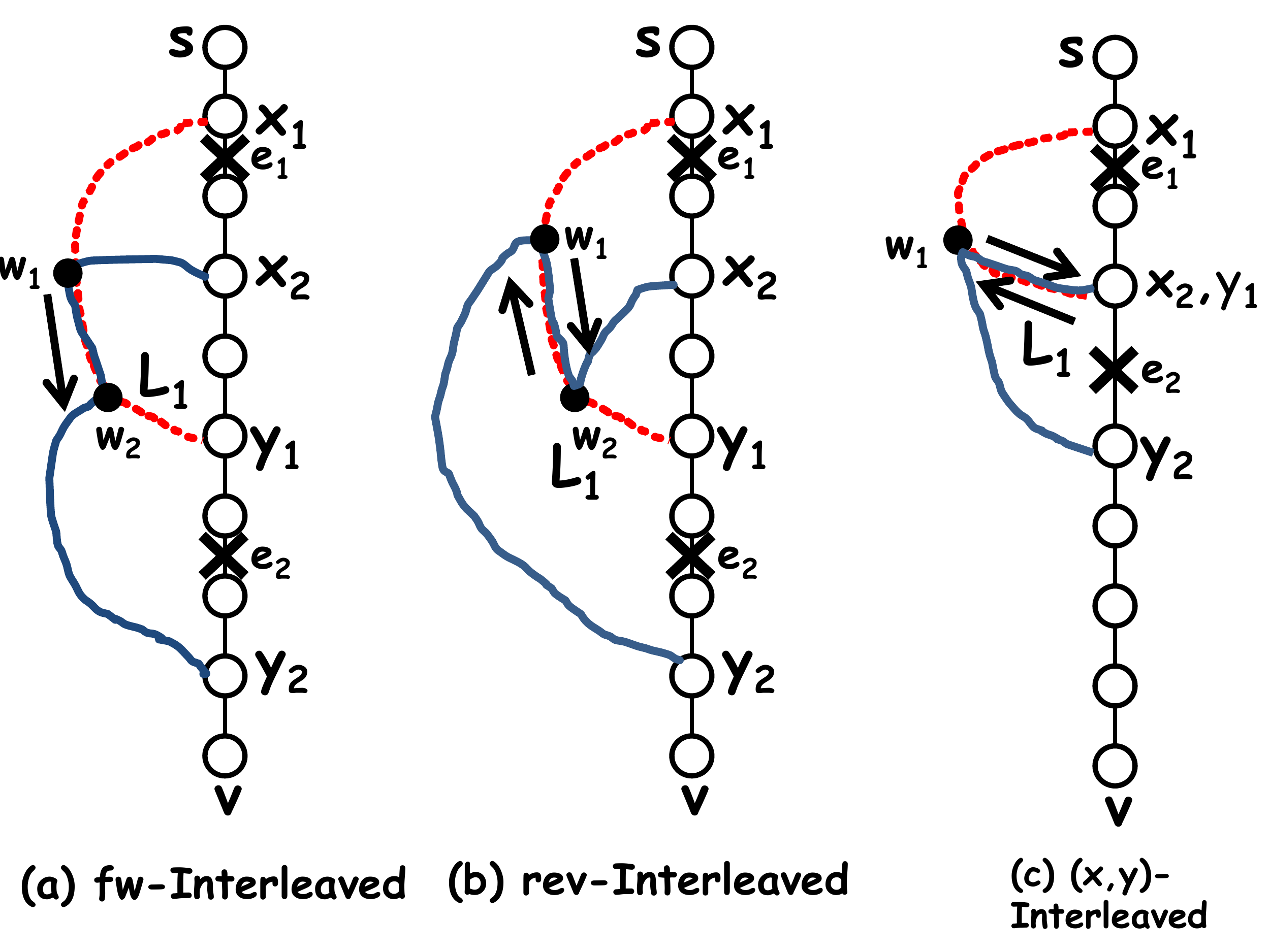}
\end{center}
\caption{Three types of dependent interleaved detours.
(a) fw-interleaved detours are dependent detour $D_1,D_2$ that use the common segment $D_1[w_1,w_2]$ in the same direction.
(b) rev-interleaved detours are dependent detour $D_1,D_2$ that use the common segment $D_1[w_1,w_2]$ in two opposite directions. (c) $(x,y)$-interleaved detours where $x_2=y_1$. The common segment $D_1[w_1,y_1]$ is used in opposite direction by $D_1$ and $D_2$.
\label{fig:fwrevdetours}}
\end{figure}

%\begin{claim}
%\label{cl:samexuy}
%If $x_i=x_j$ but $y_i<y_j$, then $e_j \in \pi(y_i,y_j)$.
%\end{claim}
%\Proof
%Assume towards contradiction that $e_j \notin \pi(y_i,y_j)$ then $e_j \in \pi(x_i, y_i)$. Overall, $e_i, e_j \in \pi(x_i,y_i)$. Let $e_i=(u_1,u_2), e_j=(u_3,u_4)$.
%Then, $P_{s,v,\{e_i\}}=SP(s, v, G(x_i,u_2),W)$ and
%$P_{s,v,\{e_i\}}=SP(s, v, G(x_i,u_4),W)$.
%Since $u$ be the deepest among $u_2$ and $u_4$ on $\pi(s,v)$, then since $e_i,e_j$ are both in $\pi(x_i,y_i)$, it holds that $P_{s,v,\{e_i\}}=SP(s,v,G(x_i,u),W)=P_{s,v,\{e_j\}}$. This leading to contradiction since $D_i \neq D_j$. The claim holds.
%\QED
\paragraph{Dependent detours.}
For dependent detours $D_1,D_2$, let $\First(D_1,D_2)$ (resp., $\Last(D_1,D_2)$) be the first (resp., last) vertex appearing on $D_1$ that is common to $D_2$.
Note that $\First(D_1,D_2)$ might not be equal to $\First(D_2,D_1)$ (in cases where the common segment $D_1 \cap D_2$ is traversed in opposite directions by the two detours).
We distinguish between two types of dependent and interleaved detours $D_1$ and $D_2$. Let $x_1 < x_2$ and let $w_1$ (resp., $w_2$) be the first (resp., last) vertex on $D_1$ that is common to $D_2$. I.e., there is no vertex in $D_1[x_1,w_1] \cup D_1[w_2,y_1]$ that is in $D_2$ as well. Note that by Cl. \ref{cl:jointdet}, we have the guarantee that $D_1[w_1,w_2]=D_2[w_1,w_2]$. Yet, since the graph is undirected, the two detour might traverse the common segment in opposite directions. If dependent and interleaved detours $x_1<x_2<y_1<y_2$ use the common segment $D_1 \cap D_2$ in the same direction (equivalently,  $\First(D_1,D_2)=\First(D_2,D_1)$) then $D_1,D_2$ are  \emph{fw-interleaved} otherwise they are \emph{rev-interleaved}. Note that dependent detours $D_1$ and $D_2$ which are $(x,y)$-interleaved, always use their common segment in opposite direction. See Fig. \ref{fig:fwrevdetours} for an illustration.
%%%%%%%%%%%%%%%%%%%
%%%%%%%%%%%%%%%%%%%
Finally, we summarize the possible configurations of dependent detours. By Cl. \ref{cl:depen_detdisjoint}
and \ref{cl:disjindep}, we have the following.
\begin{claim}
\label{cl:summ_depend}
Let $D_1$ and $D_2$ be dependent detours. Then,
(a) $D_1$ and $D_2$ are either $x$-interleaved, $y$-interleaved, $(x,y)$-interleaved, fw-interleaved or rev-interleaved.\\
(b) If $\First(D_1,D_2) \neq \First(D_2,D_1)$, then they are either rev-interleaved or $(x,y)$-interleaved.
\end{claim}
%\Proof
%Part (a) follows by Cl. \ref{cl:depen_detdisjoint}
%and \ref{cl:disjindep}.
%
%Consider part (b) and let $w_1=\First(D_1,D_2)$ and $w_2=\First(D_2,D_1)$ such that $w_1 \neq w_2$.
%
%By definition, if $D_1$ and $D_2$ are interleaved, then they are rev-interleaved. Clearly, by Cl. \ref{cl:jointdet}, $D_1$ and $D_2$ cannot be $y$-interleaved. Since the detours $D_1$ and $D_2$ intersect at some first point $w$ and collide from that point on up to $y$. Also, if $D_1$ and $D_2$ are $x$-interleaved, then the first common vertex is $x(D_1)=x(D_2)$. Finally, if $D_1$ and $D_2$ are $(x,y)$-interleaved dependent, the common segment by construction must be used in opposite directions. The claim holds.
%
%\QED
%
\paragraph{Excluded detour segment}
For detour $D_i$, the segment $\sigma \subseteq D_i$ an \emph{excluded segment} with respect to $D_i$ (or $D_i$-excluded for short) if there exists no new-ending path $P \in \mathcal{P}_v$ such that $D(P)=D_i$ and its second failing edge $F_2(P) \in \sigma$.

The next claim plays a major role in our analysis.
It concerns interleaved, $x$-interleaved and $(x,y)$-interleaved dependent detours $D_1$ and $D_2$, where $x_1\leq x_2$.
Letting $w=\Last(D_2,D_1)$ be the last point occurring on $D_2$ that is common to $D_1$, denote by $L_1=D_1[w,y_1]$ as the suffix of $D_1$ segment (see Fig. \ref{fig:fwrevdetours}(a,b,c) and Fig. \ref{fig:detours}(d)).
The claim states that this $L_1$ type segment is $D_1$-excluded segment of the detour, in the sense that there exists no $P \in \mathcal{P}_v$, such that $D(P)=D_1$ and its second failing edge $F_2(P)$ appears on $L_1$.
\begin{claim}
\label{cl:depen_detbad}
Let $D_1$ and $D_2$ be interleaved, $x$-interleaved or $(x,y)$-interleaved dependent detours, i.e., such that $x_1 \leq x_2 \leq y_1 < y_2$. Then, $D_1[w,y_1]$ is $D_1$-excluded where $w=\Last(D_2,D_1)$.
%
% and let $w$ be the last point on $D_2$ that is common to $D_1$.
%Then there is no new ending path $P \in \mathcal{P}_v$ such that $D(P)=D_1$ and $F_2(P) \in L_1=D_1[w, y_1]$.
\end{claim}
\Proof
Let $L_1=D_1[w,y_1]$ and
assume towards contradiction that there exists $P \in \mathcal{P}_v$, such that $D(P)=D_1$ and $F_2(P) \in L_1$. Let $e_1=F_1(P)$ (i.e., the edge $e_1$ is not an arbitrary edge that is protected by the detour $D_1$ but rather the first failing edge of the replacement path $P$ that is given by the contradictory assumption) and let $P_1=\pi[s, x_1] \circ D_1 \circ \pi[y_1, v] \in SP(s, v, G \setminus \{e_1\})$ be the path protecting against the failure of $e_1$. Observe that the edge $e_1$ appears on $\pi(s,v)$ \emph{before} $e_2$. This is because by Cl. \ref{cl:det_nested}(2), it holds that $e_2 \in \pi[y_1,y_2]$ and $e_1 \in \pi(x_1,y_1)$.
%and $e_2 \in \pi[y_1, y_2]$. Now, consider the case where $x_1=x_2$ and assume towards contradiction that $e_1$ appears after $e_2$ on $\pi(s,v)$. It then holds that there are two $s-v$ replacement paths in $G \setminus \{e_1,e_2\}$ given by  $P_1=\pi(s, x_1) \circ D_1 \circ \pi(y_1, v)$ and $P_2=\pi(s, x_2) \circ D_2 \circ \pi(y_2, v)$. Contradiction. The claim follows.
%\QED
%Assume, towards contradiction, that there exists a new ending path $P=P_{s,v,F}$ satisfying that $F_2(P)=t_k \in D_1[w_2, y_1]$. Hence $F_1(P)=e_1$ and $P \in SP(s, v, G \setminus \{e_1, t_k\})$.
%Let $e_1=(u_{1}, u_{2})$ and $e_2=(u_{3}, u_{4})$. By the selection of the replacement paths at step (1) of Algorithm $\ConstPath$, we have that $P_1=SP(s,v, G(x_1,u_1)\setminus \{e_1\},W)$, and $P_2=SP(s,v, G(x_2,u_3)\setminus \{e_2\},W)$.
Since $e_1$ appears on $\pi(s,v)$ before $e_2$ (i.e., $e_1$ is closer to $s$), it holds that $b(P)$, the unique $\pi$-divergence point of the new-ending path $P$ from $\pi(s,v)$, occurs above $e_1$ and hence also above $e_2$ (by Cl. \ref{cl:unique_nnew}(1) such $b(P)$ is guaranteed to exist). Since by Cl. \ref{cl:unique_nnew}(2), $P[b(P),v]$ is edge disjoint with $\pi(s,v)$, we have that $e_2 \notin P$ and overall $P \subseteq G \setminus \{e_1, e_2, F_2(P)\}$.
Consider an alternative $(\pi,\pi)$-replacement path $P'=P_{s,v,\{e_1,e_2\}}\in SP(s, v, G \setminus \{e_1,e_2\})$, i.e., both the failing edges of $P'$ occur on $\pi$.
Recall that $P'$ was added to the construction during step (2) of Alg. $\ConstPath$, i.e., before $P$ was added in step (3).  Hence, $\LastE(P) \neq \LastE(P')$ (since $P$ is new-ending).
We now consider two cases depending on whether or not the second failing edge $F_2(P)$ appears on the $(\pi,\pi)$ replacement path.
\dnsparagraph{Case (1): $F_2(P) \notin P'$}
Since $e_2 \notin P$, we get that both $P$ and $P'$ are two $s-v$ shortest paths in $G \setminus \{e_1, e_2, F_2(P)\}$. By optimality, $|P|=|P'|$. So we end with contradiction to the selection of $P$ by the algorithm.
%Contradiction to the fact that $P$ is new ending, since there exists an $s-v$ path in $G\setminus F(P)$ that uses edges from $E_2(\pi)$.

\dnsparagraph{Case (2): $F_2(P)=(q_1, q_2) \in P'$}
We now define the path $\widehat{P}=\pi(s,x_1) \circ D_1[x_1,w] \circ D_2[w,y_2]\circ \pi(y_2,v)$. Recall that $w=\Last(D_2,D_1)$ is the last point on $D_2$ and hence $F_2(P) \notin D_2[w,y_2]$.
Since by the contradictory assumption $F_2(P)$ is in the excluded region, i.e., $F_2(P)\in L_1=D_1[w,y_1]$, it holds that $F_2(P) \notin \widehat{P}$.
%Note that by the optimality of $P_2=\pi(s,x_2) \circ D_2[x_2,w_1] \circ P''[w_1,v]$, we have that
%\begin{equation}
%\label{eq:imp_eq}
%|P''[w_1,v]|=\dist(w_1,v, G \setminus \{e_2\})~.
%\end{equation}
Since step (2) of Alg. $\ConstPath$ attempts first to select $\widehat{P}$ as the replacement-path for the pair $F=\{e_1,e_2\}$, by the fact that eventually another path, namely $P'$, was selected as the replacement path $P_{s,v,\{e_1,e_2\}}$, necessarily
\begin{equation}
\label{eq:notreppp}
|\widehat{P}|>|P'|~.
\end{equation}
\par We next bound the length of $|\widehat{P}|$ and show its optimality, hence leading to contradiction.
Since $F_2(P)=(q_1,q_2) \in P'$, there exist
two $s-q_2$ shortest-paths in $G \setminus \{e_1,e_2\}$, namely,
$Q_1=\pi(s,x_1) \circ D_1[x_1,q_2]$ and $Q_2=P'[s,q_2]$.
Note that $e_2 \notin Q_1$, since the divergence point $x_1$ is above it on $\pi(s,v)$. In addition, note that $w \in Q_1$ as $q_2 \in D_1[w,y_1]$.
By optimality of the paths $P_1=P_{s,v,\{e_1\}}$ and $P'=P_{s,v,\{e_1,e_2\}}$, we get that $|Q_1|=|Q_2|$.
Hence,
\begin{eqnarray*}
|P'|&=&|Q_2|+|P'[q_2,v]|=|Q_1|+|P'[q_2,v]|
\\&=&
\dist(s,w, G \setminus \{e_1,e_2\})+\dist(w,q_2, G \setminus \{e_1,e_2\})+\dist(q_2,v, G \setminus \{e_1,e_2\})
\\&\geq&
\dist(s,w, G \setminus \{e_1\})+\dist(w,v, G \setminus \{e_2\})=|P_{s,v,\{e_1\}}[s,w]|+|P_{s,v,\{e_2\}}[w,v]|
=|\widehat{P}|~,
\end{eqnarray*}
contradiction by Eq. (\ref{eq:notreppp}). The claim follows.
\QED
Note that for rev-interleaved or $(x,y)$-interleaved dependent detours $D_1$ and $D_2$ where $x_1\leq x_2$, the excluded segment $L_1 \subseteq D_1$ contains that shared segment $D_1 \cap D_2$ (see the segments $D_1[w_1,y_1]$ in Fig. \ref{fig:fwrevdetours}(b,c)). We have the following.
\begin{corollary}
\label{cl:rev_nomutual}
Let $D_1,D_2$ be dependent $(x,y)$-interleaved or rev-interleaved detours where $x_1\leq x_2$. Then there exists no path $P \in \mathcal{P}_v$ such that $D(P)=D_1$ and $F_2(P) \in D_1 \cap D_2$.
\end{corollary}

\subsubsection{The kernel subgraph of detours}
\label{sec:tool_gradual}
For every vertex $v$, and a subset of $(\pi,\sf D)$-replacement paths $\mathcal{P} \subseteq \mathcal{P}_v$, let $\mathcal{D}=\{D(P), P \in \mathcal{P}\}$ be the set of detours of these paths. Clearly the set of relevant faulty edges is given by the subgraph $G_v(\mathcal{D})=\pi(s,v) \cup \{D_i \mid D_i \in \mathcal{D}\}$.
In this section, we show that in order to analyze the structure of the new-ending $(\pi,\sf D)$-paths, it is sufficient to consider a subgraph $\KernelGraph_v(\mathcal{D})$ of $G_v(\mathcal{D})$, denoted hereafter as the \emph{kernel subgraph of the detours}. When, $v$ is clear from the context, we simply write $\KernelGraph(\mathcal{D})$. To define the $\KernelGraph(\mathcal{D})$ subgraph, we describe a construction procedure which gradually adds segments of detours $D_i \in \mathcal{D}$ according to some predefined ordering. Essentially, from each $D_i$, only a certain segment $D[x_i,w_i]$ is added to $\KernelGraph(\mathcal{D})$.
We begin by describing the construction of $\KernelGraph(\mathcal{D})$ and then establish some of its useful properties.

\paragraph{The construction of the kernel graph $\KernelGraph(\mathcal{D})$.}
The algorithm first $(x,y)$-orders the detours, resulting with $\overrightarrow{\mathcal{D}}=\{D_1, \ldots, D_t\}$ where
$D_1 \prec D_2 \prec \ldots \prec D_t$.
Initially set $\KernelGraph^0=\emptyset$. Let $w_1=y_1$.
Add the $D_i$'s in a sequential manner: at step $i$,
we follow the detour $D_i$ and add its edges until we hit the first vertex $w_i$ on $D_i$ that was already added to the kernel graph $\KernelGraph^{i-1}$ by the previous step. We then add only $D_i[x_i,w_i]$ to the subgraph of $\KernelGraph^{i-1}$.
Formally, at step $i$, the segment of $D_i[x_i, w_i]$ is added to $\KernelGraph^{i-1}$, where $w_1=y_1$ and for every $i>1$, $w_i \in D_i$ is the first common vertex of $D_i$ and $\KernelGraph^{i-1}$. Hence, there exists some $j<i$, such that $w_i \in D_{j}[x_j, w_j]$.
Let $\KernelGraph^i=\KernelGraph^{i-1}\cup D_i[x_i, w_i]$. Finally, let $\KernelGraph(\mathcal{D})=\KernelGraph^t=\bigcup_{i=1}^{t} D_i[x_i, w_i]$.

\begin{figure}[htbp]
\begin{center}
\includegraphics[width=5in]{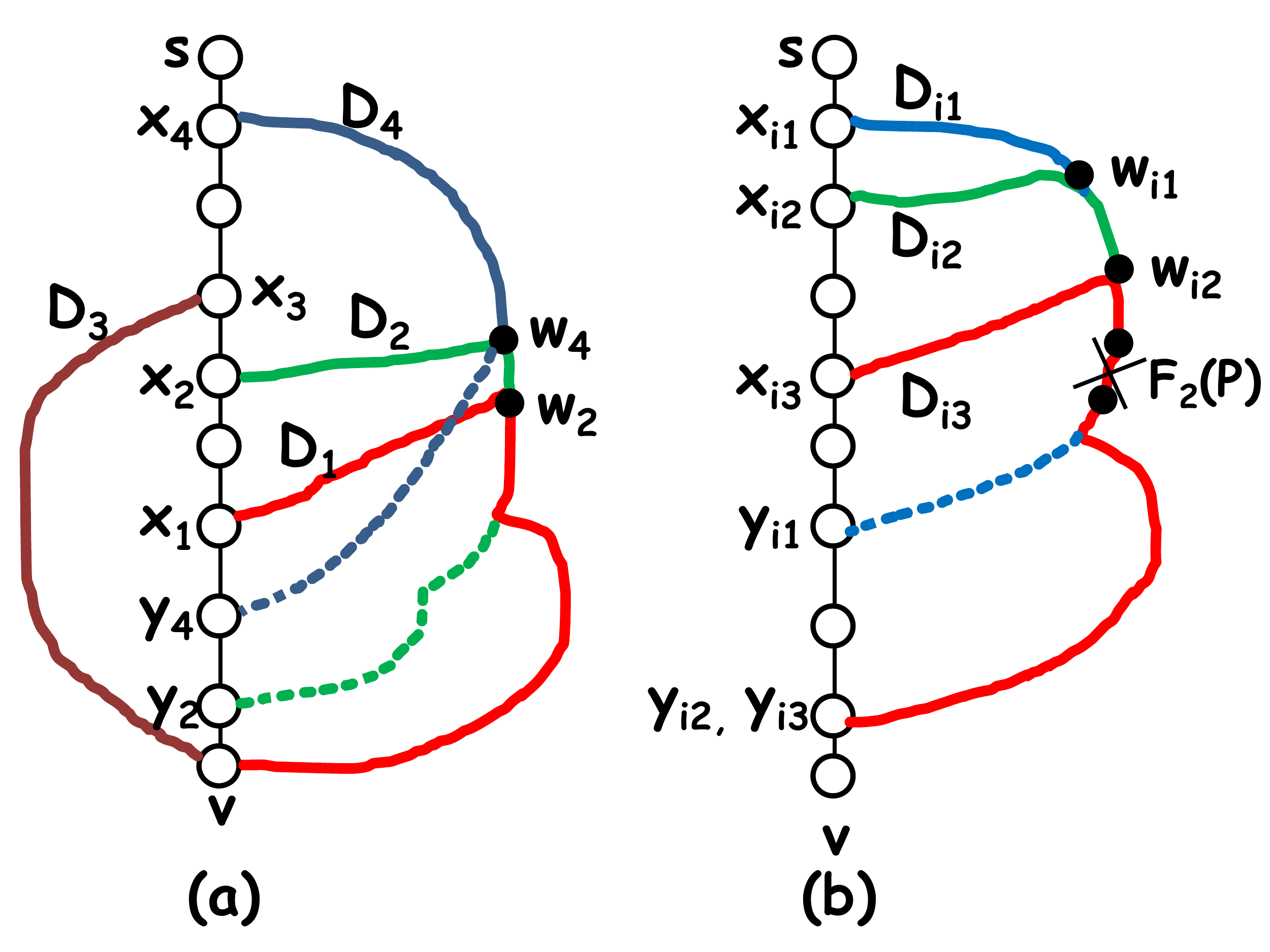}
\end{center}
\caption{Schematic illustration of the kernel graph and its useful properties. (a) The kernel graph $\KernelGraph(\mathcal{D})$ where the detours $\mathcal{D}=\{D_1,D_2,D_3,D_4\}$ are inserted in this order. For detour $D_i$, the vertex $w_i$ is the first vertex appearing in $\bigcup_{j=1}^{i-1}D_j[x_j,w_j]$.
The dotted segments are not included in the kernel, hence $D_4[w_4,y_4]$ is not taken into the kernel. The detours $D_1$ and $D_3$ are non-truncated and the detours $D_2$ and $D_4$ are truncated. The detour $D_1$ (resp., $D_2$) is the breaker of $D_2$ (resp., $D_4$). Hence, $D_1=\Breaker(D_2)$ and $D_2=\Breaker(D_4)$.
(b) Illustration for Lemma \ref{lem:detour_gradual}. Note that the sequence of vertices $w_{i_1}, \ldots, w_{i_{k-1}}$ is on $D_{i_1}$, getting closer to the failing edge $F_2(P)$.
\label{fig:kernel}}
\end{figure}

Note that some of the detours $D_i$ of $\mathcal{D}$ are added completely to $\KernelGraph(\mathcal{D})$  (i.e., $w_i=y(D_i)$) and others are added only partially, since one of their vertices has been added before. We refer to these detours as \emph{truncated} detours. Formally, a detour $D_i$ is truncated if $w_i \neq y_i$, otherwise, it is \emph{non-truncated}. For every truncated detour $D_j$, let $D_{j'} \in \mathcal{D}$ be some detour that precedes $D_j$ in the ordering, i.e., $j'<j$ and in addition, $w_j \in D_{j'}[x_{j'},w_{j'}]$. We call this detour the \emph{breaker} of $D_j$, denoted hereafter by $\Breaker(D_j)$ (the detour $D_j$ might have several breakers, in such a case one is chosen arbitrarily).
See Fig. \ref{fig:kernel}(a) for an illustration.
%
%\begin{observation}
%\label{obs:notsamex}
%Let $D_j$ be a truncated detour with $D_{j'}=\Breaker(D_j)$ %and such that $w_j \neq x_j$.
%Then $x_j < x_{j'}$.
%\end{observation}
%\Proof
%By the $(x,y)$-ordering, $x_j \leq x_{j'}$. Since $w_j \neq x_j$ is the first vertex in $D_j$ occuring on $D_{j'}[x_{j'},w_{j'}]$, the claim follows.
%\QED
%
\par The next key lemma shows that the kernel subgraph $\KernelGraph(\mathcal{D})$ consists of the faulty edges $F_2(P)$ for every new-ending path $P$ whose detour $D(P)$ is in $\mathcal{D}$.
\begin{lemma}
\label{lem:detour_gradual}
For every $(\pi,\sf D)$-replacement path $P \in \mathcal{P}_v$ with $D=D(P) \in \mathcal{D}$ and $F_2(P)=(q_1,q_2)$, it holds that
$D[x(D), q_2] \subseteq \KernelGraph(\mathcal{D})$.
\end{lemma}
\Proof
Let $P \in \mathcal{P}_v$ be such that its detour $D(P) \in \mathcal{D}$ was added to the kernel graph $\KernelGraph(\mathcal{D})$ at step $t_{i_1}$. Let $D_{i_1}=D(P)$. There are two cases. If $F_2(P) \in D_{i_1}[x_{i_1},w_{i_1}]$, then $D_{i_1}[x_{i_1},w_{i_1}] \subseteq \KernelGraph(\mathcal{D})$ and the claim holds.
Hence, it remains to consider the case where $F_2(P) \notin D_{i_1}[x_{i_1},w_{i_1}]$. See Fig. \ref{fig:kernel}(b) for an illustration.
Note that in such a case, $D_{i_1}$ is a truncated detour.
\par Consider the maximal sequence $D_{i_1}, \ldots, D_{i_k}$ where $D_{i_j}=\Breaker(D_{i_{j-1}})$ for $j \in \{2, \ldots, k\}$ such that $k$ is the first index satisfying either that $F_2(P) \in D_{i_k}[x_{i_k},w_{i_k}]$ or that $D_{i_k}$ is a non-truncated detour, i.e., it was added in its entirety to the kernel graph $\KernelGraph(\mathcal{D})$. Since the first detour $D_1$ in the $(x,y)$-ordering was added in its entirety to the kernal, the terminating element in this sequence is well defined.

We now prove by induction that for every $j \in \{1, \ldots, k\}$, the following holds.
\begin{description}
\item{(a)}
$F_2(P) \in D_{i_j}$, and for every $j \leq k-1$,
\item{(b1)}
$w_{i_j} \in D_{i_1}$, and in particular $w_{i_j}$ occurs on $D_{i_1}$ before the edge $F_2(P)$ (i.e., $w_{i_j} \in D_{i_1}[x_{i_1},q_1]$)
\item{(b2)}
$D_{i_1}[w_{i_{j-1}}, w_{i_{j}}]\subseteq \KernelGraph(\mathcal{D})$, where $w_{i_0}=x_{i_1}$.
\end{description}

The base of the induction $j=1$ holds by definition.
Assume it holds up to $j-1$ and consider $j$.
We begin with part (a) and assume towards contradiction that $F_2(P) \notin D_{i_j}$.
By part (b1) of the induction assumption for step $j-1$, it holds that $w_{i_{j-1}} \in D_{i_{1}}$ and it appears on $D_{i_1}$ before the failing edge $F_2(P)$. Since $D_{i_j}$ is the breaker of $D_{i_{j-1}}$, it holds that $w_{i_{j-1}} \in D_{i_{j}}[x_{i_j}, w_{i_j}]$. Hence, $D_{i_1}$ and $D_{i_j}$ are dependent. %Therefore, the detours $D_{i_1}$ and $D_{i_j}$ intersect before the edge $F_2(P)$.
%Therefore, the first vertex on $D_{i_1}$ that is common to $D_{i_j}$, namely, $\First(D_{i_1},D_{i_j})$, occurs on $D_{i_1}$ before the edge $F_2(P)$.
%%%%%%%%%%%%%%%%%%%%%%%%%%%%%%%%%%%%%%%%%%%%%%%%%%%%%%%%%%%%%%%%%%%
Let $w'$ be the last point on $D_{i_j}$ that is common to $D_{i_1}$. By the ordering of $\mathcal{D}$, $x_{i_j}\geq x_{i_1}$.
We distinguish between two cases.
Case (1): $x_{i_j}>x_{i_1}$. This case is further divided into 3 subcases depending on the value of $y_{i_j}$ with respect to $y_{i_1}$.
Case (1.1): $y_{i_j}=y_{i_1}$. In this case, by Cl. \ref{cl:jointdet}, $D_{i_1}[w_{i_{j-1}}, y_{i_1}]=D_{i_j}[w_{i_{j-1}}, y_{i_j}]$.
This is in contradiction by part (b1) for step $j$, as $F_2(P) \in D_{i_1}[w_{i_{j-1}},y_{i_1}]$ and hence $F_2(P) \in D_{i_j}$.\\
Case (1.2): $y_{i_j}<y_{i_1}$. Then $D_{i_j}$ is nested in $D_{i_1}$, hence by Cl. \ref{cl:depen_detdisjoint}, they are independent, in contradiction to the existence of a common vertex $w_{i_{j-1}}$.
Case (1.3): $y_{i_j}>y_{i_1}$.
First, observe that since $F_2(P) \in D_{i_1}[w_{i_{j-1}},y_{i_1}]$, by Cl.  \ref{cl:depen_detbad}
it holds that $D_{i_1}$ and $D_{i_2}$ are neither rev-interleaved not $(x,y)$-interleaved. Hence, $D_{i_1}$ and $D_{i_2}$ are fw-interleaved. By Cl. \ref{cl:depen_detbad}, there exists no $P'$ with $D(P')=D_{i_1}$ and $F_2(P') \in D_{i_1}[w',y_{i_1}]$, i.e., there are no failures in $D_{i_1}$ after $w'$. Since $F_2(P) \in D_{i_1} \setminus D_{i_j}$ and it appears on $D_{i_1}$ after the common segment $D_{i_1} \cap D_{i_j}$, it holds that $F_2(P) \in D_{i_1}[w', y_{i_1}]$, leading to contradiction.

It remains to consider Case (2) where $x_{i_1}=x_{i_j}$ (i.e., $D_{i_1}$ and $D_{i_j}$ are $x$-interleaved). By the ordering of $\mathcal{D}$, it holds that $y_{i_1}<y_{i_j}$, and hence by Cl. \ref{cl:jointdet}, $D_{i_1}[x_{i_1},w']=D_{i_j}[x_{i_j}, w']$. By the contradictory assumption, $F_2(P) \in D_{i_1}[w',y_{i_1}]$, leading to contradiction by Cl. \ref{cl:depen_detbad}.
%%%%%%%%%%%%%%%%%%%%%%%%%%%%%%%%%%%%%%%%%%%%%%%%%%%%%%%%%%%%%%55
Hence, part (a) of the induction hypothesis holds. \par We now turn to part (b1).
Since $j\leq k-1$, by the stopping criteria of the sequence, it holds that $F_2(P)$ was not included in the prefix taken from $D_{i_j}$, i.e., $F_2(P) \notin D_{i_j}[x_{i_j}, w_{i_j}]$. So by part(a), $F_2(P) \in D_{i,j}[w_{i_j}, y_{i_j}]$ (i.e., $F_2(P)$ appears on $D_{i_j}$ after $w_{i_j}$).
Since $D_{i_j}$ is the breaker of $D_{i_{j-1}}$, it holds that $w_{i_{j-1}} \in D_{i_j}[x_{i_j},w_{i_j}]$. Hence, $w_{i_j} \in D_{i_{j}}[w_{i_{j-1}}, q_1]$.
By part (b1) of the induction assumption, $w_{i_{j-1}} \in D_{i_1}$ and by definition, $q_1\in D_{i_1}$.
By Cl. \ref{cl:jointdet}, $D_{i_{1}}[w_{i_{j-1}}, q_2]=D_{i_{j}}[w_{i_{j-1}}, q_2]$, so in particular, $w_{i_j} \in D_{i_1}$ as well.
By the part (b1) of the induction assumption, $F_2(P)$ occurs on $D_{i_1}$ after $w_{i_{j-1}}$.
By the fact that $w_{i_{j-1}} \in D_{i_j}[x_{i_j},w_{i_j}]$ and $F_2(P) \in D_{i_j}[w_{i_j},y_{i_j}]$, it also holds that $F_2(P)$ occurs on $D_{i_j}$ after $w_{i_{j-1}}$.
Hence, we conclude that the common segment $D_{i_{1}}[w_{i_{j-1}}, q_2]$ of these detours is used in the same direction: from $w_{i_{j-1}}$ to $q_2$ via $w_{i_j}$. Therefore, $w_{i_j}$ appears on $D_{i_1}$ before the failing edge $F_2(P)$, so (b1) holds.

%%%%%%%%%%%%%%%%%%%%%%%%%%%%%%%%%%%%%%%%%%%%%%%%%%%%%%%%%%%%%%
Finally, consider part (b2).
By part (b1) for steps $j-1$ and $j$, we have that
$w_{i_{j-1}},w_{i_j} \in D_{i_1}$. By the definition of the detour $D_{i_j}$, it also holds that $w_{i_{j-1}},w_{i_j} \in D_{i_j}$. Hence, by Cl. \ref{cl:jointdet}, it holds that $D_{i_j}[w_{i_{j-1}}, w_{i_j}]=D_{i_1}[w_{i_{j-1}}, w_{i_j}]$. Since $D_{i_j}[w_{i_{j-1}}, w_{i_j}]\subseteq D_{i_j}[x_{i_j},w_{i_j}]$, we have that the prefix $D_{i_1}[x_{i_j},w_{i_j}]$ was taken into the kernel, i.e., $D_{i_1}[x_{i_j},w_{i_j}] \subseteq \KernelGraph(\mathcal{D})$, so (b2) holds as well.

We are now ready to complete the proof of the lemma.
Since $F_2(P) \in D_{i_k}$ and in particular $F_2(P) \in D_{i_k}[x_{i_k},w_{i_k}]$, we get that $F_2(P)$ is in $\KernelGraph(\mathcal{D})$.
By part (b2), we have that for every $j\leq k-1$,
$D_{i_1}[w_{i_{j-1}}, w_{i_j}] \subseteq \KernelGraph(\mathcal{D})$.  Combining this with the fact that $D_{i_k}[w_{i_{k-1}}, q_2] \subseteq D_{i_k}[x_{i_k}, w_{i_k}]$ was also added to $\KernelGraph(\mathcal{D})$, we get that
$$D_{i_1}[x_{i_1}, w_{i_1}] \circ D_{i_1}[w_{i_1}, w_{i_2}] \circ \ldots \circ  D_{i_1}[w_{i_{k-2}},w_{i_{k-1}}]\circ D_{i_1}[w_{i_{k-1}}, q_2]=D_{i_1}[x_{i_1},q_2] \subseteq \KernelGraph(\mathcal{D}).$$
The claim follows.
\QED

%%%%%%%%%%%%%%%%%%%
%%%%%%%%%%%%%%%%%%%

\subsection{New-ending paths protecting against two edge faults}
In this section, we turn to present several properties of new-ending paths $P_{\tau}$ (i.e., that were not contained in $G_{\tau-1}(v)$, and hence introduced $\LastE(P_{\tau})$ to the subgraph $H(v)$) and then classify the set of new-ending paths $\mathcal{P}_v$ into five path classes.

\subsubsection{Properties of new-ending replacement paths}
The following claim summarizes some basic properties of new ending $(\pi,\sf D)$-replacement paths that are useful in our analysis. It states that for every new-ending $(\pi,\sf D)$-path $P_i$, the $\pi$-divergence point $b(P_i)$ and the $\sf D$-divergence point $c(P_i)$ (if exists) are unique.
\begin{claim}
\label{cl:rp_prop1}
Let $P_{\tau}=P_{s,v,F_\tau}\in SP(s, v, G \setminus F_\tau)$ be the new-ending $(\pi,\sf D)$-replacement path added in step $\tau$ of Alg. $\ConstPath$ (i.e., it was not contained in the graph $G_{\tau-1}(v)$ defined in step 3 of the algorithm. Recall that $F_1(P_{\tau})=e_\tau \in \pi(s,v)$, $F_2(P_{\tau})=t_\tau \in D_\tau$ and $D_\tau=[x_\tau=v_0, v_1, \ldots, v_k=y_\tau]$ is the detour segment of $P_{s,v,\{e_\tau\}}$. Let $b_\tau$ be the first divergence point of $P_{s,v,F_\tau}$ from $\pi(s,v)$.
Then:\\
%I.e., and letting $w \in \pi(s,v)\setminus \{b_{i,j},v\}$ be the first common vertex of $P[b_{i,j},v]$ and $\pi(s,v)$ then $P_{s,v,F}[w,v]=\pi(w,v)$. \\
(1) there is no alternative replacement path whose first divergence point from $\pi(s,v)$ appears before $b_\tau$.\\
(2) if $b_\tau \neq x_\tau$ then $E(P_{s,v,F_\tau}) \cap E(D_\tau)=\emptyset$. \\
(3) if $b_\tau = x_\tau$, let $c_\tau$ be the first divergence point of $P_{s,v,F_\tau}$ from $D_\tau$. Then:\\
(3.1) $P_{s,v,F_\tau}=\pi(s, b_\tau) \circ D_\tau[x_\tau,c_\tau] \circ P_{s,v,F_\tau}[c_\tau,v]$ and $P_{s,v,F_\tau}[c_\tau,v]$ is edge disjoint with $D_\tau \cup \pi(s,v)$.\\
(3.2) there is no alternative replacement path with divergence point $b_{\tau}$ whose first divergence point from $D_\tau$ appears on $D_\tau$ before $c_\tau$ (i.e., closer to $x_\tau$).
\end{claim}
\Proof
Claim (1) follows immediately as the algorithm chooses the divergence point closest to $s$.
Consider claim (2). By the definition of $b_{\tau}$, $\pi(s,b_{\tau})\subseteq P_{\tau}$. Recall that in the case where $b_{\tau} \neq x_\tau$,

Assume towards contradiction that $b_{\tau} \neq x_\tau$ and yet $E(P_{\tau}) \cap E(D_\tau)\neq \emptyset$.
Let $w \in D_\tau \cap P_\tau$ be a common vertex of $P_\tau$ and $D_\tau$.
First assume that $w$ appears on $D_\tau$ above the second failing edge $t_\tau$. We show that in this case $b_{\tau}=x_\tau$. Since the algorithm defines
$P_{\tau} =SP(s,v, G(b_{\tau},v) \setminus F_{\tau},W)$, $b_{\tau}$ is the unique divergence point of $\pi(s,v)$ and $P_{\tau}$.
If $b_{\tau}$ is above $x_\tau$ on $\pi(s,v)$ then there are two distinct $s-w$ shortest paths in $G\setminus F_\tau$, , namely, $Q_1=\pi(s,x_\tau)\circ D_\tau[x_\tau,w]$ and $Q_2=P_{\tau}[s,w]$. By optimality of these subpaths, $|Q_1|=|Q_2|$, hence we end with contradiction to Claim \ref{cl:onefstru}(2) for the path $P_{s,v, \{e_i\}}$.
Similarly, if $b_{\tau}$ is below $x_\tau$ we end with contradiction to Cl. \ref{cl:rp_prop1}(1) for the path $P_{\tau}$.
Hence, $b_{\tau}=x_\tau$.
Next, consider the case where $w$ appears on $D_\tau$ \emph{after} the failing edge $t_{\tau}$. In this case, we get that there are two distinct $w-v$ paths in $G \setminus F_\tau$, namely, $Q_1=P_{\tau}[w,v] \subseteq P_{\tau}[b_{\tau},v]$ and $Q_2=D_\tau[w, y_\tau] \circ \pi(y_\tau,v)$. By the optimality of these subpaths, $|Q_1|=|Q_2|$, in contradiction to the fact that $P_\tau$ is new-ending.
Finally, consider claim (3) where $b_{\tau}=x_\tau$. Alg. $\ConstPath$ selects the replacement path $P_{\tau}$ whose closest divergence point on $D_\tau$ and by the definition of this path, this divergence point is forced to be unique i.e., the edges of $D_{\tau}[c_{\tau},y_{\tau}]$ are omitted from the graph $G_{\sf D}(c_{\tau})$ in which $P_{\tau}$ is defined. Both parts of claim (3) follows.
\QED
We conclude this section, by providing a useful property for new-ending $(\pi,\sf D)$ replacement paths $P$ that intersect their detour $D(P)$.
The next lemma states that the $\sf D$-divergence point $c(P)$ of $P$ and $D(P)$ is distinct.
\begin{lemma}
\label{lem:uniquebp}
For every $(\pi,\sf D)$-paths $P_1, P_2 \in \mathcal{P}_v$
satisfying that $E(P_i) \cap E(D(P_i)) \neq \emptyset$ for $i \in \{1,2\}$, it holds that $c(P_1) \neq c(P_2)$.
\end{lemma}
\Proof
Assume, towards contradiction, that there exists two $(\pi,\sf D)$ new-ending paths $P_1,P_2 \in \mathcal{P}_v$ such that $c(P_1) = c(P_2)$. Let $c_i=c(P_i), e_i=F_1(P_i), t_i=F_2(P_i), D_i=D(P_i)$, for $i=\{1,2\}$.
Since $P_1, P_2$ intersect with their detours $D_1$ and $D_2$ respectively, it holds that $c_1 \in D_{1}$ and $c_2 \in D_{2}$, hence $D_{1}$ and $D_{2}$ intersect at some point \emph{not after} $c_1$.
First note that $e_1 \neq e_2$, because otherwise, $D_1=D_2$ and since $t_1,t_2$ occur after $c_1$, it holds that there are two distinct new-ending $c_1-v$ shortest paths in $G \setminus \{e_1,t_1,t_2\}$, namely, $P_1[c_1,v] \neq P_2[c_1,v]$, contradiction since the selection of the latter of them by Alg. $\ConstPath$ could have been avoided.
\par From now on assume, without loss of generality, that $e_1$ is above $e_2$ on $\pi(s,v)$ (i.e., closer to $s$).  Let $w$ be the last vertex on $D_1$ that is common to $D_2$. Since $e_1$ is above $e_2$ and $D_1$ and $D_2$ are dependent  (they share a common vertex $c_1$), it holds by Cl. \ref{cl:disjindep} and \ref{cl:depen_detdisjoint} that they are neither nested nor non-nested.
Hence, we have that
\begin{equation}
\label{eq:unic}
x_1 \leq x_2~.
\end{equation}
(as otherwise by Cl. \ref{cl:det_nested}(1), $e_2 \in \pi(x_2,x_1)$ is above $e_1$).
We consider three cases.
\dnsparagraph{Case (a): $t_{1}, t_{2} \in D_{1} \cap D_{2}$}
Note that by Cor. \ref{cl:rev_nomutual}, $D_1$ and $D_2$ are neither rev-interleaved nor $(x,y)$-interleaved (i.e., in such a case, $w$ is also the last vertex on $D_2$ that is common to $D_1$).
By Cl. \ref{cl:rp_prop1}(3.1), it holds that $\pi(s, x_i) \circ D_i[x_i,c_i] \subseteq P_i$ for $i\in \{1,2\}$. Hence $t_1$ and $t_2$ occur on $D_1 \cap D_2$ after $c_1$, i.e, $t_1,t_2 \in D_1[c_1,w]=D_2[c_1,w]$. Note that since $D_1$ and $D_2$ are neither rev-interleaved nor $(x,y)$-interleaved, by Cl. \ref{cl:summ_depend}, the common segment $D_1[c_1,w]$ is used by the two detours in the same direction. There are now two distinct new-ending $c_1-v$ shortest paths in $G \setminus \{e_{1}, e_{2}, t_{1}, t_{2}\}$, namely, $P_{1}[c_1,v]$ and $P_{2}[c_1,v]$ (these paths are distinct as $\LastE(P_1) \neq \LastE(P_2)$). This is again in contradiction to the selection of $P_1$ by Alg. $\ConstPath$, since it was constructed after $P_2$, so its last edge could have been avoided.
\dnsparagraph{Case (b): $t_1 \in D_1 \setminus D_2$}
Since $t_1$ appears on $D_1$ after the common vertex $c_1$, yet it is not in $D_2$, we get that $t_1$ appears on $D_1$ after the last common vertex with $D_2$, namely, $w$.
In particular, $t_1$ appears on the non-common suffix $D_1[w,y_1]$. Hence, $D_1$ and $D_2$ are \emph{neither} $y$-interleaved nor $(x,y)$-interleaved (since in these cases, $w=y_1$).
%In addition, since $D_1$ and $D_2$ are dependent, by Cl. \ref{cl:summ_depend}
%they are either fw-interleaved, rev-interleaved or $x$-interleaved.
Let $w'$ be the last vertex on $D_2$ that is common to $D_1$. Note that $D_1[w,y_1]\subseteq D_1[w',y_1]$ (I.e., if $D_1$ and $D_2$ are rev-interleaved, $w'$ is the first vertex on $D_1$ that is common to $D_2$ and in other cases, where the common segment is used in the same direction by both detours, $w=w'$).
By Eq. (\ref{eq:unic}), Cl.  \ref{cl:depen_detbad}, and the fact that $D(P_1)=D_1$ and $t_1=F_2(P)$, we get that $t_1 \notin D_1[w,y_1]$, contradicting the fact that $t_1 \in D_1[w,y_1]$.
\dnsparagraph{Case (c): $t_{2}\in D_{2} \setminus D_{1}$ and $t_1 \in D_1 \cap D_2$}
By combining Eq. (\ref{eq:unic}), the fact that the detours are dependent and that $t_1 \in D_1 \cap D_2$ , by Cor. \ref{cl:rev_nomutual}, it holds that $D_1$ and $D_2$ are neither rev-interleaved nor $(x,y)$-interleaved.
Hence, by Cl. \ref{cl:summ_depend}(b), the common segment $D_1 \cap D_2$ is used in the same direction by both detours. In particular, $w$ is also the last vertex on $D_2$ that is common with $D_1$.
Since $t_2$ occurs after the common vertex $c_1$,  in this case, $t_2 \in D_2[w,y_2]$ and $t_1 \in D_1[c_1,w]$. Since $c_1=c_2$ appears before the failing edge $t_1$ on $D_1$, it holds that $t_1 \notin P_2$.
We now break case (c) further into two subcases.
Case (c1): $t_{2} \notin P_{1}[c_1, v]$. Then again there are two distinct $c_1-v$ shortest paths in $G \setminus \{e_{1}, e_{2}, t_{1}, t_{2}\}$, namely, $P_1[c_1,v]$ and $P_2[c_1,v]$, and we end with contradiction to the construction of these paths by Alg. $\ConstPath$.

Case (c2): $t_{2} \in P_1[c_1,v]$. Let $t_2=(z_1, z_2)$ and recall that $t_2 \in D_2[w, y_2]$.
We show that in such a case, there are two $z_2-v$ shortest-paths in $G\setminus \{e_1, e_2, t_1\}$, namely, $Q_1=P_1[z_2,v]$ and $Q_2=D_2[z_2,y_2]\circ \pi[y_2,v]$.
%First observe that since $D_1$ and $D_2$ are dependent, and since they are neither rev-interleaved or $(x,y)$-interleaved, by Cl. \ref{cl:summ_depend}, they are either fw-interleaved, $x$-interleaved or $y$-interleaved.
%
To see this, note that by Eq. (\ref{eq:unic}), $e_2 \notin Q_1$. In addition, since $t_1 \notin D_2[w,y_2]$ and $e_1 \notin \pi(y_2,v)$ (as $e_1$ is above $e_2$), it also holds that $e_1,t_1 \notin Q_2$. By optimality of these subpaths (as $Q_1 \subseteq P_{s,v,\{e_1,t_1\}}$ and $Q_2 \subseteq P_{s,v,\{e_2\}}$), we have that $|Q_1|=|Q_2|$.  We then end with contradiction to the selection of $P_1$ by Alg. $\ConstPath$ (since the algorithm could have used the alternative $s-v$ replacement path $P_1[s,z_2] \circ Q_2$ in $G \setminus F(P_1)$, which is not new-ending).
\QED

\subsubsection{New-ending path classification}
\label{sec:nw_class}
Recall that the set $\mathcal{P}_v=\{P(e) ~\mid~ e \in \New(v)\}$ contains the collection of new-ending $s-v$ replacement paths, each representing one distinct new edge from $\New(v)$.
For every new-ending path $P_i \in \mathcal{P}_v$, recall that $D(P_i)=D_i$ is the detour segment protecting the first failing edge $F_1(P_i) \in \pi(s,v)$ of $v$ and $b(P_i)$ is the unique $\pi$-divergence point of $P_i$.
\par In this section, the new-ending $s-v$ replacement path collection, $\mathcal{P}_v$, is classified into five classes. The first class $\mathcal{P}_{\pi}$ consists of new-ending $(\pi,\pi)$ paths $P_{s,v,F}$ protecting against two edge faults on $\pi(s,v)$, i.e., $F \in \mathcal{F}^2_{v}(\pi)$. The cardinality of this set is later bounded by $O(\sqrt{n})$, using an argumentation that is similar to that of the single failure case \cite{PPFTBFS13}.
The second class of paths $\mathcal{P}_{nodet}$ consists of paths $P_i \in \mathcal{P}_v$ that do not intersect the edges of their detour $D_i$ at all, namely, $P_i \cap E(D_i)=\emptyset$, as in Fig. \ref{fig:apathtype}(d).
For this class, it is shown that the first failing edge of any two paths in this class is distinct, i.e., $F_1(P_i) \neq F_1(P_j)$ for every $P_i,P_j \in \mathcal{P}_{nodet}$. This key observation is used to bound the cardinality of this class by $O(n^{2/3})$.
\par The remaining set $\mathcal{C}$ of new-ending $(\pi,\sf D)$ paths consists of paths $P_i$ for which $F(P_i) \in \mathcal{F}_v(D)$ and $P_i \cap E(D_i)\neq \emptyset$, as in Fig. \ref{fig:apathtype}(c).
This set constitutes the main technical challenge in the analysis. To bound its cardinality, we would like to employ the same high level strategy: new-ending paths consume many vertices, and since the number of vertices is limited by $n$, the number of new-ending paths is bounded as well (as a function of $n$). To do that, we would like to show that every new-ending path $P \in \mathcal{C}$ has an nonnegligible number of distinct vertices, not appearing on any other path $P' \in \mathcal{C}$. The main technical question is to identify a subpath of the new-ending path that is guaranteed to be sufficiently long and disjoint from all others. Consider the following natural approach. For every $P_i \in \mathcal{C}$, define its suffix as $P'_i=P_i[c_i,v]$ where $c_i$ is $\sf D$-divergence point of $P_i$. We then would like to claim that the $P'_i$'s are disjoint. To do that, one should prove
(by contradiction) that if there exists a common vertex $w \in P'_i \cap P'_j$, for some $P_i,P_j \in \mathcal{C}$, then one of the two suffixes from $w$ on, say $P'_i[w,v],$ could be replaced by the other suffix $P'_j[w,v]$, and hence a proper construction of the paths should have avoided the inclusion of the new edge $\LastE(P_i)$, leading to contradiction.
For such an argumentation to hold, one should show that using $P'_i[w,v]$ instead of $P'_j[w,v]$ or vice-versa is \emph{safe}, namely, that neither of these segments contains the failing edges of the other path, or more formally, $F(P_i) \cap P'_j =\emptyset$ and $F(P_j) \cap P'_i=\emptyset$. Does this statement always hold? Consider the first failing edges of these paths, namely, $F_1(P_i), F_1(P_j) \in \pi(s,v)$.
Since $P'_i \subseteq P_i[b_i,v]$ is edge disjoint with $\pi(s,v)$, it holds that $F_1(P_j) \notin P'_i$ and analogously $F_1(P_i) \notin P'_j$. So the main challenge is in showing that the second failing edge $F_2(P_j)$ does not occur on $P'_i$ and vice-versa. This, however, can be guaranteed only for the restricted case where $F_2(P_i), F_2(P_j) \in D_i \cap D_j$. Specifically, this holds as  $c_i$ (resp., $c_j$) is the unique $\sf D$-divergence point of $P_i$ (resp., $P_j$) from the detour $D_i$ (resp., $D_j$).
Hence $F_2(P_j) \notin P'_i$ and $F_2(P_i) \notin P'_j$.
The conclusion is that the main obstacle for defining a unique set of vertices for each new-ending path boils down to the cases where $P_i$ contains the second failing edge $F_2(P_j) \in D_j \setminus D_i$. This last observation motivates the definition of \emph{interference} defined next.
\paragraph{Interference and independence of replacement paths.}
For paths $P_i,P_j \in \mathcal{P}_v$, we say that
$P_i$ \emph{interferes} with $P_j$ if $F_2(P_j) \in P_i \setminus D(P_i)$. The paths $P_i,P_j \in \mathcal{P}_v$ are \emph{independent} if $P_i$ does not interfere with $P_j$ and vice-versa.
Indeed, for independent pair of paths $P_i$ and $P_j$, upon proper construction of the replacement paths, it can be shown that the segments $P'_i$ and $P'_j$ are disjoint. This leads to the definition of the third path class, $\mathcal{P}_{indep}$, consisting of all new-ending paths that do not interfere with any other new-ending path in $\mathcal{P}_v$. By exploiting the fact that these paths do not intersect after they leave their detour, namely, $P'_i$ and $P'_j$ are disjoint, we show that there are at most $O(n^{2/3})$ independent paths.

Finally, we consider the most involved case, which is that of interfering paths. The set of interfering paths is further classified into two subsets by distinguishing between two types of interference, namely, $\pi$-\emph{interference} and $\sf D$-\emph{interference}.
We proceed by giving some high level intuition for this classification.
\par
\begin{figure}[htbp]
\begin{center}
\includegraphics[width=5in]{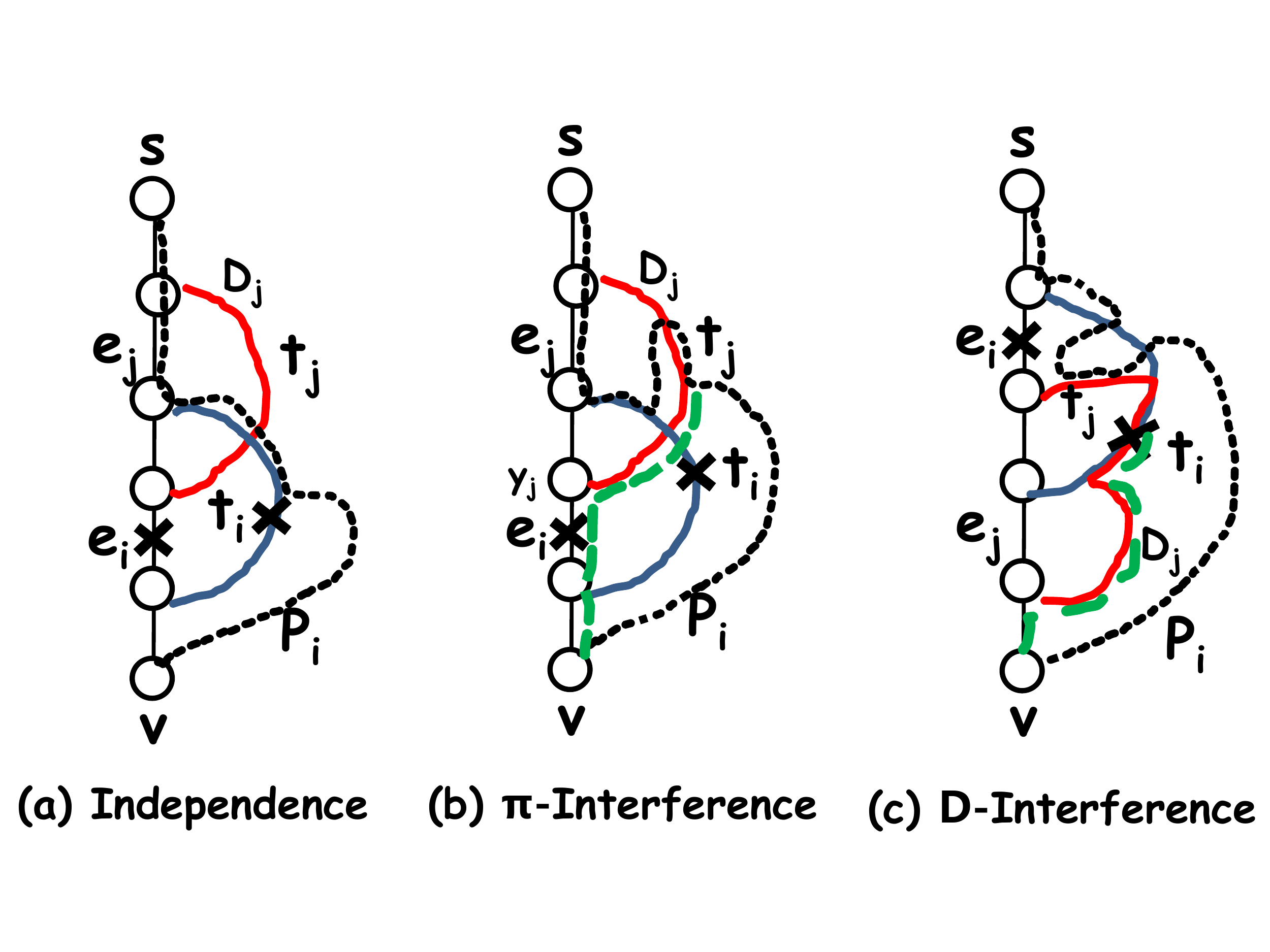}
\end{center}
\caption{Schematic illustration of independence and interference. Shown are two paths $P_i,P_j \in \mathcal{P}_v$ such that $F_1(P_k)=e_k, F_2(P_k)=t_k$ and $D_k=D(P_k)$ for $k \in \{i,j\}$. (a) The paths $P_i$ and $P_j$ are independent, since $P_i \cap \left(D_j\setminus D_i\right)=\emptyset$. (b) $P_i$ is $\pi$-interfering to $P_j$ since $e_i$ appears below $y_j=y(D_j)$ on $\pi(s,v)$. The replacement path $P_i$ traverses the failing edge $t_j \in D_j$. This path cannot proceed along the green dashed path since $e_i \in \pi(y_j,v)$ (c) $P_i$ $\sf D$-interferes with $P_j$ since $t_i \in D_j$ after the edge $t_j$, and hence this path cannot proceed along the green dashed path.
\label{fig:indep}}
\end{figure}

 Let $P_i$ be a new ending path interfering with another new-ending path $P_j$, i.e., $F_2(P_j)=(q_1,q_2)$ appears on $P_i \setminus D_i$.
On the fact of it, a natural short route from $q_2$ to $v$ in $G$ may be given by $Q=D_j[q_2,y_j] \circ \pi(y_j, v)$ (see the dashed green paths in Fig. \ref{fig:indep}(b,c)).
Note that since $Q$ is a subpath of the replacement path $P_{s,v, \{e_j\}}$, where $e_j=F_1(P_j)$, it holds that $Q \in SP(q_2,v,G \setminus \{e_j\})$.
Since $P'_i$ is a subpath of $P_i$ starting at a point that occurs on or after the $\pi$-divergence point $b_i$ of $\pi(s,v)$ and $P_i$, it holds that $P'_i \cap \pi(s,v)=\{v\}$. Hence, the alternative $q_2-v$ path $P'_i[q_2,v]$ cannot be shorter than $Q$. By the fact that Alg. $\ConstPath$ defines $P_i$ as a new-ending path, since the last edge of $Q$ was already present in the constructed structure at the time when $P_i$ was constructed, it holds that $Q \nsubseteq G \setminus F(P_i)$, i.e., the subpath $Q$, although optimal in its length, could not be used as part of the replacement path $P_i$ since it contains at least one of the two edges $F(P_i)$ against whose failure $P_i$ aims to protect. We now define two types of interference, depending on the two possible scenarios. If $Q$ contains the \emph{first} failing edge $F_1(P_i)$, i.e., $F_1(P_i) \in \pi(y_j,v)$, we say that $P_i$ $\pi$-\emph{interferes} with $P_j$. This notation indicates that the reason for not using the existing route $Q$, when considering the failing pair $F(P_i)$, is the fact that $Q$ contains the first failing edge $F_1(P_i)$, which by definition, is always in $\pi(s,v)$ (see the green dashed path in Fig. \ref{fig:indep}(b)).
Alternatively, the second optional scenario is that the $q_2-v$ route $Q$ is not used as part of  $P_i$ since it contains the second failing edge $F_2(P_i)$, which by definition (as $F(P_i) \in \mathcal{F}_v(D)$) occurs on the detour $D(P_i)$. Specifically, in such a case $F_2(P_i) \in D_j[q_2,y_j]\subseteq Q$. We then say that $P_i$ $\sf D$-\emph{interferes} with $P_j$ (see the green dashed path in Fig. \ref{fig:indep}(c)). This notation indicates that the reason for not using the existing route $Q$, is the fact that it contains the second failing edge $F_2(P_i)$ occurring on the detour $D(P_i)$.

Note that in general, these two types of interference are not exclusive and it might be the case that $P_i$ both $\pi$-interferes and $\sf D$-interferes with $P_j$.

For an interfering path $P$, let $\mathcal{I}(P)=\{P' \in \mathcal{P}_v ~\mid~ F_2(P') \in P \setminus D(P)\}$ be the set of new-ending paths interfered by $P$. We now subdivide the set of interfering paths into two classes, namely, $\mathcal{I}_{\pi}$ and $\mathcal{I}_{D}$, depending on the type of interference of $P$ on $\mathcal{I}(P)$.
If the path $P$ $\pi$-interferes with every $P' \in \mathcal{I}(P)$, then let $P \in \mathcal{I}_{\pi}$. Otherwise, if there exists at least one path $P' \in \mathcal{I}(P)$ such that $P$ $\sf D$-interferes but does not $\pi$-interfere with $P'$, then let $P \in \mathcal{I}_{D}$.
%Employed with these definitions, we define the last two remaining classes of new-ending paths.
%The first interfering class $\mathcal{I}_{\pi}$ consists of all interfering paths $P_i$ that are $\pi$-interfering to some other $P_j \in \mathcal{P}_v$. The second interfering class $\mathcal{I}_{\pi}$ consists of the remaining interfering paths $P_i$ that are $\sf D$-interfering to some $P_j \in \mathcal{P}_v$.
The cardinality of these path classes is bounded using different tools, and each is shown to contain $O(n^{2/3})$ paths.
For a schematic illustration of the different notions of interference, see Fig. \ref{fig:indep}.
%%%%%%%%%%%%%%%%%%%
%%%%%%%%%%%%%%%%%%%

\begin{figure}[htbp]
\begin{center}
\includegraphics[width=5in]{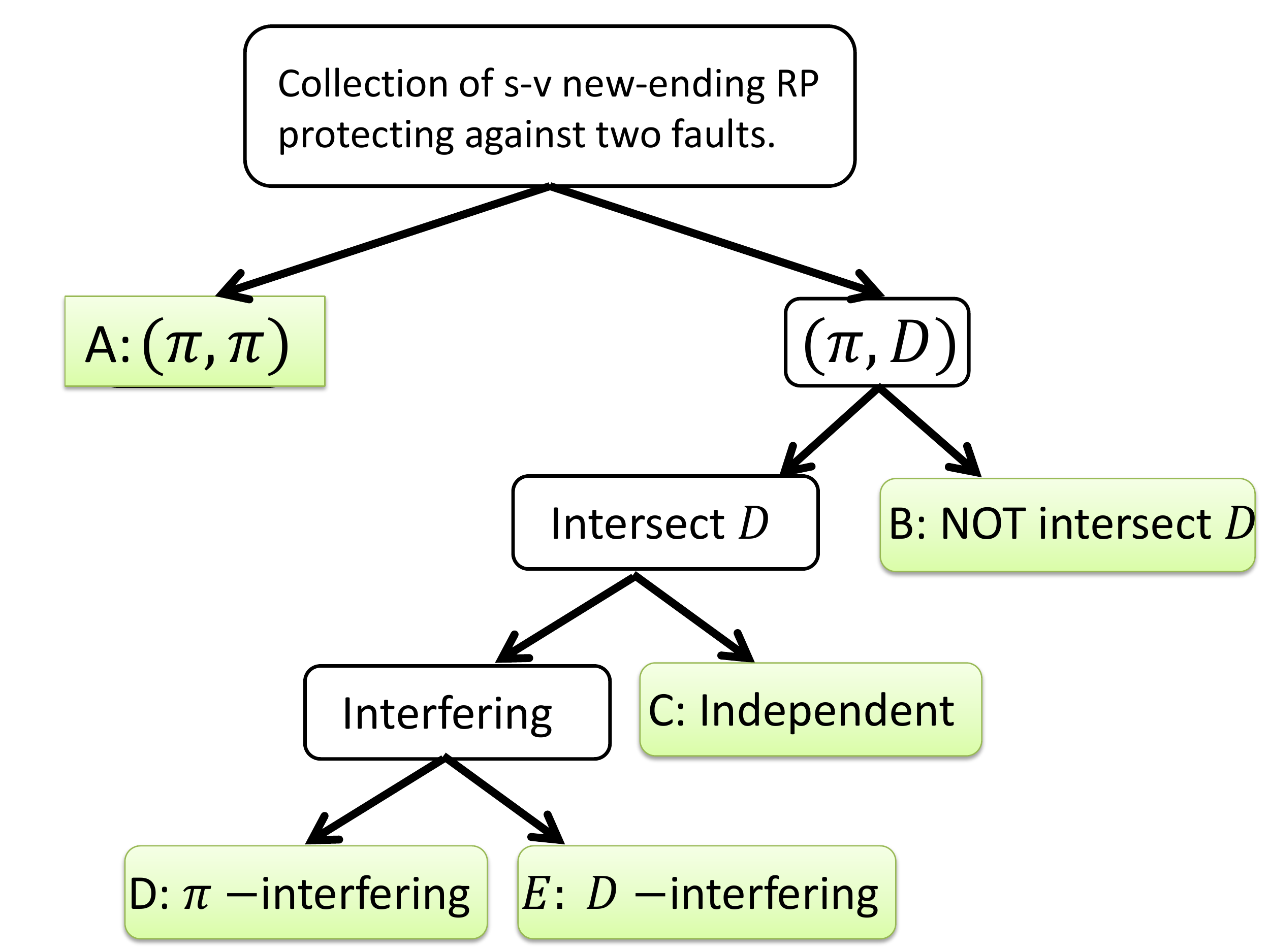}
\end{center}
\caption{Schematic illustration of replacement-path classification.
\label{fig:rpclass}}
\end{figure}

Formally, we have the following new-ending path classification.
\begin{description}
\item{(A)}
$\mathcal{P}_{\pi}=\{P_{s,v,F} ~\mid~ F \in \mathcal{F}^2_{v}(\pi)\},$
\item{(B)}
$\mathcal{P}_{nodet}=\{P=P_{s,v,F} \in \mathcal{P}_v ~\mid~ F \in \mathcal{F}_v(D) \mbox{~and~} E(P) \cap E(D(P))=\emptyset\}$,
\item{(C)}
$\mathcal{P}_{indep}=\{P=P_{s,v,F} \in \mathcal{P}_v ~\mid~ F \in \mathcal{F}_v(D)~\mbox{~and~for every~~} P'\in \mathcal{P}_v, P \mbox{~and~} P' \mbox{~are independet}\},$
\item{(D)}
$\mathcal{I}_{\pi}=\{P \in \mathcal{P}~\mid~ P \mbox{~$\pi$-interferes with every~} P' \in \mathcal{I}(P)\},$
\item{(E)}
$\mathcal{I}_{D}=\mathcal{P}_v \setminus (\mathcal{P}_{\pi} \cup \mathcal{P}_{nodet} \cup \mathcal{P}_{indep} \cup \mathcal{I}_{\pi})$.
\end{description}
For schematic illustration see Fig. \ref{fig:rpclass}.
%%%%%%%%%%%%%%%%%%%
%%%%%%%%%%%%%%%%%%%
In our argumentation, we consider these classes in different order. We first consider the class $\mathcal{P}_{nodet}$ and bound its cardinality using the kernel graph. We then use the analysis of this class to bound the cardinality of the collection of $\sf D$-interfering paths $\mathcal{I}_{\sf D}$. Next, we consider the class of independent paths $\mathcal{P}_{indep}$. The analysis of this class is completely different compared to the analysis of the previous two classes.
Finally, we consider the class of $\pi$-interfering paths, $\mathcal{I}_{\pi}$, and show that they are ``almost" independent, in the sense that interference of type $\pi$ induces only a limited amount of dependence between the replacement paths and hence the analysis for the independent case can goes through with relatively minor modifications.

%
%For an illustration of the new-ending $s-v$ replacement paths types, see Fig??
%Fig. \ref{fig:pathtype}.
%%%%%%%%%%%%%%%%%%%
%%%%\begin{figure}[htbp]
%%%%\begin{center}
%\includegraphics[width=5in]{intertype2.pdf}
%%%%\end{center}
%%%%\caption{Schematic illustration of the paths classification. Representative new ending paths in (a) $\mathcal{P}_{\pi}$, the two failing edges $F_1(P)=e_i, F_2(P)=e_j$ occur on $\pi(s,v)$.
%%%%(b) $\mathcal{P}_{nodet}$. The path $P_i$ does not intersect the detour $D_i=D(P_i)$ (b) interfering path in $\mathcal{I}_{\pi}$. The replacement path $P_{s,v, \{e_i,t_i\}}$ traverses the failing edge $t_j \in D_j$. This path cannot preceed along the $P_{s,v,\{e_j\}}$ path since $e_i \in \pi(y_j,v)$ and (c) interfering path in $\mathcal{I}_{\sf D}$. The replacement path $P_{s,v, \{e_i,t_i\}}$ traverses the failing edge $t_j \in D_j$. This path cannot preceed along the $P_{s,v,\{e_j\}}$ path since $t_i \in D_j$ below the edge $t_j$.
%%%%\label{fig:pathtype}}
%%%%\end{figure}
%%%%%%%%%%%%%%%%%%%
%%%Our goal is to bound the cardinality of the five paths classes.
%%%%%%For convenience, we consider only the largest of these classes (which contains at least $|\mathcal{P}_v|/5$ paths) and only bound that class. Since we do not know a-priori, which class is it, we bound each class separately, conditioned on the assumption that this class is large enough (i.e., contains at least $|\mathcal{P}_v|/5$ paths).
Before turning to bound the number of $(\pi,\pi)$ and $(\pi,\sf D)$ new ending paths, note that the number of last edges of replacement path $P_{s,v,\{e_i\}}$ for $e_i \in \pi(s,v)$ is bounded by $O(\sqrt{n})$ as in \cite{PPFTBFS13}.

\begin{observation}
\label{lem:single_fault}
$|E_1(\pi)|=O(\sqrt{n})$
\end{observation}
\Proof
By the uniqueness of the $\pi$-divergence point, it is required to bound the number of replacement path $P_{s,v,\{e_i\}}$ whose last edge is not in $T_0$. For every such path $P=P_{s,v,\{e_i\}}$, it holds that $P[b(P),v]$ is edge disjoint with $\pi(s,v)$, and therefore $P[b(P),v]=SP(b(P),v, G \setminus E(\pi(s,v)),W)$.
Let $P_1, \ldots, P_t$ be such that each $P_i=P_{s,v,\{e_i\}}$ ends with a different new edge of $v$ and ordered in nondecreasing distance of $\dist(b(P_i),v)$.
By the uniqueness of the weight assignment $W$, it holds that $P'_i=P_i[b(P_i),v]\setminus \{v\}$ are disjoint. Hence, $|P'_i|\geq \dist(b_i,v)-1 \geq i-2$, and $|\bigcup P'_i|=\sum_{i=1}^t |P'_i|=\Omega(t^2)$. Overall, since there are $n$ vertices in $G$, we get that $t=O(\sqrt{n})$ as required.
\QED

\subsection{Bounding the number of new-ending paths of $\mathcal{P}_{\pi}$}
We first consider the new-ending $(\pi,\pi)$-replacement paths protecting against failures on the $\pi(s,v)$ paths.
%Let $\New_{\pi}=\{\LastE(P), P \in \mathcal{P}_{\pi}\}$.
\begin{lemma}
\label{lem:not_inters}
$|E_2(\pi)|=O(\sqrt{n})$.
\end{lemma}
\Proof
By Obs. \ref{lem:single_fault}, it is sufficient to bound the edge set  $E'=E_2(\pi) \setminus \left(E_1(\pi) \cup E(v,T_0) \right)$.
For every edge $e \in E'$, select one path $P_{s,v,\{e_i,e_j\}}$ such that $\LastE(P_{s,v,\{e_i,e_j\}})=e$.
Note that it then holds that $P_{s,v,\{e_i,e_j\}}= SP(s,v,G \setminus \{e_i,e_j\},W)$ (i.e., in such a case, $P_{s,v,\{e_i,e_j\}}$ is not composed of the detours of $D_i$ and $D_j$).
Let $P_1, \ldots, P_t$ be the selected $(\pi,\pi)$ replacement paths, each ends with a distinct edge from $E'$.
Let $b_j$ be the last divergence point of $P_j \in \mathcal{P}_{\pi}$ and $\pi(s,v)$. Let $P'_j=P_j[b_j, v]$.
Note that by definition, $\LastE(P'_j) \notin T_0$ and hence $P'_j$ is edge disjoint with $\pi(s,v)$.
We now claim that $P'_{j_1}$ and $P'_{j_2}$ are vertex disjoint besides their common endpoint $v$. Assume, towards contradiction otherwise, and let $w \in  \left(P'_{j_1} \cap P'_{j_2} \right) \setminus \{v\}$ be a common vertex in the intersection.
By definition, $P'_{j_1}, P'_{j_2}$ are edge disjoint with $\pi(s,v)$. Since $\LastE(P_{j_1}) \neq \LastE(P_{j_2})$, there are two distinct $w-v$ shortest paths in $G \setminus \pi$, namely, $P'_{j_1}[w,v]$ and $P'_{j_2}[w,v]$, contradiction by the uniqueness of the shortest paths. Order the paths of detours of $\mathcal{P'}=\{P_1, \ldots, P_{t}\}$ in increasing distance of $b_j$ and $v$. Then, $|P'_i|\geq i$ hence, $n\geq|\bigcup_{i=1}^t P'_i|=\sum |P'_i|\geq \sum_{i=1}^t (i-2)=\Omega(t^2)$, concluding that $t=O(\sqrt{n})$.
\QED

\subsection{Bounding the number of paths that do not intersect their detours $\mathcal{P}_{nodet}$}

In this section, we consider the set of paths and detours
$$\mathcal{P}_{nodet}=\{P=P_{s,v,F} \in \mathcal{P}_v ~\mid~ F \in \mathcal{F}_v(D) \mbox{~and~} E(P) \cap E(D(P))=\emptyset\} \mbox{~and~} \mathcal{D}_{nodet}=\{D(P) ~\mid~ P \in \mathcal{P}_{nodet}\}.$$
and bound its cardinality. Our strategy is as follows. We first show that the first failing edge of every path $P$ in this set is distinct. Then, Lemma \ref{lem:bound_no_touch} uses this property and the kernel subgraph $\KernelGraph(\mathcal{D}_{nodet})$ of the detours of $\mathcal{D}_{nodet}$ to bound the cardinality of this set.
\begin{observation}
\label{lem:not_inters}
$F_1(P_1) \neq F_1(P_2)$ for every $P_1, P_2 \in \mathcal{P}_{nodet}$.
\end{observation}
\Proof
Towards contradiction assume otherwise, that $F_1(P_1)= F_1(P_2)$ and hence also $D(P_1)=D(P_2)$. Without loss of generality, assume that $P_1$ was constructed by Alg. $\ConstPath$ before $P_2$.
Since both are $(\pi,\sf D)$ paths, we have that the second faults are on the same detour $F_2(P_1), F_2(P_2) \in D_1(P_1)$. Since $P_1, P_2 \in \mathcal{P}_{nodet}$, we have that $F_2(P_1), F_2(P_2) \notin P_1,P_2$. This implies that there are two shortest $s-v$ paths in $G \setminus \left(F(P_1) \cup F(P_2)\right)$, namely $P_1$ and $P_2$.
By the optimality of these paths $|P_1|=|P_2|$, in contradiction to the selection of the last edge of $P_2$ by Alg. $\ConstPath$.
\QED
The next lemma bounds the number of paths in any collection of new-ending replacement paths $\mathcal{P}\subseteq \mathcal{P}_v$ satisfying that $F_1(P_1)\neq F_1(P_2)$ for every $P_1,P_2 \in \mathcal{P}$.
\begin{lemma}
\label{lem:bound_no_touch}
Every collection of new-ending replacement paths $\mathcal{P} \subseteq \mathcal{P}_v$ satisfying that  $F_1(P_1)\neq F_1(P_2)$ for every $P_1,P_2 \in \mathcal{P}$, is of size $|\mathcal{P}|=O(n^{2/3})$.
\end{lemma}
\Proof
Let $N=|\mathcal{P}|$. Order the paths of $\mathcal{P}=\{P_1, \ldots, P_{N}\}$ in increasing distance of $s$ and $F_1(P_j)$, i.e., $\dist(s, F_1(P_1))<\ldots<\dist(s, F_1(P_{N}))$. Let $e_i=F_1(P_i)$ and $M=\lfloor N/2 \rfloor$.
We now restrict attention to the set of first $M$ paths $\mathcal{P}_M=\{P_{j} ~\mid~ 1 \leq j \leq M\}$.
Note that for each path $P_j \in \mathcal{P}_M$, the distance from $v$ to the first failing edge $e_j=F_1(P_j)$ is $\dist(e_j,v,\pi(s,v))\geq M$. Let $D_i=D(P_i)$. We now construct the kernel subgraph $\KernelGraph(\mathcal{D}_M)$ of the corresponding $M$ detours $\mathcal{D}_M=\{D_1, \ldots, D_{M}\}$. %Let $V' \subseteq \bigcup_{i=1}^M D_i$ be the total set of vertices in $\KernelGraph^v$.
Recall that $D_i$ contributed only its prefix $D_i[x_i, w_i]$ to the kernel subgraph $\KernelGraph(\mathcal{D}_M)$.
\par From now on, let $V'(\KernelGraph(\mathcal{D}_M))$ refer to the vertices of the kernel graph $V(\KernelGraph(\mathcal{D}_M))$ excluding the vertices appearing on $\pi(s,v)$, i.e., $V'(\KernelGraph(\mathcal{D}_M)) =V(\KernelGraph(\mathcal{D}_M))\setminus \pi(s,v)$.

Let $\mathcal{P}^1_M=\{P_i \in \mathcal{P}_M ~\mid~ P_i \cap V'(\KernelGraph(\mathcal{D}_M))=\emptyset\}$ be paths in $\mathcal{P}_M$ that have no common vertex with $V'(\KernelGraph(\mathcal{D}_M))$ and let $\mathcal{P}^2_M=\mathcal{P}_M \setminus \mathcal{P}^1_M$ be the remaining paths. We begin by bounding the cardinality of $\mathcal{P}^1_M$.

\begin{observation}
\label{obs:nondet_det}
$|\mathcal{P}^1_M|=O(\sqrt{n})$.
\end{observation}
\Proof
Let $b_i$ be the first divergence point of $P_i$ from $\pi(s,v_i)$. By Cl. \ref{cl:unique_nnew}, $b_i$ is a unique divergence point and hence $P_i[b_i, v]$ and $\pi(s,v)$ are edge disjoint. Without loss of generality, assume that $P_i$ was constructed by Alg. $\ConstPath$ before $P_j$. We now claim that $P'_i=P_{i}[b_i, v]$ and $P'_j=P_{j}[b_j, v]$ are vertex disjoint, except for their common endpoint $v$, for every $P_i,P_j \in \mathcal{P}^1_M$. By Cl. \ref{lem:detour_gradual}, the second failing edges $F_2(P_i), F_2(P_j)$ appears in the kernel subgraph $\KernelGraph(\mathcal{D}_M)$ as $D(P_i),D(P_j) \in \mathcal{D}_M$. Since $P_i,P_j$ do not intersect with $V'(\KernelGraph(\mathcal{D}_M))$, it holds that $F_2(P_i),F_2(P_j) \notin P_i,P_j$. Assume, towards contradiction, that there exists a common vertex $w \neq v$ in the intersection of $P'_i$ and $P'_j$. It implies that there are two $w-v$ paths in $G \setminus \left(F(P_i) \cup F(P_j)\right)$, namely
$P_i[w,v] \neq P_j[w,v]$, contradiction to the selection of $P_j$ by Alg. $\ConstPath$, since its new edge could have been saved.

Order the paths $\mathcal{P}^1_M=\{P_1, \ldots, P_t\}$ in decreasing distance of $\dist(s, b_i)$, i.e., $\dist(s, b_1)> \ldots >\dist(s, b_t)$. It holds that $|P'_i|\geq |\pi(b_i, v)|\geq i-1$. Hence, $|\bigcup_{i=1}^t V(P'_i) \setminus \{v\}|=\sum_{i=1}^t |V(P'_i)\setminus \{v\}|\geq (t-1)^2$. As there are $n$ vertices in $G$, we get that $t=O(\sqrt{n})$. The claim follows.
\QED

So, it remains to bound the set $\mathcal{P}^2_M$ of paths
that intersect the vertex set $V'(\KernelGraph(\mathcal{D}_M))$.
For every path $P_i \in \mathcal{P}^2_M$, let $a_i$ be the last common vertex of $P_i$ and $V'(\KernelGraph(\mathcal{D}_M))$ on $P_i$, and define its suffix $\widetilde{P}_i=P_i[a_i,v]$. We now show that the $\widetilde{P}_i$ segments are disjoint, and using the structure of the kernel subgraph, we also show that the number of vertices in these subpaths is rapidly increasing with $n$.
\begin{claim}
\label{cl:disjoint_nondet}
$\widetilde{P}_{i} \cap \widetilde{P}_{j}=\{v\}$.
\end{claim}
\Proof
The proof is similar to that of Obs. \ref{obs:nondet_det}.
Since $\widetilde{P}_{i}$ and $\widetilde{P}_{j}$ are edge disjoint with $\pi(s,v)$ and by definition they are vertex disjoint with $V'(\KernelGraph(\mathcal{D}_M))$, by Lemma \ref{lem:detour_gradual}, it holds that $F(P_{i}), F(P_{j})$ are not in $\widetilde{P}_i$ and $\widetilde{P}_j$. If there is a common vertex $w \in (\widetilde{P}_{i} \cap \widetilde{P}_{j}) \setminus \{v\}$, then there are two distinct $w-v$ paths in $G \setminus \left(F(P_{i}) \cup F(P_{j})\right)$, and we end with contradiction to the construction of these paths by Algorithm $\ConstPath$.
\QED
We now classify the detours $D_i \in \mathcal{D}_M$ according to the length of the prefix $D_i[x_i,w_i]$ that was taken into the kernel.
A detour $D_j$ is \emph{expensive} if $|D_j[x_j, w_j]|\geq M/2$, otherwise it is \emph{cheap}.
Next, the new-ending paths $P_i$ of $\mathcal{P}^2_M$ are classified according to the first detour $D(a_i)$ in the $(x,y)$-ordering $\overrightarrow{\mathcal{D}}_M$ on which $a_i$ (the last common vertex of $P_i$ and $V'(\KernelGraph(\mathcal{D}_M))$) appears on its prefix, i.e., $D(a_i)=D_j$ is the first detour in $\overrightarrow{\mathcal{D}}_M$ satisfying that $a_i \in D_j[x_j,w_j]$.
Then $P_i$ is expensive (resp., cheap) if $D(a_i)$ is expensive (resp., cheap).
Let $\mathcal{P}_{cheap}=\{P_i \in \mathcal{P}^2_M~\mid~ D(a_i) \mbox{~is cheap~}\}$ and $\mathcal{P}_{expen}=\{P_i  \in \mathcal{P}^2_M~\mid~ D(a_i) \mbox{~is expensive~}\}$, where $\mathcal{P}^2_M=\mathcal{P}_{cheap} \cup \mathcal{P}_{expen}$.
To bound $|\mathcal{P}^2_M|$, we separately bound $|\mathcal{P}_{cheap}|$ and $|\mathcal{P}_{expen}|$.

We first consider the cheap paths.
\begin{claim}
\label{cl:cheap}
$|\mathcal{P}_{cheap}|=O(\sqrt{n})$.
\end{claim}
\Proof
Let $V_{cheap}=\bigcup_{P_i \in \mathcal{P}_{cheap}}V(P_i[a_i, v])\setminus \{v\}$.
By Cl. \ref{cl:disjoint_nondet}, since $\widetilde{P}_i=P_i[a_i, v]$ are disjoint (except for the common endpoint $v$), hence
$|V_{cheap}|=\sum_{P_i \in \mathcal{P}_{cheap}} (|\widetilde{P}_i|-1)$.
We now focus on some $P_i$ and show that $|\widetilde{P}_i|\geq M/2$.
First note that $\widetilde{P}_i$ and $\pi(s,v)$ are vertex disjoint (except for the common endpoint $v$), as $a_i$ occurs after $b_i$, the unique $\pi$-divergence point of $P_i$ from $\pi(s,v)$. Hence,
\begin{equation}
\label{eq:nodet1}
|\widetilde{P}_i|\geq \dist(a_i, v, G \setminus \pi(s,v))~.
\end{equation}
Let $D_j=D(a_i) \in \mathcal{D}_M$ be the detour protecting against the failing of the edge $e_j$. Then,
\begin{eqnarray}
\label{eqn:nondet2}
\dist(x_j, v, G \setminus \{e_j\})&\geq& \dist(x_j, v, G)
\geq \dist(e_j, v) \geq M~,
\end{eqnarray}
where the penultimate inequality follows as $x_j$ appears above the failing edge on $\pi(s,v)$ and last inequality follows by the fact that $D_j \in \mathcal{D}_M$.
Since $a_i$ appears on a cheap detour $D_j$, we get that
\begin{eqnarray}
\label{eqn:nondet33}
\dist(x_j,a_i, G \setminus \{e_j\})\leq \dist(x_j,w_j, G \setminus \{e_j\})=|D_j[x_j,w_j]|\leq M/2~,
\end{eqnarray}
and hence by Eq. (\ref{eqn:nondet2}) and Eq. (\ref{eqn:nondet33}), we get that $\dist(a_i, v, G \setminus \{e_j\})\geq M/2$.

Overall, by combining with Eq. (\ref{eq:nodet1}), we get that $|\widetilde{P}_i|\geq M/2$. We therefore have that $M/2 \cdot |\mathcal{P}_{cheap}| \leq |V_{cheap}|\leq n$. It follows that $|\mathcal{P}_{cheap}|\leq 2n/M$. Since clearly, also $|\mathcal{P}_{cheap}|\leq M$, we have $|\mathcal{P}_{cheap}|\leq \min\{M, 2n/M\}\leq \sqrt{2n}$. The claim follows.
\QED

\begin{claim}
\label{cl:expens_detours}
$|\mathcal{P}_{expen}|=O(n^{2/3})$.
\end{claim}
\Proof
Let $\mathcal{D}_{expen}=\{D_j \in \mathcal{D}_M ~\mid~ |D_j[x_j,w_j]|\geq M/2\}$ be the collection of
expensive detours, $z=|\mathcal{D}_{expen}|$.
We now classify the expensive paths of $\mathcal{P}_{expen}$ into $z$ classes where each path $P_i$ is mapped to the class of the detour $D_j \in \mathcal{D}_{expen}$ on which $a_i$ appears.

For every $D_j \in \mathcal{D}_{expen}$,
let $\mathcal{P}_j=\{P_i \in \mathcal{P}_{expen} ~\mid~ D(a_i)=D_j\}$, and let $N_j=|\mathcal{P}_j|$ be the cardinality of this set.
\par We begin by bounding the number of vertices appearing in the expensive detours, let $V_{\sf D}=\bigcup_{D_j \in \mathcal{D}_{expen}} V(D_j[x_j, w_j])$ be the vertices appearing on the expensive
detours. By the construction of the kernel graph, the sets $D_j[x_j, w_j]$ are disjoint except for the point $w_j$ (in cases where $D_j$ is truncated).  Hence, since every $D_j \in \mathcal{D}_{expen}$ is expensive, we get that
\begin{equation}
\label{eq:nodet_2}
|V_{\sf D}| \geq z \cdot (M-1)/2~.
\end{equation}
We now proceed by bounding the number of vertices appearing on the expensive replacement paths,
$V_P=\bigcup_{P_i \in \mathcal{P}_{expen}}V(\widetilde{P}_i)\setminus \{a_i, v\}.$
Note that for every expensive path $P_i$, its segment $\widetilde{P}_i$ is vertex disjoint (expect for its endpoints $a_i$ and $v$) with the vertex set $V_{\sf D}$ since $V_{\sf D} \subseteq \KernelGraph(\mathcal{D}_M)$.

Fix some $j \in \{1, \ldots, z\}$, with $N_j$ expensive paths $\mathcal{P}_j$. We now claim that $V_j=\bigcup_{P_i \in \mathcal{P}_j} \widetilde{P}_i$ contains $\Omega(N_j^2)$ vertices.
By Cl. \ref{cl:disjoint_nondet}, the $\widetilde{P}_{i}$ segments are disjoint. Order the paths of $\mathcal{P}_j$ in increasing distance of $a_i$ from $v$. Since $a_i \in D_j$ for every $P_i \in \mathcal{P}_j$ and the $a_i$'s are distinct it holds that $|V_j|\geq (N_j-1)^2/2$ and summing over all $j$ (as the $\widetilde{P}_i \setminus \{v\}$ are disjoint) and using the Cauchy-Schwarz inequality, we get that
\begin{equation}
\label{eq:nodetimcs}
|V_P|\geq \sum_{j=1}^z (N_j-1)^2/2~.
\end{equation}
Recall that the sets $V_P$ and $V_{\sf D}$ are disjoint, and thus by Eq. (\ref{eq:nodet_2}) and (\ref{eq:nodetimcs}),
$$n \geq |V_P \cup V_{\sf D}|=|V_P|+|V_{\sf D}|\geq\sum_{j=1}^z (N_j-1)^2/2+ z \cdot M/2=\Omega(M^{3/2})~.$$
We get $M = O(n^{2/3})$, as required.
\QED
Lemma \ref{lem:bound_no_touch} now follows by combining Obs. \ref{obs:nondet_det}, Cl. \ref{cl:cheap} and Cl. \ref{cl:expens_detours}.
\QED
By combining Lemma \ref{lem:not_inters} with
Lemma \ref{lem:bound_no_touch}, we get the following.
\begin{corollary}
\label{cor:bound_no_touch}
$|\mathcal{P}_{nodet}|=O(n^{2/3})$.
\end{corollary}

\subsection{Bounding the number of $\sf D$-interfering paths $\mathcal{I}_{\sf D}$}
In this section, we consider the set of $\sf D$-interfering paths
$$\mathcal{I}_{\sf D}=\mathcal{P}_v \setminus (\mathcal{P}_{\pi} \cup \mathcal{P}_{nodet} \cup \mathcal{P}_{indep} \cup \mathcal{I}_{\pi}).$$

For every path $P \in \mathcal{I}_{\sf D}$, recall that $\mathcal{I}(P)=\{P' \in \mathcal{P}_v ~\mid~ F_2(P') \in P\setminus D(P)\}$ is the set of paths to which $P$ interferes. Since $P \in \mathcal{I}_{\sf D}$, the set of interfered paths $\mathcal{I}(P)$ is non-empty.
%
%Recall that for any pair of non disjoint detours $D_1,D_2$, $\First(D_1,D_2)$ denote their first common vertex.
%
By the definition of $\mathcal{I}_{\sf D}$, a path
$P \in\mathcal{I}_{\sf D}$ satisfies $P \notin (\mathcal{P}_{nodet} \cup \mathcal{I}_{\pi})$, so we have the following.
\begin{observation}
\label{obs:typee}
For every $P \in \mathcal{I}_{\sf D}$:
(1) $E(P) \cap E(D(P)) \neq \emptyset$,\\
(2) there exists $P' \in \mathcal{I}(P)$, such that $F_2(P')=(q_1,q_2) \in P \setminus D(P)$ and $F_2(P) \in D'[q_1,y'] \cap D(P)$ and $F_1(P) \notin \pi(y',v)$ where $D'=D(P')$ and $y'=y(D')$.
\end{observation}

Let $\mathcal{D}_v=\{D(P) ~\mid~ P \in \mathcal{P}_v\}$ be the collection of detours corresponding to new-ending $s-v$ paths $\mathcal{P}_v$.
Let $N_P=|\mathcal{I}_{\sf D}|$ be the number of the $\sf D$-interfering paths and let $N_{\sf D}=|\mathcal{D}_v|$ be the total number of detours.
The main challenge in this section is to show that $N_P=O(N_{\sf D})$ and hence Lemma \ref{lem:bound_no_touch} can be applied to bound from above the cardinality of $\mathcal{I}_{\sf D}$. We begin by stating a useful claim for the paths in $\mathcal{I}_{\sf D}$.
\begin{claim}
\label{cl:interd}
For every $P \in \mathcal{I}_{\sf D}$ and for every $P' \in \mathcal{I}(P)$, $D=D(P)$ and $D'=D(P')$, it holds that
$D$ and $D'$ are dependent and not $x$-interleaved (i.e., $x(D)\neq x(D')$).
\end{claim}
\Proof
Since the path $P$ $\sf D$-interferes with the path $P'$, it holds that the failing edge $F_2(P) \in D \cap D'$, i.e., the detours $D'$ and $D'$ are dependent. In particular, the failing edge $F_2(P)$ appears on $D'$ after the edge $F_2(P')$. We now show that $x\neq x'$ where $x=x(D)$ and  $x'=x(D')$.
Towards contradiction, assume otherwise. Then since
$F_2(P)=(q_1,q_2)$ is common with $D$ and $D'$, by Cl. \ref{cl:jointdet}, $D[x,q_2]=D'[x,q_2]$.
Since $F_2(P')$ occurs on $D'$ before the edge $F_2(P)$, it holds that $F_2(P') \in D[x,q_2]$, contradiction to the fact that $F_2(P') \in P \setminus D$. The claim follows.
\QED
\par In view of Cl. \ref{cl:interd}, the interfering paths $P$ of $\mathcal{I}_{\sf D}$ are now subdivided into two subsets depending on the relation between their detour $D(P)$  and detour $D(P')$ of the interfered paths $P' \in \mathcal{I}(P)$.
Let $\mathcal{I}^1_{\sf D}$ be the set of interfering paths $P$ that interfere with a least one path $P'$ whose detour $D(P')$ and $D(P)$ are $y$-interleaved (i.e., ends at the same $y$-point), i.e., $\mathcal{I}^1_{\sf D}=\{P \in \mathcal{I}_{\sf D} ~\mid~ \exists P' \in \mathcal{I}(P) \mbox{~such that~} y(D(P'))=y(D(P))\}$.
Let $\mathcal{I}^2_{\sf D}=\mathcal{I}_{\sf D} \setminus \mathcal{I}^1_{\sf D}$ be the complementary set of $\sf D$ interfering paths.
%By Cl. \ref{cl:interd}, for a $P \in \mathcal{I}^2_{\sf D}$, the detour $D(P')$ of every interfered path $P' \in \mathcal{I}(P)$ is interleaved with $D(P)$.
%
To bound the cardinality of $\mathcal{I}_{\sf D}$ as a function of the total number of detours, we bound separately $\mathcal{I}^1_{\sf D}$ and $\mathcal{I}^2_{\sf D}$. We begin with bounding $\mathcal{I}^1_{\sf D}$.

\paragraph{Bounding the number of paths in $\mathcal{I}^1_{\sf D}$.}
For every $P$, let $I(P)$ be some path $P'$ interfered by $P$ whose detour is $y$-interleaved with $D(P)$, i.e., $P'  \in \mathcal{I}(P)$ and $y(D(P'))=y(D(P))$.
Since $P \in \mathcal{I}^1_{\sf D}$, such $I(P)$ is guaranteed to exist.

Our strategy now is to classify the paths $P$ of $\mathcal{I}^1_{\sf D}$ according to the $y$-value of their detour $D(P)$ and consider each class separately.
%\textbf{MP: I think that $\mathcal{D}(y')$ should contain all detours ending at $y'$}.\\

For every vertex $y' \in \pi(s,v)$, let $\mathcal{D}(y')$ be the set of detours in whose $y$-value is $y'$, i.e., $\mathcal{D}(y')=\{D(P) ~\mid~ P \in \mathcal{P}_v \mbox{~and~} y(D(P))=y'\}$.

Let $\mathcal{P}(y')$ be the set new-ending paths in $\mathcal{I}^1_{\sf D}$ whose detours are in $\mathcal{D}(y')$, i.e., $\mathcal{P}(y')=\{P \in \mathcal{I}^1_{\sf D} ~\mid~ D(P) \in \mathcal{D}(y')\}$.

We now fix some $y' \in \pi(s,v)$ and bound the number of paths $N_P(y')=|\mathcal{P}(y')|$  by the number of $y$-interleaved detours $N_{\sf D}(y')=|\mathcal{D}(y')|$.
Our goal is to show that for every $y' \in \pi(s,v)$, $N_P(y')=O(N_{\sf D}(y'))$. To show this, we consider $y' \in \pi(s,v)$ and construct the kernel graph $\KernelGraph(\mathcal{D}(y'))$ on the subset of $y$-interleaved detours $\mathcal{D}(y')=\{D_1,\ldots,D_{\ell}\}$.
Note that whereas in general $\KernelGraph(\mathcal{D})$ is a \emph{subgraph} of the graph $G(\mathcal{D})$ obtained by the union of the detours in $\mathcal{D}$, in this specific case ,where all the detours are $y$-interleaved (end in the same vertex), the kernel graph coincides with the whole graph, i.e.,
$\KernelGraph(\mathcal{D})=G(\mathcal{D})$. This is proven formally in Obs. \ref{obs:all}.
Let $\mathcal{X}_1=\{x_i ~\mid~D_i \in \mathcal{D}(y')\}$ and $\mathcal{W}_1=\{w_i ~\mid~ D_i \in \mathcal{D}(y')\}$ denote the endpoints of the detour fragments taken into the kernel graph $\KernelGraph(\mathcal{D}(y'))$.
%Let $W_1=\{x_i,y_i ~\mid~ D_i \in \mathcal{D}(y')\} \cap \bigcup_{D_1,D_2 \in \mathcal{D}(y')} \First(D_1,D_2)$ be the total number of intersection points and the $x_i,y_i$ points.
We begin by claiming that for every two $y$-interleaved detours $D_1,D_2 \in \mathcal{D}(y')$, their first common intersection point $\First(D_1,D_2)$ is in the endpoint set $\mathcal{W}_1$.
(This means that the number of first common intersection points among detour pairs in $\mathcal{D}(y')$ is only $O(\ell)$ rather than $\Omega(\ell^2)$.)
Recall that by Cl. \ref{cl:summ_depend}, since $D_1$ and $D_2$ are $y$-interleaved, it holds that the first common vertex is the same, i.e., $\First(D_1,D_2)=\First(D_2,D_1)$.
\begin{claim}
\label{cl:w}
For every $D_1,D_2 \in \mathcal{D}(y')$, $\First(D_1,D_2)\in \mathcal{W}_1$.
\end{claim}
\Proof
Let $\mathcal{D}(y')=\{D_1, \ldots, D_\ell\}$ be ordered by $(x,y)$-ordering $\overrightarrow{\mathcal{D}}(y')$, corresponding to their addition into the kernel subgraph $\KernelGraph(\mathcal{D}(y'))$.
We prove by induction on $i \in \{1, \ldots, \ell\}$, that $\First(D_i,D_j) \in \mathcal{W}_1$ for every $j \leq i-1$.
The base of the induction $i=1$ holds vacuously.
Assume the claim holds up to $i-1$ and consider $i$.
There are two cases. Case (1): $D_i$ is a non-truncated detour (i.e., $D_i \cap D_j=\{y'\}$ for every $j \leq i-1$). In this case the claim holds vacuously again.
Case (2): $D_i$ is truncated. Let $D_{j'}=\Breaker(D_i)$ be the breaker detour of $D_i$ for some $j' \leq i-1$.
In other words, $D_{j'}$ is the detour satisfying that $w_i=\First(D_i, D_{j'})$. By definition of the kernel graph, the vertex $w_i=\First(D_i, D_{j'})$ is included in the endpoint set $\mathcal{W}_1$. Note that for every other $D_k$, for $k\leq i-1$, the first common vertex $\First(D_i, D_k)$ appears on $D_i$ not before $\First(D_i, D_{j'})$.
Since $D_i$ and $D_{j'}$ are $y$-interleaved, by Cl. \ref{cl:jointdet}, $D_i[w_i, y']=D_{j'}[w_i,y']$.
The remaining detours $D_k$ for $k  \leq i-1$ can now be divided into two types. The first type consists of detours $D_k$ whose first common vertex with $D_{j'}$, namely, $q_k=\First(D_k,D_{j'})$ appears on the detour $D_{j'}$ not after $w_i$. In this case, since $D_k$ and $D_{j'}$ are $y$-interleaved, by  Cl. \ref{cl:jointdet}, we get again that
$D_k[q_k, y']=D_{j'}[q_k,y']$, and hence the first common vertex of these detours with $D_i$ is exactly $w_i$, which was added, that is $\First(D_i,D_k)=\First(D_i,D_{j'})\in \mathcal{W}_1$.
The second type consists of detours $D_k$, for $k \leq i-1$, whose first common vertex with $D_{j'}$, namely, $q_k=\First(D_k,D_{j'})$, appears on $D_{j'}$ after $w_i$. In such a case, since $D_i$ and $D_{j'}$ are $y$-interleaved, by Cl. \ref{cl:jointdet}, we get that $D_{i}[w_i, y']=D_{j'}[w_i, y']$. Therefore, the first common vertex of these detours with $D_i$ is exactly the same as their first common vertex with $D_{j'}$, i.e., $\First(D_k,D_{i})=\First(D_k,D_{j'})$. By the induction assumption for $j'\leq i-1$, it holds that $\First(D_k,D_{i})\in \mathcal{W}_1$. The claim holds.
\QED

\begin{figure}[htbp]
\begin{center}
\includegraphics[width=5in]{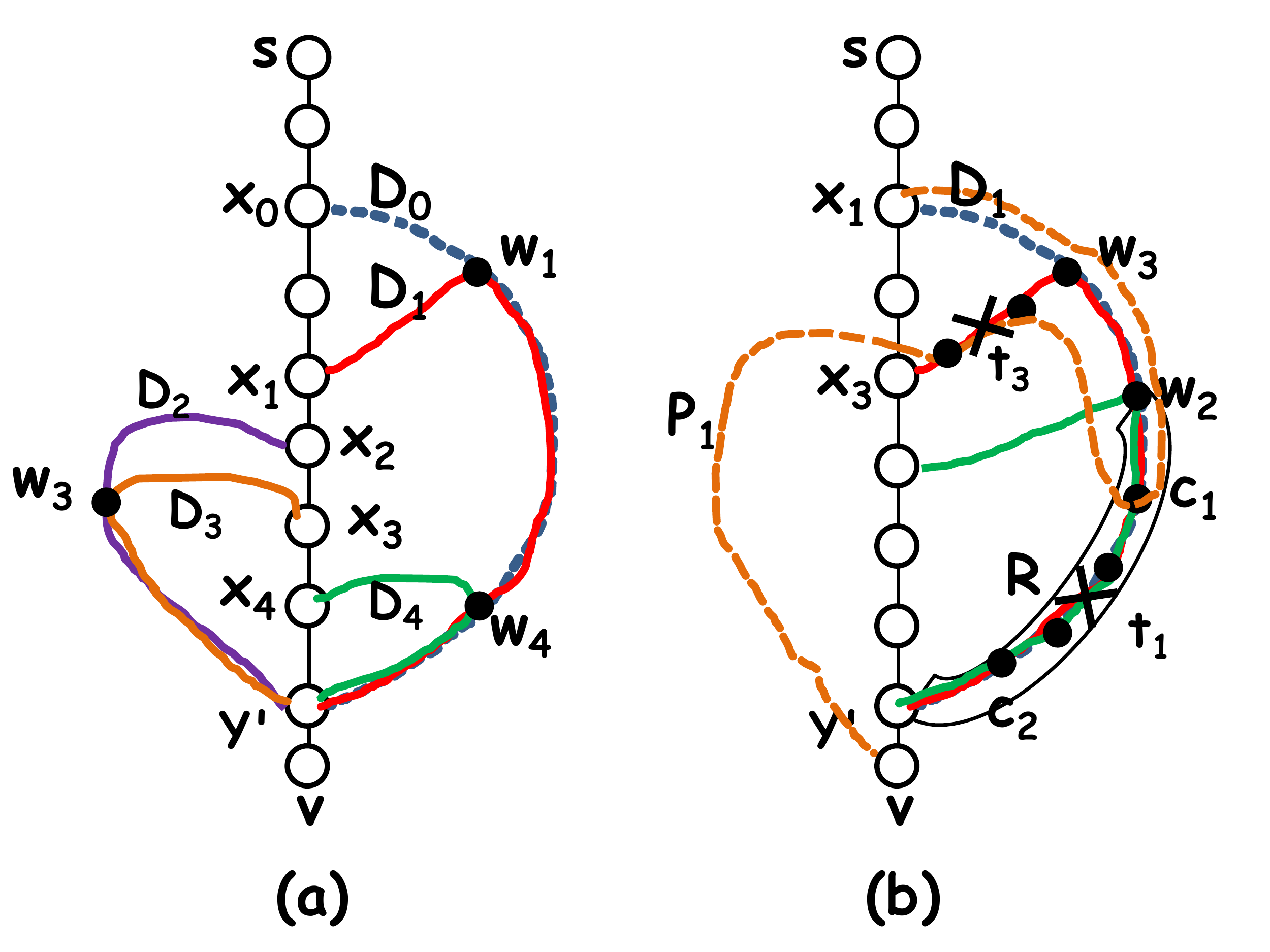}
\end{center}
\caption{(a) Illustration of the kernel graph $\KernelGraph_v(\mathcal{D}(y'))$, the segment $D_1[w_1,w_4]$ is a region in $\mathcal{R}(y')$.
(b) Illustration for Lemma \ref{lem:struc_r}. Shown is a region $R=D_1[w_2,y']$ with two $\sf D$-divergence points $c_1$ and $c_2$ such that $F_2(P_1)=t_1$ is in between them. Since $P_1$ $\sf D$-interferes with $P_3$, it visits $F_2(P_3)=t_3$ after departing from its detour at the point $c_1$. Using the route from $w_3$ to $t_2$ provided by $D_3$ is strictly better.
\label{fig:interdyy}}
\end{figure}

%Let $\mathcal{R}$ be a collection of detour fragments of $\mathcal{D}(y')$ segmented by points in $\mathcal{W}_1$.
\paragraph{Regions.}
Observe that in $\KernelGraph(\mathcal{D}(y'))$, the vertices of $\mathcal{W}_1$ have degree at least $3$, those of $\mathcal{X}_1$ have degree $1$, and all other vertices have degree $2$. Hence, $\KernelGraph(\mathcal{D}(y'))$ can be decomposed into a collection of maximal paths that are
fragments of detours referred to hereafter as \emph{regions}.
A subpath $R \subseteq \KernelGraph(\mathcal{D}(y'))$ is a \emph{region} if is satisfies the following two properties: (1) the endpoints of the subpath $R$ are in $\mathcal{W}_1$, i.e, $R$ is a $u_1-u_2$ path in the kernel graph, for some $u_1,u_2 \in \mathcal{X}_1 \cup \mathcal{W}_1$ and (2) $R$ contains no other points in $\mathcal{X}_1 \cup \mathcal{W}_1$, i.e., $R \cap \left( \mathcal{X}_1 \cup \mathcal{W}_1 \right)=\{u_1,u_2\}$. See Fig. \ref{fig:interdyy}(a) for an illustration.
\par Let $\mathcal{R}(y')$ be the collection of regions in $\KernelGraph(\mathcal{D}(y'))$.
Note that the union of regions in $\mathcal{R}(y')$ covers the kernel graph $\KernelGraph(\mathcal{D}(y'))$, i.e., $\KernelGraph(\mathcal{D}(y'))=\{R \in \mathcal{R}(y')\}$.
Let $N_R(y')=|\mathcal{R}(y')|$. We now bound $N_R(y')$ by the number of detours $N_{\sf D}(y')$.
\begin{claim}
\label{cl:nregion}
(1) $N_R(y')\leq 2 \cdot N_{\sf D}(y')$ regions.
(2) For every region $R \in \mathcal{R}(y')$, there exists  a detour $D_i \in \mathcal{D}(y')$ that contains it (i.e., $R \subseteq D_i$).
\end{claim}
\Proof
The two claim are shown by induction on the iterative process that constructs the kernel graph $\KernelGraph_v(\mathcal{D}(y'))$, analyzing the regions induced at each step.
Let $\mathcal{D}_\tau=\{D_1, \ldots, D_\tau\} \subseteq \mathcal{D}(y')$ and $N_\tau(y')$ be the number of regions induced up to step $\tau$ in  $\KernelGraph^\tau_v=\bigcup_{i\leq t}D_i[x_i, w_i]$.
For the induction base consider $D_1$.
The detour $D_1$ is non-truncated and hence the graph $\KernelGraph^1_v=D_1$ consists of a single region and $N_1(y')=1$, so (1) holds. In addition, the single region is $D_1$ and hence (2) holds as well.
\par Now assume that the two claims holds up to step $\tau-1$, and consider step $\tau$ when the detour $D_\tau$ is added to the kernel. If the current detour $D_\tau$ is non-truncated, then only one new region is added, namely, $D_\tau$, so $N_{\tau}(y')=N_{\tau-1}(y')+1 \leq 2\tau-1$ by the induction assumption, so (1) holds. In addition, since the new region is exactly $D_{\tau}$, part (2) holds as well.
Else, if $D_\tau$ is a truncated detour, let $R'$ be the region in the current kernel graph $\KernelGraph^{\tau-1}_v$ that contains the vertex $w_\tau$ (i.e., the first common vertex of $D_\tau$ and the current kernel graph $\KernelGraph^{\tau-1}_v$).
As $w_{\tau}$ joins the set $\mathcal{W}_1$, this region $R'$ is bisected into two regions, namely, before and after the vertex $w_\tau$, and there is an additional new region corresponding to the fragment $D_\tau[x_\tau, w_\tau]$. By the induction assumption, part (2) holds for the region $R'$ and hence it also holds for its two new fragments. The new region $D_{\tau}[x_{\tau},w_{\tau}]$ clearly satisfies part (2) as well.
Note that the remaining regions $R'' \neq R'$ in $\KernelGraph^{\tau-1}_v$ are unaffected by the addition of the detour $D_{\tau}$.
Overall, after this step $N_{\tau}=N_{\tau-1}+2$ so (1) holds. The claim follows.
\QED
The following claim shows a useful property of a regions.
\begin{claim}
\label{cl:region_useful}
If $D' \cap R \neq \emptyset$, then $R \subseteq D'$.
\end{claim}
\Proof
Let $R=D_{\ell}[w_1,w_2]$ for some $D_{\ell} \in \mathcal{D}(y')$ and $w_1,w_2 \in \mathcal{W}_1$. (By Cl. \ref{cl:nregion}(2) and by the definition of a region this is well defined.)
Assume, towards contradiction, that there exists a detour $D_1 \in \mathcal{D}(y')$ intersecting with $R$ but not containing it. Let $p \in D_1 \cap R$ be the first common vertex of $D_1$ in the region $R$. Since the detours $D_1$ and $D_\ell$ are $y$-interleaved, by Cl. \ref{cl:jointdet}, $D_{\ell}[p,y']=D_1[p,y']$ and hence $R[p,w_2] \subseteq D_1$. So by the contradictory assumption, there exists some vertex $p_2\in R[w_1,p_1] \setminus D_1$.
Let $w'=\First(D_1,D_\ell)$ be the first common vertex of $D_1$ and $D_\ell$.
By Cl. \ref{cl:w}, $w' \in \mathcal{W}_1$.
Hence, there are two cases to consider.
If the $w_1$ endpoint of $R$, is $w'$, the claim follows by Cl. \ref{cl:jointdet}, since $D_1[w_1,p_1]=D_{\ell}[w_1,p_1]$.

Otherwise, consider the case where $w_1\neq w'$.
By the definition of the region $R$, $w' \notin R[w_1, p_1]$. Since $w'$ appears on $D_{\ell}$ not after the common vertex $p_1$, it holds that it must appear \emph{before} $w_1$. The claim follows by Cl. \ref{cl:jointdet} again.
\QED
%\textbf{MP: Not really needed eventually.}
\begin{observation}
\label{obs:all}
$\KernelGraph(\mathcal{D}(y'))=G(\mathcal{D}(y'))$.
\end{observation}
\Proof
Assume towards contradiction otherwise and let
$D_t$ be the first detour in the ordering for which $D_t \nsubseteq \KernelGraph(\mathcal{D}(y'))$. Clearly, $t>1$. Let $D_{t'}$ for $t'<t$ be the first detour that $D_t$ intersected with when added to $\KernelGraph(\mathcal{D}(y'))$. Hence, $D_{t}[x_t,w_t] \subseteq \KernelGraph(\mathcal{D}(y'))$. Since $w_t, y' \in D_t \cap D_{t'}$, by Cl. \ref{cl:jointdet}, $D_{t}[w_t,y']=D_{t'}[w_t,y']$. Since $D_{t'} \subseteq \KernelGraph(\mathcal{D}(y'))$, we end with contradiction the claim holds.
\QED

For every replacement path $P_i \in \mathcal{P}(y')$, let $c_i \in D(P_i)$ be the unique $\sf D$-divergence point of $P_i$ and $D_i=D(P_i)$. Recall that since $P_i \notin \mathcal{P}_{nodet}$, such $c_i$ exists.
Let $\mathcal{P}^1(y')=\{P_i \in \mathcal{P}(y')~\mid~ c_i \in \mathcal{X}_1 \cup \mathcal{W}_1\}$ be those paths $P_i$ whose divergence points $c_i$ is an endpoint vertex in the $\mathcal{X}_1 \cup \mathcal{W}_1$ and let $\mathcal{P}^2(y')=\mathcal{P}(y') \setminus \mathcal{P}^1(y')$ be the remaining paths, whose divergence point from their detour is strictly inside a region.
\par We first bound the number of paths in $\mathcal{P}^1(y')$.
Since $|\mathcal{X}_1 \cup \mathcal{W}_1|\leq 2|N_{\sf D}(y')|$, by the distinctness of the $\sf D$-divergence points established in Lemma \ref{lem:uniquebp}, we have the following.
\begin{observation}
\label{obs:boundd1}
$|\mathcal{P}^1(y')|\leq 2|N_{\sf D}(y')|$.
\end{observation}
It remains to consider the replacement paths in $\mathcal{P}^2(y')$.
The goal of constructing the kernel graph $\KernelGraph(\mathcal{D}(y'))$ and its decomposition into regions, is the following key lemma.
\begin{lemma}
\label{lem:struc_r}
In any region $R \in \mathcal{R}(y')$, there exists at most one $\sf D$-divergence point $c_i$ for a unique $P_i \in \mathcal{P}^2(y')$.
\end{lemma}
\Proof
Assume, towards contradiction, that there is a region $R$ with two distinct divergence points $c_1 \neq c_2$ for two $(\pi,\sf D)$-paths $P_i \in SP(s, v, G\setminus \{e_i,t_i\})$, $i \in \{1,2\}$. Let $D_i=D(P_i)$ for $i \in \{1,2\}$. Note that it might be the case that $D_1=D_2$.
By Cl. \ref{cl:region_useful}, we get that $R \subseteq D_1,D_2$.
Since $P_i \notin \mathcal{P}_{nodet}$, it holds $c_i \in D_i$ for $i \in \{1,2\}$. Without loss of generality, assume that $c_1$ appears on $D_{1}$ (and $D_2$) before $c_2$.
Then by  Cl. \ref{cl:rp_prop1}(3.1), it holds that $P_2[s,c_2]=\pi(s,x_2) \circ D_2[x_2,c_2]$ and since $c_1 \in D_2[x_2,c_2]$ we have that $c_1 \in P_{2}$.

Let $P_3=I(P_1)$ be the path to which
$P_{1}$ interferes and let $D_3=D(P_3)$. By the selection of the interfered path $I(P_1)$, the detours $D_3$ and $D_1$ are $y$-interleaved ($y_3=y_1=y'$). Hence $D_1,D_2,D_3 \in \mathcal{D}(y')$. Let $t_3=(q_1,q_2)=F_2(P_3)$ and by Obs. \ref{obs:typee}(2), $F_2(P_1)$ appears on $D_3$ strictly after $t_3$, i.e., $F_2(P_1) \in D_3[q_1,y_3]$.
We now distinguish between two cases depending on the location of second failing edge $t_1=F_2(P_1)$ of $P_1$.
\dnsparagraph{Case (1): the edge $t_{1} \notin D_{1}[c_1, c_2]$}
Since $c_1, y' \in D_1, D_2$, by Cl. \ref{cl:jointdet}, $D_1[c_1, y']=D_2[c_1,y']$. Since $F_2(P_1), F_2(P_2) \in D_2[c_2,y']$ and $c_1,c_2$ are the unique divergence points of $P_1$ and $P_2$ from their detours $D_1$ and $D_2$ respectively, we get that there are two distinct $c_1-v$ shortest paths in $G \setminus (F(P_1) \cup F(P_2))$, namely, $P_1[c_1,v]$ and $P_2[c_1,v]$. By the optimality of $P_1$ and $P_2$, these subpaths are of the same length. Hence assuming, without loss of generality, that $P_1$ was constructed before $P_2$ by Alg. $\ConstPath$, we end with contradiction to the selection of $P_2$ (i.e., the last new edge of $P_2$ could have been avoided).

\dnsparagraph{Case (2): $t_{1} \in D_{1}[c_1, c_2]$}
Since $P_{1}$ interferes with $I(P_1)=P_3$, it follows that $y(D_3)=y'$ and the edge $t_3=F_2(P_3) \notin D_{1}$ appears on the suffix $P_{1}[c_1, v]$.
See Fig. \ref{fig:interdyy}(b) for an illustration.
Recall that $t_3=(q_1,q_2)$ where $q_1$ appears on $D_3$ before $q_2$.
By Obs. \ref{obs:typee}, the failing edge $t_1$ appears in $D_3$ after $t_3$. First note that since $D_1$ and $D_3$ are $y$-interleaved, it holds that $x_1 \neq x_3$ (as otherwise, $D_1=D_3$). In addition, since $D_3$ has a non-empty intersection with the region $R$ as $t_1 \in R \cap D_3$, by Cl. \ref{cl:region_useful}, we have that $R \subseteq D_3$.
Let $w=\First(D_1,D_3)$ be the first common vertex of the detours $D_1$ and $D_3$. By Cl. \ref{cl:w}, $w\in \mathcal{W}_1$.
As $c_1$ is an internal point in the region $R$, it does not belong to $\mathcal{W}_1$, so $w \neq c_1$, and moreover $w$ appears on $D_3$ and $D_1$ strictly before the detour divergence point $c_1$.

Note that $F_2(P_1) \notin D_3[t_3,c_1]$ since $F_2(P_1)$ occurs on $D_1,D_3$ only after $c_1$. Hence
by the optimality of the replacement path $P_{s,v,\{e_3\}}$ (where $e_3=F_1(P_3)$), we have that $P_3[q_2,c_1]=D_3[q_2,w] \circ D_3[w,c_1]$, and therefore
\begin{equation}
\label{eq:ineq111}
\dist(q_2,w, G \setminus (F(P_1) \cup \{e_3\}))<\dist(q_2, c_1, G \setminus (F(P_1) \cup \{e_3\}))~.
\end{equation}
On the other hand, since $P_1[b_1, v]$ is edge disjoint with $\pi(s,v)$ where $b_1$ is the unique $\pi$-divergence point of $P_1$ from $\pi(s,v)$, by the optimality of $P_1$, we have the following $w-q_2$ shortest path in $G \setminus (F(P) \cup \{e_3\})$, namely, $P_1[w,c_1] \circ P[c_1,q_2]$, since $w \neq c_1$, we have that
\begin{equation*}
\label{eq:ineq21}
\dist(q_2,w, G \setminus (F(P_1) \cup \{e_3\}))>\dist(q_2, c_1, G \setminus (F(P_1) \cup \{e_3\}))~,
\end{equation*}
which contradicts Eq. (\ref{eq:ineq111}). Note that, indeed, by the structure of the new-ending path $P_1 \notin \mathcal{P}_{nodet}$, it visits an edge $F_2(P_3)$ which is not on its detour, only \emph{after} leaving its detour. The claim follows.
\QED

Since every region contains exactly one $\sf D$-divergence point, by the distinctness of these points(see Lemma \ref{lem:uniquebp}), and by Lemma \ref{cl:nregion}, we have the following.
\begin{corollary}
\label{lem:onebp_inregion}
For every $y' \in \pi(s,v)$, $N_P(y')=O(N_{\sf D}(y'))$.
\end{corollary}

By Obs. \ref{obs:boundd1} and Cor. \ref{lem:onebp_inregion} we now have:
\begin{corollary}
\label{cor:nidonerelate}
$|\mathcal{I}^1_{\sf D}|=O(|\mathcal{D}_v|)$.
\end{corollary}
\Proof
By definition, the sets $\mathcal{D}(y_1),\mathcal{D}(y_2)$ are disjoint and thus $\mathcal{P}(y_1),\mathcal{P}(y_2)$ are disjoint as well. By Cor. \ref{lem:onebp_inregion}, we have that
$|\mathcal{I}^1_{\sf D}|=\sum_{y' \in \pi(s,v)} N_P(y')\leq c \cdot \sum_{y' \in \pi(s,v)}N_{\sf D}(y')=c \cdot |\mathcal{D}_v|$. The corollary holds.
\QED

%%%%%%%%%%%%%%%%%%%
%%%%%%%%%%%%%%%%%%%

\paragraph{Bounding the number of paths in $\mathcal{I}^2_{\sf D}$.}

We begin by defining for every path $P \in \mathcal{I}^2_{\sf D}$, a special interfered path $I(P)\in \mathcal{I}(P)$ such that $D_1=D(P)$ and $D_2=D(I(P))$ are fw-interleaved and moreover, $x(D_1)<x(D_2)$. The next observation justifies the existence of such a path.
\begin{observation}
\label{obs:interd1}
For every $P \in \mathcal{I}^2_{\sf D}$, there exists $I(P) \in \mathcal{I}(P)$ satisfying that $D_1=D(P)$ and $D_2=D(I(P))$ are fw-interleaved such that $x(D_1) < x(D_2)$.
\end{observation}
\Proof
Consider some path $P \in \mathcal{I}^2_{\sf D}$.
For ease of notation, for every $(\pi,\sf D)$ new-ending path $P'$, let $x(P')=x(D(P'))$ and $y(P')=y(D(P'))$.
We first show that there exists $P'\in \mathcal{I}(P)$ such that $y(P)\leq y(P')$.
Assume towards contradiction that for every $P' \in \mathcal{I}(P)$, $y(P)>y(P')$. Since $D(P)$ and $D(P')$ are dependent (see Cl. \ref{cl:interd}), by Cl. \ref{cl:det_nested}(b), $F_1(P) \in \pi(y(P'),y(P))$, hence $P$ $\pi$-interferes with every $P'\in \mathcal{I}(P)$, leading to contradiction by the fact that $P \in \mathcal{I}_{\sf D}$ (i.e., in such a case, $P \in \mathcal{I}_{\pi}$). Hence, there exists $P'$, for which $y(P) \leq y(P')$.
Since $P \notin \mathcal{I}^1_{\sf D}$, necessarily $y(P)\neq y(P')$, concluding that $y(P) < y(P')$.
\par We now show that also $x(P) < x(P')$.
By Cl. \ref{cl:interd}, we have that $x(P)\neq x(P')$.
So it remains to disqualify the possibility that $x(P)> x(P')$. Indeed if $x(P)> x(P')$, then combining with the fact that $y(P) < y(P')$, we get that $D(P)$ is nested in $D(P')$ and since the detours are dependent, we end with contradiction by Cl. \ref{cl:depen_detdisjoint}.
\par So far, we have that $x(P)<x(P')\leq y(P)<y(P')$.
Note that since $F_2(P) \in D(P) \cap D(P')$ and $x(P)<x(P')$, by Cor. \ref{cl:rev_nomutual}, $D(P)$ and $D(P')$ are neither $(x,y)$-interleaved nor rev-interleaved. Hence, by Cl. \ref{cl:summ_depend}(a), it must hold that
$D(P)$ and $D(P')$ are fw-interleaved.
The claim follows.
\QED
Let $\mathcal{D}_2=\{D(P) ~\mid~ P \in \mathcal{I}^2_{\sf D}\}$ be the set of detours corresponding to the paths in $\mathcal{I}^2_{\sf D}$.
For every detour $D \in \mathcal{D}_2$, define the set of detours $\mathcal{I}(D)$ as the collection of detours $D(P')$ such that $P'=I(P)$ for some path $P \in \mathcal{I}^2_{\sf D}$ whose detour $D(P)$ is $D$, i.e.,
$$\mathcal{I}(D)=\{D(P') ~\mid~ \mbox{~there exists~} P \in \mathcal{I}^2_{\sf D} \mbox{~such that~} D(P)=D, I(P)=P' \}.$$
By Obs. \ref{obs:interd1}, we have the following.
\begin{observation}
\label{obs:interd2}
For every $D \in \mathcal{D}_2$ and every $D' \in \mathcal{I}(D)$, it holds that $D'$ and $D$ are dependent and fw-interleaved such that $x(D)<x(D')<y(D)<y(D')$.
\end{observation}
To bound the number of paths in $\mathcal{I}^2_{\sf D}$, we define a prefix $D[x(D),w(D)]$ for every detour $D \in \mathcal{D}_2$ and show that every such prefix contains at most one $\sf D$-divergence point of some path in $\mathcal{I}^2_{\sf D}$. In this sense, the prefix is the analogue of the region, used in the analysis of the paths in $\mathcal{I}^1_{\sf D}$.
\par
\begin{figure}[htbp]
\begin{center}
\includegraphics[width=5in]{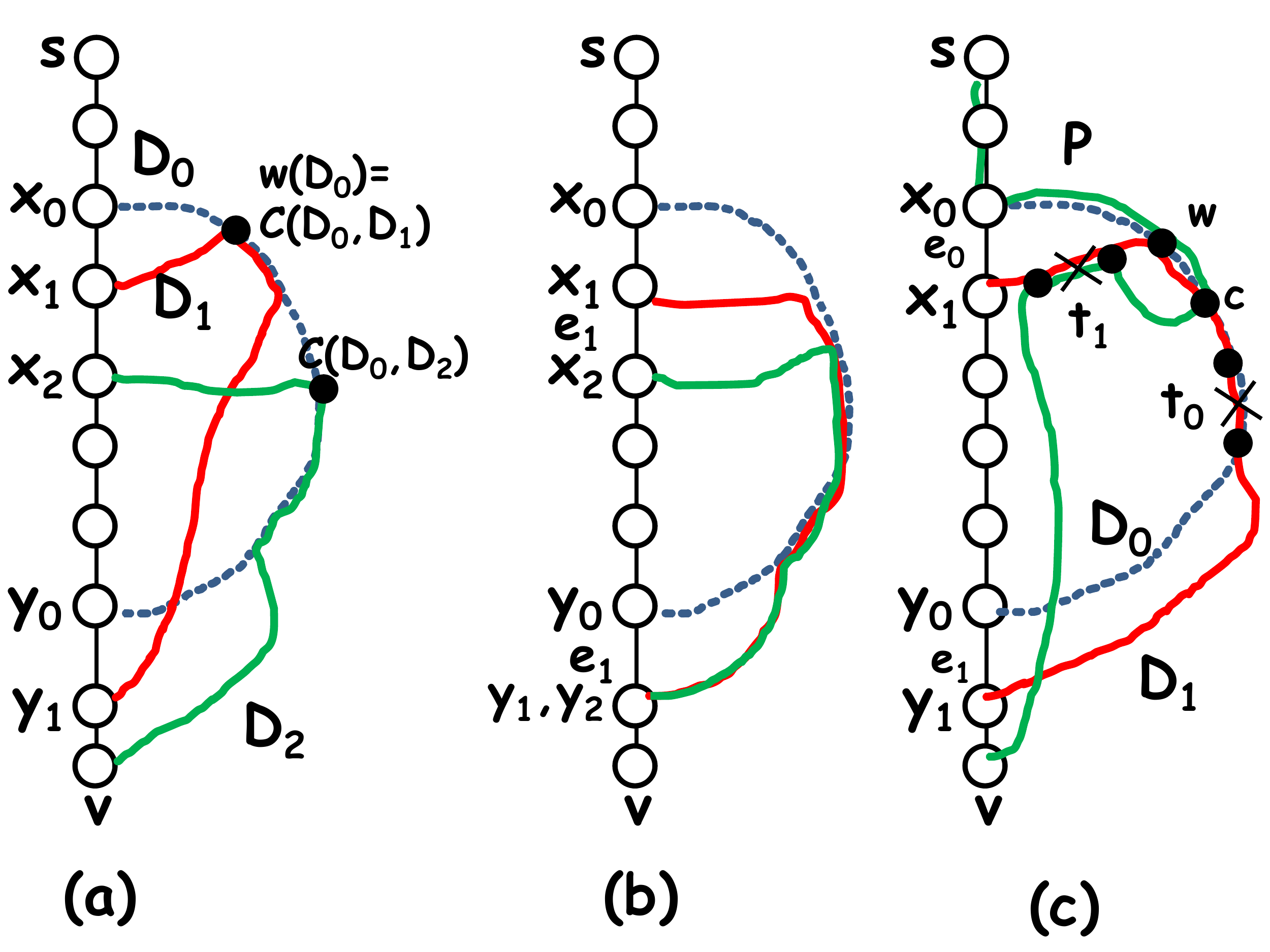}
\end{center}
\caption{(a) The detours $D_1,D_2 \in \mathcal{I}(D_0)$. The prefix $D_0[x_0,w(D_0)]$ is determined by the earliest first common vertex with the detours of $\mathcal{I}(D_0)$. In this case, $w(D_0)=\First(D_0,D_1)$ since $\First(D_0,D_2)$ appears on $D_0$ strictly after it.
(b) Illustration for Lemma \ref{lem:disjoint_inter}.
The detours $D_0$ and $D_1$ are interleaving and also $D_0$ and $D_2$ are interleaving. The assumption that $D_1$ and $D_2$ are dependent, implies contradicting positions for the location of $e_1$. Since $y_0<y_1$ and $D_0$ and $D_1$ are dependent, we have that $e_1 \in \pi(y_0,y_1)$. On the other hand, $x_1<x_2$ and $D_1$ and $D_2$ are dependent, we have that $e_1 \in \pi(x_1,x_2)$, leading to contradiction since $y_0\geq x_2$.
(c) Illustration for Lemma \ref{lem:wherep}.
Shown is a replacement path $P=P_{s,v,\{e_0,t_0\}}$ where $t_0 \in D_0$ and hence $D(P)=D_0$. The path $P$ $\sf D$-interferes with a path $P'=P_{s,v,\{e_1,t_1\}}$, hence $P$ visits the edge $t_1 \in P \setminus D_0$ after diverging from its detour at the point $c$. Since $c \notin D_0[x_0,w(D_0)]$, a strictly better route from $w$ to $t_1$ exists using the subpath provided by the detour $D_1$.
\label{fig:interd}}
\end{figure}

 For every detour $D \in \mathcal{D}_2$, define
the unique point $w(D) \in D$ in the following manner.
Let $W(D)$ be the collection of first common vertices of the detours $D$ and $D' \in \mathcal{I}(D)$, i.e., $W(D)=\{\First(D,D') ~\mid~ D' \in \mathcal{I}(D)\}$.
Note that by Obs. \ref{obs:interd2}, $D$ and $D'$ are dependent and fw-interleaved, hence by Cl. \ref{cl:summ_depend}, $\First(D,D')=\First(D',D)$.
Then, let $w(D) \in W(D)$ be the point whose distance from $x(D)$ on $D$ is \emph{minimal}. In other words, $w(D)$ is the earliest first common vertex of the detour $D$ with any of the detours in $\mathcal{I}(D)$. See Fig. \ref{fig:interd}(a) for a schematic illustration. Our next goal is to show that every $\sf D$-divergence point $c(P)$ must occur on the prefix $\Psi(D)=D[x(D),w(D)]$ and in addition, for each $D \in \mathcal{D}_2$, there exists at most one path $P \in \mathcal{I}^2_{\sf D}$ whose divergence point $c(P)$ is in $\Psi(D)$. This implies the stronger conclusion that $D(P)\neq D(P')$ for every $P,P' \in \mathcal{I}^2_{\sf D}$ and hence $|\mathcal{D}_2|=|\mathcal{I}^2_{\sf D}$.

The next lemma is crucial for analyzing the set $\mathcal{I}^2_{\sf D}$.
%Consider a detour $D_1 \in \mathcal{D}_2$ and $D_2,D_3$ such that there exists $P_1,P_2 \in \mathcal{I}^2_{\sf D}$, $D(P_1)=D(P_2)=D_1$ and $D(I(P_1))=D_2, D(I(P_2))=D_3$ where
%$w_1=\First(D_1,D_2), w_2=\First(D_1,D_3) \in W(D)$, $w_1$ appears strictly before $w_2$ on $D_1$.
\begin{lemma}
\label{lem:disjoint_inter}
Let $P_1,P_2 \in \mathcal{I}^2_{\sf D}$ be such that $D(P_1)=D(P_2)=D_0$. Let $I_1=I(P_1)$ and $I_2=I(P_2), D_1=D(I_1), D_2=D(I_2)$. Hence $D_1,D_2\in \mathcal{I}(D)$.
If $\First(D_0,D_1) \neq \First(D_0,D_2)$, then $D_0 \cap D_1 \cap D_2=\emptyset$.
\end{lemma}
\Proof
Let $x_i=x(D_i)$ and $y_i=y(D_i)$ for $i \in \{0,1,2\}$, $w_1=\First(D_0,D_1), w_2=\First(D_0,D_2)$ and assume without loss of generality that $w_1$ appears  on $D_0$ strictly before $w_2$.
Note that by Obs. \ref{obs:interd2}, $D_0$ is fw-interleaved with both $D_1$ and $D_2$. Hence, $\First(D_0,D_1)=\First(D_1,D_0)$ and also $\First(D_0,D_2)=\First(D_2,D_0)$.
Assume towards contradiction that there exists a common vertex $\ell \in D_0 \cap D_1 \cap D_2$.
We check two cases.\\
Case (a): $x_1=x_2$. By Cl. \ref{cl:jointdet}, $D_1[x_1,\ell]=D_2[x_2,\ell]$. By definition, the common vertex $\ell$ appears not before $w_1$ on $D_1$, hence $w_1 \in D_1[x_1,\ell]=D_2[x_2,\ell]$, contradicting the fact that $w_2$ is the first common vertex of $D_0$ and $D_2$.
\\
Case (b) $x_1 \neq x_2$.
In this case, we show that the detours $D_1$ and $D_2$ are independent, i.e., $D_1 \cap D_2=\emptyset$, which implies the claim. See Fig. \ref{fig:interd}(b) for an illustration.
Assume towards contradiction that $D_1 \cap D_2 \neq \emptyset$. Assume first that $x_1<x_2$.
By Obs. \ref{obs:interd2}, $D_0$ and $D_1$ are dependent and $y_0<y_1$. Hence, by Cl. \ref{cl:det_nested}(b),
$F_1(I_1) \in \pi(y_0,y_1)$.
On the other hand, since by the contradictory assumption
$D_1$ and $D_2$ are dependent, by Cl. \ref{cl:det_nested}(a),
$F_1(I_1) \in \pi(x_1,x_2)$.
Overall, we have that $F_1(I_1) \in \pi(x_1,x_2)\cap  \pi(y_0,y_1)$. Finally, by Obs. \ref{obs:interd2}, $D_0$ and $D_2$
are interleaved such that $x_2<y_0$, leading to contradiction as $\pi(x_1,x_2)\cap  \pi(y_0,y_1)=\emptyset$. The case where $x_1>x_2$ is analogous. The claim follows.
%
%Recall that $x_0<x_1,x_2$. Assume first that $x_1<x_2$ and recall that $y_1,y_2>y_0$.
%Let $e_0=F_1(P_1), e_1=F_1(I_1), e_2=F_1(I_2)$.
%Since $I_1 \in \mathcal{I}(P_1)$ by Obs. \ref{obs:typee}(2), it holds that $F_2(P_1) \in D_1 \cap D_0$, i.e., the detours $D_0$ and $D_1$ are dependent.
%By Cl. \ref{cl:det_nested}, it holds that $e_0 \in \pi(x_0,x_1)$ and $e_1 \in \pi(y_0,y_1)$. In addition, by Cl. \ref{cl:disjindep}, it holds that the detours $D_0$ and $D_i$ are interleaved for $i \in \{1,2\}$, hence $x_1,x_2<y_0$. Analogously, for the detours $D_0$ and $D_2$, it holds that they are interleaved and dependent and hence $e_2 \in \pi(y_0,y_2)$.
%Consider now the detours $D_1$ and $D_2$ under the contradictory assumption that these detours are dependent.
%Since $x_1 \neq x_2$, there are two subcases. If $x_1<x_2$, then by the proof of Cl. \ref{cl:det_nested}, it holds that $e_1 \in \pi(x_1,x_2)$. Contradiction to the fact that $e_1 \in \pi(y_0,y_1)$. Analogously, if $x_1>x_2$ then $e_2 \in \pi(x_2,x_1)$. Contradiction to the fact that $e_2 \in \pi(y_0,y_2)$. The claim follows.
\QED

\begin{lemma}
\label{lem:wherep}
For every $P_0 \in \mathcal{I}^2_{\sf D}$:\\
(1) the unique $\sf D$-divergence point $c=c(P_0)$ with $D_0=D(P_0)$ is in $D_0[x_0,w(D_0)]$ where $x_0=x(D_0)$ and (2) $F_2(P_0) \in D_0[w(D_0),y(D_0)]$.
\end{lemma}
\Proof
Begin with (1) and assume towards contradiction otherwise. Consider $P_0\in \mathcal{I}^2_{\sf D}$ such that $D(P_0)=D_0$ and $c\notin D_0[x_0,w(D_0)]$, let $P_1=I(P_0)$, $D_1=D(P_1)$.

We first claim that $w(D_0) \in P_0$.
To see this observe that since $P_0 \notin \mathcal{P}_{nodet}$, by Cl. \ref{cl:rp_prop1}(3.1), $P_0[s,c]=\pi(s, x_0)\circ D_0[x_0, c]$. Since (by the contradictory assumption) $c$ appears on $D_0$ strictly after $w(D_0)$, it holds that $w(D_0) \in D_0[x_0, c]$ and hence $w(D_0) \in P_0$.
We next distinguish between two cases depending on the value of $\First(D_0,D_1)$.
\dnsparagraph{Case (1): $w(D_0)=\First(D_0,D_1)$}
In this case, $\First(D_0,D_1)$ is selected as the vertex $w(D_0)$ that defines the prefix $D_0[x_0,w(D_0)]$.
By Obs. \ref{obs:typee}, $F_2(P_0) \in D_0 \cap D_1$, and hence $F_2(P_0)$ appears on $D_1$ \emph{after} the first common vertex $w(D_0)$. In addition, $F_2(P_1)$ appears on $D_1$ before $F_2(P_0)$ and since  $F_2(P_1)\in D_1 \setminus D_0$ (by the definition of interference), it holds that $F_2(P_1)$ appears on $D_1$ \emph{before} the common vertex $w(D_0)$.

Let $F_2(P_0)=(q_1,q_2)$. We then have that $w(D_0), q_2 \in D_0 \cap D_1$ and hence by Cl. \ref{cl:jointdet}, $D_0[w(D_0),q_2]=D_1[w(D_0),q_2]$.
Since $c$ appears on $D_0$ after $w(D_0)$ but before the failing edge $F_2(P_0)$, we get that $c \in D_0[w(D_0),q_2]=D_1[w(D_0),q_2]$. Let $F_2(P_1)=(a_1,a_2)$.
We now consider the path $Q=D_1[a_1,w(D_0)] \circ D_1[w(D_0),c]$. We claim that $Q$ is an $a_1-c$ shortest-path in $G \setminus \left(F(P_0) \cup \{F_1(P_1)\} \right)$, since (1) clearly, $D_1[a_1,c] \in SP(s,v, G \setminus \{F_1(P_1)\})$ and (2) $F_2(P_0)$ occurs on $D_1$ only after the $\sf D$-divergence point $c$, so $F_2(P_0) \notin D_1[a_1,c]$, and hence $D_1[a_1,c] \in SP(s,v, G \setminus F(P_0))$.
\par Since the path $D_1[a_1,c]$ visits $w(D_0)$, we have that
\begin{equation}
\label{eq:ineq1}
\dist(a_1,w(D_0), G \setminus \left(F(P_0) \cup \{F_1(P_1)\} \right))<\dist(a_1, c, G \setminus \left(F(P_0) \cup \{F_1(P_1)\}\right))~.
\end{equation}
We now use the path $P_0$ to present a $w(D_0)-a_1$ path in $G \setminus \left(F(P_0) \cup \{F_1(P_1)\}\right)$ that goes through $c$. Recall that $w(D_0) \in P_0$.
Note that by  Cl. \ref{cl:rp_prop1}(3.1), since the interfered edge $F_2(P_1)$ is not on $D(P_0)$ (by the definition of interference), it holds that $P_0$ visits this edge only after leaving its detour $D_0$, i.e., after visiting the $\sf D$-divergence point $c$.
Hence, the route in $P_0$ from $w(D_0)$ to $a_1$ is given by $P_0[w(D_0),c] \circ P_0[c,a_1]$.
Note that since $c$ appears on $P_0$ after the $\pi$-divergence point $b(P_0)$, indeed the subpath $P_0[w(D_0),a_1]$ does not contain the edge $F_1(P_1)$, implying that $|P_0[w(D_0),a_1]|=\dist(w(D_0),a_1, G \setminus \left(F(P_0) \cup \{F_1(P_1)\} \right))$.
Since $P_0[w(D_0),a_1]$ visits $c$, we have that
\begin{equation}
\label{eq:ineq2}
\dist(a_1,w(D_0), G \setminus \left(F(P_0) \cup \{F_1(P_1)\} \right))>\dist(a_1, c, G \setminus \left(F(P_0) \cup \{F_1(P_1)\}\right))~,
\end{equation}
contradiction by Eq. (\ref{eq:ineq1}).
See Fig. \ref{fig:interd}(c) for an illustration.

\dnsparagraph{Case (2): $w(D_0) \neq \First(D_0,D_1)$}
Let $D^* \in \mathcal{I}(D_0)$ be the detour satisfying that $w(D_0)=\First(D_0,D^*)$, i.e., its first common vertex with $D_0$ appears earlier than all other detours $D' \in \mathcal{I}(D)$. Hence, there exists a path $\widehat{P} \in \mathcal{I}^2_{\sf D}$ such that $D(\widehat{P})=D_0$ and $D(\widehat{I})=D^*$ where $\widehat{I}=I(\widehat{P})$.

\par Consider now the $w(D_0)-v$ path $Q=D^*[w(D_0), y(D^*)] \circ \pi(y(D^*),v)$. We have that $Q \in SP(w(D_0), v, G \setminus \{F_1(\widehat{I})\})$. We now show that $F_1(P_0),F_2(P_0) \notin Q$. By Obs. \ref{obs:interd2}, $y(D^*)>y(D_0)$ and hence $F_1(P_0) \notin \pi(y(D^*),v)$.
Clearly, also $F_1(P_0) \notin D^*$, hence $F_1(P_0)\notin Q$. In addition, since $\First(D_0,D^*) \neq \First(D_0,D_1)$ and $F_2(P_0) \in D_0 \cap D_1$, by Cl. \ref{lem:disjoint_inter}, $D_0 \cap D_1 \cap D^*=\emptyset$, and hence $F_2(P_0) \notin D^*$.
Clearly, $F_2(P_0) \notin \pi(s,v)$, hence $F_2(P_0) \notin Q$. Overall, we have that $Q$ is a $w(D_0)-v$ shortest-path
in $G'=G\setminus \left(F(P_0) \cup \{F_1(\widehat{I})\}\right)$.

We now use $P_0$ to present an alternative $w(D_0)-v$ shortest-path in $G'$, namely, $P_0[w(D_0),v]$.
Since $P_0$ visits $w(D_0)$ only after leaving the shortest-path $\pi(s,v)$, it holds that $F_1(\widehat{I}) \notin P_0[w(D_0),v]$, hence $P_0[w(D_0),v] \in SP(w(D_0),v, G')$ as well.
By the optimality of $P_0[w(D_0),v]$ and $Q$, we get that there are of the same lengths, leading to contradiction in the selection of $P_0$ by Alg. $\ConstPath$ (i.e., the path $P_0[s,w] \circ Q$ is optimal in length and it is not new-ending, so the new-edge of $P_0$ could have been avoided). \par Now, consider part (2). Since $F_2(P_0) \in D_0 \cap D_1$, it holds that $F_2(P_0)$ occurs on $D_0$ after $\First(D_0,D_1)$. Since $w(D_0)$ is the earliest intersection point with some $D' \in \mathcal{I}(D_0)$, the claim holds.
\QED

%%%%%%%%%%%%%%%%%%%
%%%%%%%%%%%%%%%%%%%
\begin{lemma}
\label{lem:counti2}
$|\mathcal{I}^2_{\sf D}|=|\mathcal{D}_2|$~.
\end{lemma}
\Proof
We show that $D(P) \neq D(P')$ for every $P,P' \in \mathcal{I}^2_{\sf D}$.
Assume towards contradiction that there exists at least two paths $P,P'\in \mathcal{I}^2_{\sf D}$ such that $D(P)=D(P')=D$.
Since $P,P' \notin \mathcal{P}_{nodet}$, their unique $\sf D$-divergence points $c=c(P)$ and $c'=c(P')$ respectively, appear on the common detour $D$.
In particular, by Lemma \ref{lem:wherep}, it holds that $c,c' \in  D[x,w(D)]$.
By the distinctness of the $\sf D$ divergence points of Lemma \ref{lem:uniquebp}, $c\neq c'$, so without loss of generality, assume that $c$ occurs on $D$ before $c'$.
We now show that the failing edges $F(P)$ and $F(P')$ do not occur on both of the paths $P$ and $P'$.
First, note that since $P,P' \notin  \mathcal{P}_{nodet}$, their $\pi$-divergence point is
$b(P)=b(P')=x(D)$, and thus $F_1(P),F_1(P') \notin P,P'$.
We now show that also $F_2(P),F_2(P') \notin P,P'$.
By Lemma \ref{lem:wherep}, $F_2(P),F_2(P') \notin D[x,w(D)]$ hence it also holds that  $F_2(P),F_2(P') \notin D[c,c']$.
Finally, since $F_2(P),F_2(P') \notin D[c,c']$ and $P[c,v]$ and $P[c',v]$ are disjoint with the detour $D$, the claim holds.
So, we have two distinct $c-v$ paths in $G \setminus (F(P) \cup F(P'))$, namely, $P[c,v]$ and $P'[c,v]$, leading to contradiction to the selection of the latter paths by Algorithm $\ConstPath$. The claim holds.
\QED

\begin{corollary}
\label{cor:ninterb}
(1) $N_P=O(N_{\sf D})$ and
(2) $N_P=O(n^{2/3})$.
\end{corollary}
\Proof
Part (1) follows immediately by Obs. \ref{obs:boundd1} and Lemma \ref{lem:onebp_inregion}. To prove part (2),
select a subset $\mathcal{P'}$ of $N_{\sf D}$ new-ending paths, such that for every pair of paths $P,P' \in  \mathcal{P'}$, $D(P) \neq D(P')$ and hence also $F_1(P)\neq F_1(P')$. That is for every detour $D \in \mathcal{D}_v$ we take one path representor $P \in \mathcal{P}_v$ satisfying that $D(P)=D$.
Note that $\mathcal{P'}$ is not necessarily related to the set of interfering paths $\mathcal{I}_{\sf D}$. By Lemma \ref{lem:bound_no_touch}, it holds that $|\mathcal{P'}|=N_{\sf D}=O(n^{2/3})$, hence combining with by part (1), $N_P=O(n^{2/3})$. Part (2) follows.
\QED

\subsection{Bounding the number of independent paths $\mathcal{P}_{indep}$}
\label{sec:indeppaths}
Let
$\mathcal{P}_{indep}$ denote the set of indpendent new-ending paths where $$\mathcal{P}_{indep}=\{P \in \mathcal{P}_v
~\mid~ \mbox{~there exists no~} P' \in \mathcal{P}_v \mbox{~such that~} F_2(P') \in P \setminus D(P)\}$$
and let $N_{indep}=|\mathcal{P}_{indep}|$ denote their number.
In this subsection, we bound $N_{indep}$ by $O(n^{2/3})$.
%
%Order the paths of $\mathcal{P}_{indep}$ in increasing order for every $e_i \in \pi(s,v)$ and $t_j\in D_i$.
%Let $H_0=BFS(s,G)$ and $\mathcal{P}_0=\emptyset$. At step $\ell\geq 1$,  consider the pair $\langle e_i,t_j\rangle$ and add $P_{i,j} \in SP(s, v, G \setminus \{e_i, t_j\})$ to $H_{\ell-1}$ and to $\mathcal{P}_{\ell-1}$ only if
%$\dist(s, v, H_{\ell-1} \setminus \{e_i, t_j\})>\dist(s, v, G \setminus \{e_i, t_j\})$. Let $\mathcal{P}'$ be the final collection of paths.
%
%
% For every path $P \in \mathcal{P}_{indep}$, such that $P \in SP(s,v, G \setminus \{e_i,t_j\})$, let $e_k=\widehat{F}_1(P)$ be the lowest edge on $\pi(s,v)$ such that $t_j \in D_k$ and $P \in SP(s,v, G\setminus \{e_k,t_j\},W)$. Hence, $F_1(P)$ appears not below $\widehat{F}_1(P)$ on $\pi(s,v)$.
%
%
%
Recall that for every $(\pi,\sf D)$ new-ending path $P_i \notin \mathcal{P}_{nodet}$, $b_i$ is its $\pi$-divergence point and $c_i$ is its $\sf D$-divergence point and by Cl. \ref{cl:rp_prop1}, these divergence points are unique (i.e., $P_i$ does not return to $\pi(s,v)$ after $b_i$, and does not return to $D(P_i)$ after $c_i$).
We begin by showing that the suffix of every independent path $P_i$ starting from its $\sf D$-divergence point $c_i$ (i.e., $P'_i=P_i[c_i,v]\setminus \{v\}$) is  disjoint from the suffix $P'_j$ of any other $P'_j \in \mathcal{P}_{indep}$.
%The next observation follows immediately by the independence.

\begin{observation}
\label{obs:dist_indep}
For every two paths $P_i,P_j \in \mathcal{P}_{indep}$,
$P'_i=P_i[c_i, v]\setminus \{v\}$ and $P'_j=P_j[c_j,v]\setminus \{v\}$ are vertex disjoint.
\end{observation}
\Proof
%By the construction of Alg. $\ConstPath$, $P_{k}[c_k, v] \in SP(s,v, G \setminus \left(D(P_k) \cup \pi(s,v) \right),W)$ and hence the uniqueness of shortest paths can be applied on these segments.
%\textbf{MP: an alternative argumentation.}
%Consider a path $P_i \in \mathcal{P}_{indep}$ and let $b_i$ be the first divergence point of $P_i$ and $\pi(s,v)$.
Without loss of generality, assume the $P_i$ was constructed by Alg. $\ConstPath$ before $P_j$.
Since $P'_k \subseteq P_k[b_k,v]$ for $k \in \{i,j\}$, by the uniqueness of the $\pi$-divergence point it holds that $P'_k$ is edge disjoint with $\pi(s,v)$ for .
Hence, $F_1(P_i) \notin P'_j$ and $F_1(P_j) \notin P'_i$.
We now consider the second faults.
By the definition of $\mathcal{P}_{indep}$, it holds that $F_2(P_j) \notin P_i \setminus D(P_i)$. Hence, by the uniqueness of the $\sf D$-divergence points $c_i$ and $c_j$, it holds that also $F_2(P_j) \notin P'_i$ and similarly $F_2(P_i) \notin P'_j$.

Assume, towards contradiction, that there exists a common vertex $w \neq v$ in the intersection of the suffixes $P'_i$ and $P'_j$. By the above, we get that there are two distinct $w-v$ paths in $G \setminus \left(F(P_i) \cup F(P_j)\right)$, namely, $P'_i[w,v] \neq P'_j[w,v]$, leading to contradiction by the selection of $P_j$ by Algorithm $\ConstPath$ (i.e., the last edge of $P_j$ could have been avoided).
\QED
%\begin{claim}
%\label{cl:nodisb}
%For every two new-ending paths $P_1,P_2 \in \mathcal{P}_{indep}$ with $D_i=D(P_i), i \in \{1,2\}$.
%If $x(D_1)=x(D_2)$, then $D_1$ and $D_2$ are \emph{dependent}.
%\end{claim}
%\Proof
%Without loss of generality, assume that $P_1$ was constructed by Alg. $\ConstPath$ before $P_2$.
%Assume towards contradiction that $D_1$ and $D_2$ are independent. Then since $P_1,P_2 \notin \mathcal{P}_{nodet}$, it holds that the unique $\pi$-divergence point is the first point of the corresponding detour, hence $b(P_1)=x(D_1)=x(D_2)$, and $F_1(P_2) \notin P_1$. In addition, since $P_1$ does not interfere with $P_2$, it holds that $F_2(P_2) \notin P_1 \setminus D_1$. Finally, as $P_1[x_1, c_1] \subseteq D_1$ contains no common edge with $D_2$, we have that also $F_2(P_2) \notin P_1$. In the same manner, it also holds that $F_1(P_1),F_2(P_1) \notin P_2$. Hence, there are two $s-v$ paths in $G \setminus \left(F(P_1) \cup F(P_2) \right)$, and by optimality of these paths, they  are of the same lengths. We end with contradiction to the selection of $P_2$ by Alg. $\ConstPath$. The claim holds.
%\QED
\begin{claim}
\label{cl:allpoint_ondetour}
For every two independent paths $P_1,P_2 \in \mathcal{P}_{indep}$ with $D_i=D(P_i), i \in \{1,2\}$, and $y_2>y_1$, if $b_1=b_2$ then $c_1 \in  D_2$ where $y_2>y_1$.
\end{claim}
\Proof
Let $w=\Last(D_1,D_2)$ be the last common vertex of $D_1$ and $D_2$ (since $D_1$ and $D_2$ are $x$-interleaved, such $w$ exists). Hence, by Cl. \ref{cl:jointdet}, $D_1[x_1,w]=D_2[x_2,w]$.
Assume towards contradiction that $c_1 \in D_1[w,y_1]$. Since $P_1 \notin \mathcal{P}_{nodet}$, by Cl. \ref{cl:rp_prop1}(3.1), it holds that $P_1[s,c_1]=\pi(s,x_1) \circ D_1[x_1,c_1]$, concluding that $F_2(P_1)$ must fall in the $D_1$-excluded region $D_1[w,y_1]$, leading to contradiction by Cl. \ref{cl:depen_detbad}.
The claim follows.
\QED

Equipped with Cl. \ref{cl:allpoint_ondetour}, we now induce a $(b,c)$-ordering on the independent paths $\mathcal{P}_{indep}$, which can be viewed as based on treating $b_i$ and $c_i$ lexicographically.
Recall that for two vertices $u_1,u_2 \in \pi(s,v)$, we denote $u_1<u_2$ if $\dist(s,u_1,\pi(s,v))<\dist(s,u_2,\pi(s,v))$.
For $b_i\neq b_j$, we say that $(b_i,c_i)<(b_j,c_j)$ if $b_i < b_j$. For $b_i=b_j$, we use the second coordinate $c_i$ to break the tie. Let $D_k \in \{D(P_i),D(P_j)\}$ be the detour with the lower $y$-value among the two (i.e., closer to $v$ on $\pi(s,v)$). Note that by Cl. \ref{cl:allpoint_ondetour}, both $c_i,c_j \in D_k$. In addition, by Lemma \ref{lem:uniquebp}, $c_i \neq c_j$.
This allows us to define, in this case, that $(b_i,c_i) < (b_i,c_j)$ iff $\dist(b_i,c_i, D_k)<\dist(b_i,c_j, D_k)$.

We now order the independent paths in increasing $(b,c)$ order of their values.
Let $\overrightarrow{\mathcal{P}}_{indep}=\{P_1, \ldots, P_{\ell}\}$ where $(b_1,c_1)<(b_2,c_2)< \ldots<(b_\ell,c_\ell)$.
Our next goal is to show that the lengths of the paths in the ordered set $\overrightarrow{\mathcal{P}}_{indep}$ is strictly monotone decreasing, i.e., $|P_1| > \ldots > |P_{\ell}|$. Towards this, we establish the next important lemma.
\begin{lemma}
\label{cl:nofailedg}
Let $P_i,P_j \in \mathcal{P}_v\setminus \mathcal{P}_{nodet}$ such that $P_i$ does not interfere with $P_j$ and $b_i<b_j$. Then, $F_1(P_j), F_2(P_j) \notin P_i$.
\end{lemma}
\Proof
Let $e_k=F_1(P_k)$, $t_k=F_2(P_k)$, $D_k=D(P_k)$ for $k \in \{i,j\}$.
Note that since $P_k \notin \mathcal{P}_{nodet}$, it holds that $x_k=b_k$ (where $x_k=x(D_k)$) for $k \in \{i,j\}$.
Since $P_i$ is a new-ending replacement path, it diverges from $\pi(s,v)$ above $b_j$ and hence also above $e_j$. Therefore, $e_j \notin P_i$. So it remains to consider $t_j$. Assume towards contradiction that $t_j \in P_i$.  Since $P_i$ does not interfere with $P_j$, it holds that $t_j \in D_i \cap D_j$ (i.e., $D_i$ and $D_j$ are dependent). In particular, since $P_i \notin \mathcal{P}_{nodet}$, by Cl. \ref{cl:rp_prop1}(3.1), it holds that $t_j \in D_i[x_i, c_i]$. Hence, $t_j$ appears on $D_i$ strictly above $t_i$.
Note that since $D_i$ and $D_j$ are dependent and $x_i<x_j$, by Cl. \ref{cl:det_nested}(a), we have that
$e_i\in \pi(x_i,x_j)$.
%Note that by Cor. \ref{cor:rev_nomutual}, $D_i$ and $D_j$ are \emph{not} rev-interleaved, hence the common segment of these dependent detours is used in the same direction.
%
%
%We first claim that $e_i = F_1(P_i) \in \pi(b_i,b_j)$.
%\begin{claim}
%\label{cl:locateedge}
%$e_i = F_1(P_i) \in \pi(b_i,b_j)$.
%\end{claim}
%\Proof
%Let $w \in D_i \cap D_j$ be a first common vertex of $D_j$ and $D_i$. Assume towards contradiction that $e_i \notin \pi(b_i,b_j)$.
%In this case, there are two distinct $b_i-w$ paths in $G \setminus \{e_i,e_j\}$ given by $Q_1=\pi(b_{i},b_j) \circ D_j[b_j,w]$ and $Q_2=D_i[b_i,w]$. By optimality of $P_{s,v,\{e_i\}}$ and $P_{s,v,\{e_j\}}$, it holds that both paths are of the same lengths $|Q_1|=|Q_2|$.
%Hence, we end with contradiction to the construction of $P_{s,v,\{e_j\}}$ by Alg. $\ConstPath$, since it should have chosen $b_i$ as the closest divergence point to $s$. The claim holds.
%\QED
So far, we have the following: $e_i$ is above $e_j$ on $\pi(s,v)$ and $t_j$ is above $t_i$ on $D_i$. We first claim that in a such a case $D_i$ and $D_j$ are neither rev-interleaved nor $(x,y)$-interleaved.
We prove this by contradiction. Let $w$ be the first point on $D_i$ that is common with $D_j$.
Since $t_j \in D_i \cap D_j$, it holds that $t_j \in D_i[w,y_1]$. Since $t_i$ appears after $t_j$ on $D_i$, it also holds that $t_i \in D_i[w,y_1]$. But if $D_i$ and $D_j$ are rev-interleaved or $(x,y)$-interleaved, $D_i[w,y_1]$ is part of the $D_i$-excluded region, leading to contradiction by Cl. \ref{cl:depen_detbad}.

\par We next claim that $t_i \in D_j$.
To see this, assume towards contradiction otherwise. First, observe that in such a case, $D_i$ and $D_j$ are also not $y$-interleaved (as otherwise $D_i[w,y_i]=D_j[w,y_j]$ and since $t_i \in D_i[w,y_i]$, it holds that also $t_i \in D_j$). Hence by Cl. \ref{cl:summ_depend}(a), $D_i$ and $D_j$ must be fw-interleaved. Since $t_i \in D_i \setminus D_j$, we end with contradiction by Cl. \ref{cl:depen_detbad}.

Hence, we have that both second failing edges are common to the two detours, i.e., $t_i,t_j \in D_i \cap D_j$.
Let $Q=P_{s,v,\{e_j,t_i\}}$ be the $s-v$ replacement path concerning the pair $e_j$ and $t_i \in D_j$. Note that $Q$ is not necessarily a new-ending path. We have the following.
\begin{claim}
\label{cl:eithercl}
$t_j \in Q$.
\end{claim}
\Proof
Assume towards contradiction that $t_j \notin Q$.
Since $P_j$ is a new-ending path, it diverges from $D_j$ before the failing edge $t_j$ and hence also above the failing edge $t_i$, concluding that $t_i \notin P_j$. Combining with the contradictory assumption, it holds that $e_j, t_i,t_j \notin Q,P_j$. By the optimality of $Q$ and $P_j$, $|Q|=|P_j|$.
By the ordering of Alg. $\ConstPath$, $Q$ was selected \emph{before} $P_j$, since $t_i$ is strictly below $t_j$. Hence, we end with contradiction to the construction of $P_j$ by Alg. Algorithm $\ConstPath$, as the new edge $\LastE(P_j)$ could have been avoided.
\QED
We therefore have that the failing edge $t_j=(q_1,q_2)$ is common with both of the replacement paths $Q$ and $P_i$, i.e., $t_j \in Q \cap P_i$.
We proceed by showing that this implies the existence of two $q_2-v$ shortest paths in $G \setminus \{e_i,e_j,t_i\}$, $Z_1=P_i[q_2,v]$ and $Z_2=Q[q_2,v]$.

To prove this, it remains to show that $e_i \notin Z_2$ (we have already shown that $e_j \notin P_i$). By Claim \ref{cl:unique_nnew}, the replacement path $Q$ has a unique divergence point $b$ from $\pi(s,v)$. Since $q_2$ is an endpoint of an edge on the detour $D_j$, it implies that $q_2$ appears on $Q$ strictly after it diverges from $\pi(s,v)$. Let $b'$ be the first point on $Q[q_2,v]$ that is common with $\pi(s,v)$.
By the uniqueness of the divergence point $b$, the point $b'$ is not a divergence point, and hence $Q[b',v]=\pi(b',v)$.
Hence, assuming that $e_i \in Z_2$ it holds that $e_i \in Q[b',v]$, but as $e_j$ is below $e_i$ on $\pi(s,v)$, we get that $e_j \in Q$, contradiction to the fact that $Q  \in SP(s,v, G \setminus \{e_j,t_i\})$.
Hence, by the optimality of these subpaths, $|Z_1|=|Z_2|$. Finally, note that by the ordering of the construction of Alg. $\ConstPath$, the pair edge $(e_j,t_i)$ was considered before $(e_i,t_i)$, as $e_j$ is below $e_i$. Hence, $P_i$ was constructed after $Q$. Contradiction to the selection of $P_j$ by Alg. $\ConstPath$ (as its last new edge could have been avoided). The lemma follows.
\QED
We then have the following.
\begin{lemma}
\label{lem:ordering}
$|P_1| >  \ldots >|P_\ell|$
(or alternatively, if $(b_i,c_i)<(b_j,c_j)$ then $|P_i|>|P_j|$).
\end{lemma}
\Proof
Assume towards contradiction that there exist two paths $P_i, P_j \in \overrightarrow{\mathcal{P}}_{indep}$ such that $i<j$ and $|P_i| \leq |P_j|$. First, consider the case where $b_i=b_j$. Let $D_i=D(P_i)$ and $D_j=D(P_j)$. Let $D_k \in \{D_i,D_j\}$ be the detour whose $y$-value is deeper on $\pi(s,v)$. By Cl. \ref{cl:allpoint_ondetour}, $c_i,c_j \in D_k$. In addition, by the ordering, $c_i$ appears on $D_k$ \emph{before} $c_j$. We now claim that $F_1(P_j), F_2(P_j) \notin P_i$. By the uniqueness of the $\pi$-divergence point $b_i=x_i$, it holds that $F_1(P_j)$ (that appears below $x_i$ on $\pi(s,v)$) is not in $P_i$.
Next, assume towards contradiction that $F_2(P_j) \in P_i$. Since $P_i$ does not interfere with $P_j$, it holds that $F_2(P_j) \in D_i \cap D_j$. Let $F_2(P_j) =(q_1,q_2)$. By Cl. \ref{cl:jointdet}, $D_i[x_i,q_2]=D_j[x_j,q_2]$.
Since $c_i$ appears on $D_k$ above $c_j$, it holds that it is also above $F_2(P_j)$, leading to contradiction as
$P_i[c_i,v]$ is edge disjoint with $D_i$.
Hence, we have that $F_1(P_j),F_2(P_j) \notin P_i$. Finally, since we assume that $|P_i|\leq |P_j|$, we end with contradiction to the selection of $P_j$ by Alg. $\ConstPath$, which selects the replacement path whose $\sf D$-divergence point from $D_j$ is as closest to $x_j$ as possible (and $c_i$ is strictly closer).
\par Next, consider the complementary case where $b_i<b_j$.  By Cl. \ref{cl:nofailedg}, $F_1(P_j),F_2(P_j)$ are not on $P_i$.
Since $|P_i| \leq |P_j|$, we end with contradiction to the construction of $P_j$ since Alg. $\ConstPath$ selects the replacement path whose $\pi$-divergence point from $\pi(s,v)$ is as closest to $s$ as possible. The lemma holds.
\QED

Towards bounding the number of independent paths, we classify them into $n'$ classes $\mathcal{P}_1, \ldots, \mathcal{P}_{n'}$ for some parameter $n'$ to be revealed later. These classes cover all the independent paths $\mathcal{P}_{indep}$. In each $\mathcal{P}_{i}$ class, there is a special path representor $P^*_i$, that guaranteed to be sufficiently long.
We now describe the path classification in details.
Initially, set $\overrightarrow{\mathcal{P}}^1_{indep}=\overrightarrow{\mathcal{P}}_{indep}$, the $(b,c)$ ordered set $\mathcal{P}_{indep}$.
At step $\tau \geq 1$, we are given a $(b,c)$ ordered set $\overrightarrow{\mathcal{P}}^\tau_{indep}$ consisting of the remaining independent paths that have not been yet assigned to any of the classes. Let $P^*_\tau$ be the first path in this increasing  $(b,c)$ ordering $\overrightarrow{\mathcal{P}}^\tau_{indep}$. The class $\mathcal{P}_{\tau}$ consists of the representor $P^*_\tau$ and the paths $P$ in $\mathcal{P}^\tau_{indep}$ that satisfy at least one of the two conditions:
\begin{description}
\label{desc:indep}
\item{(O1)}
$c(P) \in D(P^*_\tau)$.
\item{(O2)}
$D(P)$ and $D(P^*_\tau)$ are $x$-interleaved.
\end{description}
I.e., if $P$ satisfies (O1), (O2) or both, it is added to the class $\mathcal{P}_{\tau}$.
This process continues, until all independent paths are assigned to some class. Let $n'$ be the last time step of the classification process, where
$\mathcal{P}_{indep}=\bigcup_{i=1}^{n'} \mathcal{P}_i$.
For every $\tau \in \{1,\ldots,n'\}$, let $L_\tau=|\mathcal{P}_{\tau}|$ be the number of paths in the $\tau^{th}$ class.
We now establish several useful properties about these classes and then use it to bound the total cardinality of the independent set $\mathcal{P}_{indep}$.

\begin{claim}
\label{cl:oder}
For every $\tau$ and for every $j \in \{\tau, \ldots,n'\}$, it holds that
$|P^*_\tau|>|P'|$ for every $P'  \in \mathcal{P}_j \setminus \{P^*_\tau\}$.
\end{claim}
\Proof
Note that $\mathcal{P}^\tau_{indep}=\bigcup_{j=\tau}^{n'}\mathcal{P}_k$.
Since $P^*_\tau$ is the first path in the $(b,c)$ ordered set $\overrightarrow{\mathcal{P}}^\tau_{indep}$, the claim holds by Lemma \ref{lem:ordering}.
\QED
For every $i \in \{1,\ldots, n'\}$, we next define two subsets of vertices $V_1(i)$ and $V_2(i)$ appearing on the paths of the class $\mathcal{P}_i$. The first set $V_1(i)$ consists of the suffixes $P_k[c_k,v]\setminus \{v\}$ for every $P_k \in \mathcal{P}_i\setminus \{P^*_i\}$.
The second set $V_2(i)$ consists of the suffix of the representor $P^*_{i}[b_i,v]$ where $b_i$ is its $\pi$-divergence point. Formally, let
\begin{equation}
\label{eq:v12sets}
V_1(i)=\{P_k[c_k, v]\setminus \{v\} ~\mid~ P_k \in \mathcal{P}_i \setminus \{P^*_i\}\} \mbox{~and~} V_2(i)=P^*_i[b_i,v] \setminus \{v\}.
\end{equation}
Our goal is to show that the union of these sets, namely, $\mathcal{V}_{k}=\bigcup_{i=1}^{n'} V_k(i)$ for $k \in \{1,2\}$, is sufficiently large.
We first consider the sets $V_1(i)$.
\begin{claim}
\label{cl:v1i}
For every $i \in \{1,\ldots,n'\}$:\\
(a) $V_1(i) \cap V_1(j)=\emptyset$ for every $j \neq i$  and \\
(b) $|V_1(i)|=\Omega(L_i^2)$.
\end{claim}
\Proof
Part (a) follows immediately by Obs. \ref{obs:dist_indep}.
Consider part (b). We classify the $L_i$ paths in $\mathcal{P}_i$ into two sets depending on the condition they satisfy when joining the class. Let $\mathcal{P}^1_i$ be the set of paths in $\mathcal{P}_i$ that satisfy (O1) and let $\mathcal{P}^2_i$ be the complementary set of paths.

First, assume that the majority of the $L_i$ paths belongs to the first class $\mathcal{P}^1_i$ (i.e., satisfy condition (O1)). Hence, $|\mathcal{P}^1_i|\geq L_i/2$. Note that the $\sf D$-divergence point of each path in this class appears on $D_i=D(P^*_i)$.
By the uniqueness of the $\sf D$-divergence point (see  Lemma \ref{cl:rp_prop1}), the paths in $\mathcal{P}^1_i=\{P_1, \ldots, P_k\}$ can be ordered in increasing distance from $y(D_i)$. Hence, $|P_j[c_j,v]\setminus \{v\}| \geq j-1$ for every $j \in \{1,\ldots, k\}$.
By applying Obs. \ref{obs:dist_indep} again, we get that
$|V_1(i)|\geq |\bigcup_{j=1}^k P_j[c_j,v]\setminus\{v\}|=\sum_{j=1}^k |P_j[c_j,v]\setminus \{v\}| \geq (k-1)^2/2$. Since $k \geq L_i/2$, the claim holds.

\par Next, consider the complementary case, where the majority of the paths in this class are in $\mathcal{P}^2_i$, hence they all satisfied the condition (O2).  Note that for every $P,P' \in \mathcal{P}^2_i$, it holds that $D(P)$ and $D(P')$ are $x$-interleaved (since they are $x$-interleaved with $D(P^*_i)$).
Let $D^*_i$ be the detour of some path $P'$ in the set $\mathcal{P}^2_i$ whose $y$-value is the deepest on $\pi(s,v)$.
Since for every $P \in \mathcal{P}^2_i$, it holds that $D(P)$ and $D^*_i$ are $x$-interleaved, by Cl. \ref{cl:allpoint_ondetour}, it holds that the $\sf D$-divergence point $c(P)$ appears on $D^*_i$ for every $P \in \mathcal{P}^2_i$. By applying the uniqueness of the $\sf D$-divergence point (Cl. \ref{cl:rp_prop1}) and the disjointness of the segments $P[c(P),v]\setminus \{v\}$ (Obs. \ref{obs:dist_indep}), the argument follows the exact same line as for the $\mathcal{P}^1_i$ class. Part (b) holds.
%Finally, (c) follows by combining part (a) and (b).
\QED
We proceed by analyzing the sets $V_2(i)$.
\begin{claim}
\label{cl:boundvi2}
$|V_2(i)|\geq \left(\sum_{j=i}^{n'} L_j\right)-2$.
\end{claim}
\Proof
Recall that $\overrightarrow{\mathcal{P}}^i_{indep}=\bigcup_{j=i}^{n'} \mathcal{P}_j$.
Since $P^*_i$ was the first path in the increasing
$(b,c)$-ordering of $\overrightarrow{\mathcal{P}}^i_{indep}$, it holds that $b(P^*_i)\leq b(P)$ for every $P \in \overrightarrow{\mathcal{P}}^i_{indep}$.

Since every path $P \in \widehat{\mathcal{P}}_{i}$ is not in $\mathcal{P}_{nodet}$, we have that $P[s,b(P)]=\pi(s,b(P)]$ and hence letting $b^*=b(P^*_i)$, we get that $P^*_i[s,b^*]=P[s,b^*]$ for every $P \in \widehat{\mathcal{P}}_{i}$.

By Lemma \ref{lem:ordering}, it holds that the paths of the $(b,c)$ ordered set $\overrightarrow{\mathcal{P}}^i_{indep}$ correspond to a strictly monotone increasing in lengths sequence $\overrightarrow{\mathcal{P}}^i_{indep}=\{P_1=P^*_i, P_2, \ldots, P_{L'}\}$ of $L'=\sum_{j=i}^{n'} L_j$ paths such that $|P_1|> |P_2| >\ldots>|P'_{L'}|$.
Since all these paths share the prefix $P^*_i[s,b^*]$, it also holds that
$|P'_1|> |P'_2| >\ldots >|P'_{L'}|$  where $P'_{k}=P_k[b^*,v]$ for every $k \in \{1,\ldots,L'\}$.
Concluding that $|V_2(i)|=|P^*_i[b^*,v] \setminus \{v\}|=|P'_1|-1\geq L'-2$. The claim follows.
\QED
The next claim is useful for bounding the cardinality of the $V_2(i)$ sets.
\begin{claim}
\label{cl:piindep_det}
Let $P^*_{i_1},P^*_{i_2},P^*_{i_3}$ such that $i_1<i_2<i_3$. Let $b_{j}=b(P^*_{i_j})$, $c_{j}=c(P^*_{i_j})$  and $D_{j}=D(P^*_{i_j})$ for $j \in \{1,2,3\}$, then $\bigcap_{k=1}^3 D_k[b_k,c_k]=\emptyset$.
\end{claim}
\Proof
Assume towards contradiction that there exists a vertex $w \in \bigcap_{k=1}^3 D_k[b_k,c_k]$.
Let $x_j=x(D_j)$ and $y_j=y(D_j)$ for $j \in \{1,2,3\}$. Since $P^*_{i_j} \notin \mathcal{P}_{nodet}$, $x_j=b_j$ for $j \in \{1,2,3\}$.
We first claim that $x_1<x_2<x_3$.
Since $P^*_2$ does not satisfy condition (O2) for the class $\mathcal{P}_{i_1}$, it holds that $D_1$ and $D_2$ are not $x$-interleaved, i.e, $x_1\neq x_2$.
Combining this with the fact that $P^*_{i_1}$ precedes $P^*_{i_2}$ in the increasing $(b,c)$ ordering, we get that $x_1 < x_2$. In the same manner, we also have that $x_2<x_3$.

We next claim that $y_1<y_2<y_3$.
Consider first $y_1$ and $y_2$. There are two alternative cases.
Case (a): $y_1>y_2$. Then $D_2$ is nested in $D_1$, since these detours are dependent, we end with  contradiction to Cl. \ref{cl:depen_detdisjoint}.\\
Case (b): $y_1=y_2$. Since $w \in D_1[x_1,c_1]\cap D_2[x_2,c_2]$, it holds that $c_2 \in D_2[w,y_2]=D_1[w,y_1]$, where the last equality holds by Cl. \ref{cl:jointdet}. We therefore have that $P^*_{i_2}$ satisfies condition (O1) for the class $\mathcal{P}_{i_1}$. Leading to contradiction to the selection of $P^*_{i_2}$ by the classification procedure.
Hence, we conclude that $y_1<y_2$. By applying the same argument for $P^*_{i_2}$ and $P^*_{i_3}$, it also holds that $y_2<y_3$.

Since $D_1,D_2$ are dependent and interleaved such that $y_1<y_2$, by Cl. \ref{cl:det_nested}(b), $e_2 \in \pi(y_1,y_2)$. I.e., $e_2$ appears below $y_1$ on $\pi(s,v)$. In the same manner, since $D_2$ and $D_3$ are dependent and interleaved such that $x_2<x_3$, by Cl. \ref{cl:det_nested}(a), $e_2 \in \pi(x_2,x_3)$, i.e., $e_2$ appears above $x_3$ on $\pi(s,v)$. We now claim that $y_1\geq x_3$ and hence establishing the claim as $\pi(y_1,y_2)\cap\pi(x_2,x_3)=\emptyset$.
To see why $y_1\geq x_3$, assume towards contradiction that $y_1<x_3$, then $D_1$ and $D_3$ are non-nested and by Cl. \ref{cl:disjindep}, they are independent, in contradiction to the existence of $w$ in the intersection. The claim holds.
\QED
We are now ready to bound the number of independent paths.
\begin{claim}
\label{cor:sum}
(a) $|\mathcal{V}_{1} \cup \mathcal{V}_{2}|= \Omega\left(\sum_{i=1}^{n'}\left( L_i^2+ i \cdot L_i \right) \right)$.\\
(b) $N_{indep}=O(n^{2/3})$.
\end{claim}
\Proof
Consider (a). Our strategy is follows. We consider some vertex $u \in \mathcal{V}_{1} \cup \mathcal{V}_{2}$ and bound the number of sets $V_1(j),V_2(j)$, $i \in \{1,\ldots,n'\}$ in which it appears. Let $\mathcal{J}^1(u)=\{j ~\mid~ u \in V_1(j)\}$ and $\mathcal{J}^2(u)=\{j ~\mid~ u \in V_2(j)\}$ be the set of indices corresponding to the $V_1(j),V_2(j)$ sets in which $u$ appears (respectively).

First observe that by the disjointness of the $V_1(j)$ sets (established in Cl. \ref{cl:v1i}(a)), it holds that $u$ can appear in at most one set $V_1(j)$, hence $|\mathcal{J}^1(u)|\leq 1$.

We next claim that $u$ can appear in at most two additional sets $V_2(i_1), V_2(i_2)$ for some $i_1,i_2 \in \{1, \ldots, n'\}$. I.e., $|\mathcal{J}^2(u)|\leq 2$.
To see this, assume towards contradiction that $u$ appears in three sets $V_2(i_1), V_2(i_2), V_2(i_3)$.
Without loss of generality, assume that $i_1<i_2<i_3$.
Let $D_k=D(P^*_{i_k})$,  $b_k=b(P^*_{i_k})$ and $c_k=c(P^*_{i_k})$ for $k \in \{1,2,3\}$.
%Recall that since $P^*_{i_k} \notin \mathcal{P}_{nodet}$, it holds that $b_k=x(D_k)$ for $k =\{1,2,3\}$. In addition, recall that $V_2(i_k)=P^*_{i_k}[b_k,v]\setminus \{v\}$.
%
Since $u \in \bigcap_{k=1}^3 P^*_{i_k}[b_k,v]$, by the disjointness of the suffixes $P^*_{i_k}[c_k,v]$ (see Obs. \ref{obs:dist_indep}), it holds that $u$ must appear in the detour segment of these paths, namely, that
$u \in \bigcap_{k=1}^3 P^*_{i_k}[b_k,c_k]=\bigcap_{k=1}^3  D_k[b_k,c_k]$, where the last equality holds by Cl. \ref{cl:rp_prop1}(3.1), leading to contradiction by Cl. \ref{cl:piindep_det}.
We therefore have the following.
\begin{eqnarray*}
3 \cdot |\mathcal{V}_{1} \cup \mathcal{V}_{2}| &\geq&
\sum_{u \in  \mathcal{V}_{1} \cup \mathcal{V}_{2}}\left( |\mathcal{J}^1(u)|+|\mathcal{J}^2(u)|\right)
=\sum_{i=1}^{n'} \sum_{k=1}^2 |V_k(i)|~.
\end{eqnarray*}
Part (a)  follows by combining this with Cl. \ref{cl:v1i}(b) and Cl. \ref{cl:boundvi2}.
Finally, since $|\mathcal{V}_{1} \cup \mathcal{V}_{2}|\leq n$, by using Lagrange multiplier, we get that $N_{indep}= \sum_{i=1}^{n'} L_i= O(n^{2/3})$ as required.
\QED

\subsection{Bounding the number of $\pi$-interfering paths $\mathcal{I}_{\pi}$ }
In this section, we bound the number of interfering paths of type $\pi$. Recall that
$$\mathcal{I}_{\pi}=\{P \in \mathcal{P}_v~\mid~ \mbox{if~} F_2(P') \in P\setminus D(P) \mbox{~for~} P' \in \mathcal{P}_v \mbox{~then~} F_1(P) \in
\pi(y(D(P')),v)\}.$$
Let $P_1 \in \mathcal{I}_{\pi}$ be such that $P_1$ $\pi$-interferes with $P_2$. Let $e_1=F_1(P_1), e_2=F_1(P_2)$ and let $x_1, y_1$ (resp., $x_2,y_2$) be the first and last point of the detour $D(P_1)$ (resp., $D(P_2)$). \par To bound the number of $\pi$-interfering paths, we show that an interference of type $\pi$ induces a strict detour configuration which implies that two paths $P_1,P_2 \in \mathcal{I}_{\pi}$ that are $\pi$-interfered by a third path $P_3$, are \emph{independent}. This key observation enables us to treat this class as a nearly independent set of paths. In particular, equipped with this observation, only minor modifications are required to employ the quantitative analysis of Sec. \ref{sec:indeppaths} to the setting of $\pi$-interference.
\begin{observation}
\label{obs:char_intera}
If $P_1$ $\pi$-interferes with $P_2$, then:
(a) $e_1 \in \pi(y_2,y_1)$ and \\
(b) $e_2\in \pi(s, x_1)$. Hence, in particular $x_2<x_1$.
\end{observation}
\Proof
Consider (a). By the definition, $e_1 \in \pi(y_2,v)$ and $e_1 \in \pi(x_1,y_1)$ hence we conclude that $e_1 \in \pi(y_2,y_1)$. The claim holds. We now turn to consider part (b).
Assume towards contradiction that $e_2 \notin \pi(s,x_1)$.
Let $F_2(P_2)=(q_1,q_2)$.
We distinguish between two cases.
\dnsparagraph{Case (1): $D_1$ and $D_2$ are independent}
There are two $s-q_2$ paths in $G \setminus \{e_1,e_2\}$, namely, $Q_1=\pi(s,x_1) \circ D_1[x_1,c_1]\circ P_1[c_1,q_2]$ and $Q_2=\pi(s,x_2) \circ D_2[x_2,q_2]$.
By the optimality of $Q_1$, it holds that $Q_1\in SP(s, q_2, G \setminus F(P_1))$. Since $e_2 \notin \pi(s,x_1)$, it also holds that $e_2 \notin Q_1$. In addition, by the optimality of $Q_2$, it holds that $Q_2 \in SP(s,q_2, G \setminus \{e_2\})$. By part (a), $e_1 \in \pi(y_2,y_1)$ and hence $e_1 \notin Q_2$. Since $D_1$ and $D_2$ are independent, we also have that $F_2(P_1) \notin Q_2$. Hence, overall we have that $Q_1, Q_2 \in  SP(s, q_2, G \setminus \left(F(P_1) \cup \{e_2\}\right))$, concluding that $|Q_1|=|Q_2|$.
This case is further divided into two subcases.\\
Case (1.1): $x_1 \neq x_2$. If $x_1<x_2$ (resp., $x_2<x_1$) , then we end with contradiction to the selection of $P_{s,v,\{e_2\}}$ (resp., $P_{s,v, F(P_1)}$) by Alg. $\ConstPath$, since there exists an alternative shortest-path whose $\pi$-divergence point from $\pi(s,v)$ is strictly closer to $s$. \\
Case (1.2): $x_1=x_2$. Note that since $P_1 \notin \mathcal{P}_{nodet}$, it holds that $c_1 \neq x_1$ and specifically, $c_1$ is strictly inside $D_1$.
Since $|Q_1|=|Q_2|$, we end with contradiction to the selection of $P_1$ by Alg. $\ConstPath$, since there exists an alternative shortest-path, namely, $Q_2 \circ P_1[q_2,v]$ whose $\sf D$-divergence point from $D_1$ is $x_1$, i.e., strictly above $c_1$ on $D_1$.
\dnsparagraph{Case (2): $D_1$ and $D_2$ are dependent}
Case (2.1): $x_1=x_2$. Let $w$ be the last common vertex of $D_1$ and $D_2$. Since $F_2(P_2) \in D_2 \setminus D_1$, so $F_2(P_2) \in D_2[w,y_2]$, leading to contradiction by Cl. \ref{cl:depen_detbad}.
Case (2.2): $x_1<x_2$. Then, since $y_2<y_1$ by part (a), $D_2$ is nested in $D_1$, leading to contradiction by Cl. \ref{cl:depen_detdisjoint}. Case (2.3): $x_1>x_2$. By claim Cl. \ref{cl:det_nested}(a), we have that $e_2 \in \pi(x_2,x_1)$, contradiction to the fact that $e_2 \notin \pi(s,x_1)$.
The claim follows.
\QED

By Obs. \ref{obs:dist_indep}, we have the following.
\begin{observation}
\label{obs:inter_explain}
If $P'_i=P_i[c_i,v]$ and $P'_j[c_i,v]$ intersect, then $P_i$ interferes with $P_j$ or vice-versa, namely either $F_2(P_i) \in P'_j$ or $F_2(P_j) \in P'_i$.
\end{observation}
Recall that we consider only $(\pi, \sf D)$-replacement paths for which $F_2(P_i) \in D(P_i)$ and hence $F_2(P_i) \notin \pi(s,v)$. We now provide the key lemma which enables us to bound from above the set of $\pi$-interfering path. It states that the suffix $P'_{j}=P_j[c_j,v]$ of two paths $P_{j_1}$ and $P_{j_2}$ that are $\pi$-interfered by the same path $P_i \in \mathcal{I}_{\pi}$, are \emph{disjoint}.
\begin{lemma}
\label{lem:disjoint_neig}
Let $P_{j_1}, P_{j_2} \in \mathcal{I}_{\pi}$ be two paths such that there exits $P_i\in \mathcal{I}_{\pi}$ that interferes with   $P_{j_1}, P_{j_2}$. Then, $P_{j_1}$ and $P_{j_2}$ are independent and hence $P'_{j_1}=P_{j_1}[c_{j_1},v]$ and $P'_{j_2}=P_{j_2}[c_{j_2},v]$ are disjoint (besides the common vertex $v$).
\end{lemma}
\Proof
Note that by Lemma \ref{obs:inter_explain}, if
$P_{j_1}$ and $P_{j_2}$ are independent then the segments $P'_{j_1} \setminus\{v\}$ and $P'_{j_2} \setminus\{v\}$ are disjoint.
Assume towards contradiction that $P_{j_1}$ and $P_{j_2}$ are not independent and without loss of generality assume that $P_{j_1}$ interferes with $P_{j_2}$. Since $P_{j_1} \in \mathcal{I}_{\pi}$, it must hold that $P_{j_1}$ $\pi$-interferes with $P_{j_2}$.
For ease of notation, let $P_1=P_i, P_2=P_{j_1}$ and $P_{3}=P_{j_2}$. Let $e_i=F_1(P_i)$ and $x_i,y_i$ denote the first (resp., last) point of the detour $D_i=D(P_i)$ for $i=\{1,2,3\}$.
By Obs. \ref{obs:char_intera}(1), since $P_2$ $\pi$-interferes with $P_3$, it must hold that $e_2 \in \pi(y_3,y_2)$.
On the other hand, since $P_1$ $\pi$-interferes with $P_2$, by Obs. \ref{obs:char_intera}(2), $e_2 \in \pi(x_2,x_1)$.
We now show that these two requirements contradict each other by showing that $\pi(y_3,y_2) \cap \pi(x_2,x_1)=\emptyset$. Specifically, we show that $x_1$ is \emph{not} below $y_3$ on $\pi(s,v)$, i.e., $x_1\leq y_3$.

Assume towards contradiction
that $x_1>y_3$, i.e., $D_1$ and $D_3$ are non-nested. By Cl. \ref{cl:disjindep}, we then have that $D_1$ and $D_3$ are also independent. Let $F_2(P_3)=(q_1,q_2)$, since $P_1$ interferes with $P_3$, it visits the failing edge $F_2(P_3)$, i.e., $F_2(P_3) \in P_1$. We now present two $y_3-q_2$ shortest-paths in $G \setminus \{e_1, e_3,F_2(P_1)\}$, namely, $Z_1=P_1[y_3,q_2]=\pi(y_3,x_1) \circ D_1[x_1,c_1] \circ P_1[c_1, q_2]$ and $Z_2=D_3[y_3,q_2]$.
Since $e_3$ appears above $y_3$ on $\pi(s,v)$ and $P_1[x_1,v]$ is edge disjoint with $\pi(s,v)$, we get that $e_3 \notin Z_1$. In addition, since $D_3$ and $D_1$ are independent and $Z_2$ is edge disjoint with $\pi(s,v)$, we have that $F_1(P_1), F_2(P_1) \notin Z_2$. By optimality of the replacement paths $P_1$ and $P_{s,v,\{e_3\}}$, we have that $|Z_1|=|Z_2|$. Hence, we end with contradiction to the selection of $P_1$ by Alg. $\ConstPath$, as there exists an alternative shortest path $\pi(s,y_3) \circ Z_2 \circ P_1[q_2,v]$ in $G \setminus F(P_1)$, whose $\pi$ divergence-point is strictly above $x_1$. The claim follows.
\QED
Using Lemma \ref{lem:disjoint_neig}, the analysis of Sec. \ref{sec:indeppaths} extends to the setting of $\pi$-interference with only minor modifications.
We now briefly sketch the main steps of the analysis and highlight the require modifications.
%First, note that by Obs. \ref{obs:char_intera}, for every pair of paths $P$ and $P'=I(P)$, it holds that $D(P)$ and $D(P')$ are \emph{not} $x$-interleaved.
%We then induce a $(b,c)$-ordering $\overrightarrow{\mathcal{I}}_{\pi}=\{P_1, \ldots, P_{\ell}\}$ on the set of $\pi$-interfered paths $\mathcal{I}_{\pi}$.
%
%We hence have that Lemma \ref{lem:ordering} immediately hold for this case as well.
\par Let $\overrightarrow{\mathcal{I}}_{\pi}=\{P_1,\ldots,P_{\ell}\}$ be the increasing $(b,c)$-ordered set of $\mathcal{I}_{\pi}$ paths as in Sec. \ref{sec:indeppaths}, where $\ell=|\mathcal{I}_{\pi}|$. Hence, $b_1 \leq b_2 \leq \ldots \leq b_{\ell}$. Since $P_i \notin \mathcal{P}_{nodet}$, it holds that $b_i=x(D(P_i))$.
Note that if a path $P_i$ $\pi$-interferes with $P_j$
then by Obs. \ref{obs:char_intera}, $b_i$ is necessarily below $b_j$ on $\pi(s,v)$. Also note that since $P_i$ is in $\mathcal{I}_{\pi}$, if $P_i$ interferes with $P_j$ then necessarily it is an interference of type $\pi$. We have the following.
\begin{observation}
\label{obs:ordernointer}
$P_i$ does not interferes with $P_j$ for every $j>i$.
\end{observation}
The last observation implies that the proof of Lemma \ref{lem:ordering} established for the case of independent paths, extends as is to the case of $\pi$-interfering sets $\mathcal{I}_{\pi}$.
\begin{lemma}
\label{lem:orderingpi}
$|P_1| >  \ldots  >|P_\ell|$
(or alternatively, if $(b_i,c_i)<(b_j,c_j)$ then $|P_i|>|P_j|$).
\end{lemma}
Towards bounding the number of independent paths, we classify them into $n'$ classes $\mathcal{P}_1, \ldots, \mathcal{P}_{n'}$ for some parameter $n'$ to be revealed later. These classes cover all the independent paths $\mathcal{I}_{\pi}$. In each class, $\mathcal{P}_{i}$, there is a special representor $P^*_i$. The classification procedure is identical to that of Sec. \ref{sec:indeppaths}.
For every $i \in \{1,\ldots, n'\}$, we define two subsets of disjoint vertices $V_1(i)$ and $V_2(i)$ according to Eq. (\ref{eq:v12sets}).
\par The next auxiliary claim extends Cl.  \ref{cl:piindep_det} to the case of $\pi$-interference.
Let $P^*_{i_1},P^*_{i_2},P^*_{i_3}$ such that $i_1<i_2<i_3$. Let $b_{j}=b(P^*_{i_j})$, $c_{j}=c(P^*_{i_j})$  and $D_{j}=D(P^*_{i_j})$ for $j \in \{1,2,3\}$.
\begin{claim}
\label{cl:piindep_detpi}
For every triple of path $P^*_1,P^*_2,P^*_3$, we have that:\\
(1) $\bigcap_{k=1}^3 D_k[b_k,c_k]=\emptyset$.\\
(2) $\bigcap_{k=1}^3 P_k[c_k,v]=\emptyset$.
\end{claim}
\Proof
Part (1) follows immediately by Cl. \ref{cl:piindep_det} (the proof of this claim did not use the fact that given paths are independent). Consider Part (2). Assume towards contradiction that there exists a vertex $u \in \bigcap_{k=1}^3 P_k[c_k,v]$.

We now claim that the existence of the common vertex $u$, implies that $P_3$ $\pi$-interferes with both $P_1$ and $P_2$. Assume towards contradiction, that $P_3$ does not $\pi$-interfere with $P_j$ for $j \in \{1,2\}$.
By Obs. \ref{obs:ordernointer}, $P_j$ does not interfere with $P_3$. Hence, $P_3$ and $P_j$ are independent. By Obs. \ref{obs:dist_indep}, we then have that $P_3[c_3,v]$ and $P_j[c_j,v]$ are disjoint,  leading to contradiction that $u$ is a common vertex in the intersection.
Hence, $P_3$ interferes with $P_1$ and $P_2$.
Since $P_3 \in \mathcal{I}_{\pi}$, it holds that this interference is of type $\pi$. By
Lemma \ref{lem:disjoint_neig}, we get that $P_1[c_1,v]$ and $P_2[c_2,v]$ are disjoint, leading to contradiction again to the existence of the common vertex $u$. The claim follows.
\QED

Using Cl. \ref{cl:piindep_detpi}, we are now ready to bound the cardinality of the vertex sets $V_1(i)$ and $V_2(i)$.

\begin{claim}
\label{cl:v1ipi}
$|V_1(i)|=\Omega(L_i^2)$.
\end{claim}
\Proof
We classify the $L_i$ paths in $\mathcal{P}_i$ into two sets depending on the condition they satisfy when joining the class. Let $\mathcal{P}^1_i$ be the set of paths in $\mathcal{P}_i$ that satisfy (O1) and let $\mathcal{P}^2_i$ be the complementary set of paths.

First, assume that the majority of the $L_i$ paths belongs to the first class $\mathcal{P}^1_i$ (i.e., satisfy condition (O1)). Hence, $|\mathcal{P}^1_i|\geq L_i/2$. Note that the $\sf D$-divergence point of each path in this class appears on $D_i=D(P^*_i)$.
By the uniqueness of the $\sf D$-divergence point (see  Lemma \ref{cl:rp_prop1}), the paths in $\mathcal{P}^1_i=\{P_1, \ldots, P_k\}$ can be ordered in increasing distance from $y(D_i)$. Hence, $|P_j[c_j,v]\setminus \{v\}| \geq j-1$ for every $j \in \{1,\ldots, k\}$.
By Cl. \ref{cl:piindep_detpi}(2),
every vertex $u \in V_1(i)$ is counted at most twice by
$P'_j$ and $P'_k$ where $P'_k=I(P'_j)$ or vice-versa.
Hence,
$|V_1(i)|=|\bigcup_{j=1}^k P_j[c_j,v]\setminus\{v\}|\geq \frac{1}{2} \cdot \sum_{j=1}^k |P_j[c_j,v]\setminus \{v\}| \geq (k-1)^2/4$. Since $k \geq L_i/2$, the claim holds.
The  complementary case, where the majority of the paths in this class are in $\mathcal{P}^2_i$, holds analogously.
\QED
Let $\mathcal{V}_{k}=\bigcup_{i=1}^{n'} V_k(i)$ for $k \in \{1,2\}$. We next bound the number of independent paths by showing that $|\mathcal{V}_{k}|$ is large for $k \in \{1,2\}$.
\begin{claim}
\label{cor:sumpi}
(a) $|\mathcal{V}_{1} \cup \mathcal{V}_{2}|= \Omega\left(\sum_{i=1}^{n'}\left( L_i^2+ i \cdot L_i \right) \right)$.\\
(b) $|\mathcal{I}_{\pi}|=O(n^{2/3})$.
\end{claim}
\Proof
Consider (a). Our strategy is follows. We consider some vertex $u \in \mathcal{V}_{1} \cup \mathcal{V}_{2}$ and bound the number of sets $V_1(i),V_2(i)$, $i \in \{1,\ldots,n'\}$ in which it appears. By Cl. \ref{cl:piindep_detpi}(a) and (b), a vertex $u$ may appear in at most two $P_i[b_i,c_i]$ segments and in at most two $P_i[c_i,v]$ segments, hence overall it may be re-counted four times by the sets of $V_1(i),V_2(i)$.
The claim follows now the exact same line as the proof of Cl. \ref{cor:sum}.
\QED
\par We are now ready to complete the proof for Thm. \ref{thm:upper}.
\Proof[Thm. \ref{thm:upper}]
By Obs. \ref{lem:single_fault}, Lemma \ref{lem:not_inters}, Cor. \ref{cor:bound_no_touch},
Cor. \ref{cor:ninterb}, Cl. \ref{cor:sum}(b) and Cl. \ref{cor:sumpi}(b), we have that for every $v$, the number of new-ending paths in $\mathcal{P}_v$ is $O(n^{2/3})$.
Overall, we have the following.
$|E(H)|=|T_0 \cup \bigcup_{v \in V} \bigcup_{P \in \mathcal{P}_v}\LastE(P)|=O(n^{5/3})$. The theorem follows.
\QED

%%%%%%%%%%%%%%%%%%%%%%%%%%%%%%%%%%%%%%%%%%%%%
\section{Lower bound for $f$-failure FT-BFS structure}
\label{sec:lb}
%%%%%%%%%%%%%%%%%%%%%%%%%%%%%%%%%%%%%%%%%%%%%
In this section, we consider a lower bound constructions for $\FTMBFS$ structures resilient to up to $f$-faults for general $f \geq 2$ and for every number of sources $\NSource$. These construction extends the construction of \cite{PPFTBFS13} for the single failure case.
\begin{theorem}
\label{thm:lowerbound_f}
For every $n\geq o(1)$ and $1 \geq f$, there exists an $n$-vertex graph $G^*_f(V, E)$
and a source set $S  \subseteq V$ such that any $f$-failure \FTMBFS\ structure with respect to $S$ has $\Omega(|S|^{1-1/(f+1)} \cdot n^{2-1/(f+1)})$ edges.
\end{theorem}
Note that for $f=\Omega(\log n)$ the claimed bound becomes trivial.
Hence we will assume that $f=o(\log n)$. We begin by showing the construction for the single source case and then extend it to the case multiple sources.
Our construction is based on the graph $G_f(d)=(V_f,E_f)$, defined inductively.
For $f=1$, $G_1(d)=(V_1, E_1)$ consists of three components:
(1) a set of vertices $U=\{u^1_1,\ldots,u^1_d\}$ connected by a path
$P_1=[u^1_1, \ldots, u^1_d]$,
(2) a set of terminal vertices $Z=\{z_1,\ldots,z_d\}$
(viewed by convention as ordered from left to right),
and
(3) a collection of $d$ vertex disjoint paths $Q^1_{i}$ of length
$|Q^1_i|=6+2 \cdot (d-i)$ connecting $u^1_i$ and $z_i$
for every $i \in \{1, \ldots, d\}$.
%Thus $|Q^1_1|> \ldots> |Q^1_d|$.
The vertex $\Root(G_1(d))=u^1_d$ is fixed as the root of $G_1(d)$, hence
the edges of the paths $Q^1_i$ are viewed as directed away from $u^1_i$,
and the terminal vertices of $Z$ are viewed as the \emph{leaves} of the graph,
denoted $\Leaf(G_1(d))=Z$.
%\commabs
See Fig. \ref{fig:lowerboundg1} for illustration.
%\commabsend

Overall, the vertex and edge sets of $G_1(d)$ are
$V_1=U \cup Z \cup \bigcup_{i=1}^d V(Q^1_i)$ and
$E_1=E(P_1) \cup \bigcup_{i=1}^d E(Q^1_i)$.

For ease of future analysis, we assign labels to the leaves
$z_i \in \Leaf(G_1(d))$.
Let $\LAB_f: \Leaf(G_f(d))  \to E(G_1(d))^f$.
The label of each leaf corresponds to a set of edge faults under which
the path from root to leaf is still maintained (this will be proved later on).
%Label the $Z$ vertices in the following manner
Specifically, $\LAB_1(z_i, G_1(d))=(u^1_i,u^1_{i+1})$ for $i \leq d-1$ and
$\LAB_e(z_i, G_1(d))=\emptyset$. In addition, define
$P(z_i,G_1(d)) = P_1[\Root(G_1(d)),u^1_i] \circ Q^1_i$
%$P(z_i,G_1(d))=\pi[\Root(G_1(d)),u^1_i] \circ Q^1_i$
to be the path from the root $u^1_1$ to the leaf $z_i$.

To complete the inductive construction, let us describe the construction
of the graph $G_{f}(d)=(V_{f}, E_{f})$, for $f\ge 2$,
given the graph $G_{f-1}(d)=(V_{f-1}, E_{f-1})$.
% and evaluate its measures.
The graph $G_{f}(d)=(V_{f}, E_{f})$ consists of the following components.
First, it contains a path $P=[u^f_1, \ldots, u^f_d]$, where
the node $\Root(G_{f}(d))=u^f_1$ is fixed to be the root.
In addition, it contains $d$ disjoint copies of the graph $G'=G_{f-1}(d)$,
denoted by $G'_1, \ldots, G'_d$
(viewed by convention as ordered from left to right),
where each $G'_i$ is connected to $u^f_i$ by a collection of $d$
vertex disjoint paths $Q^f_i$, for $i \in \{1, \ldots, d\}$,
connecting the vertices $u^f_i$ with $\Root(G'_i)$.
The length of $Q^f_i$ is $|Q^f_i|=(d-i)\cdot \depth(G_{f-1}(d))$.
The leaf set of the graph $G_{f}(d)$ is the union of the leaf sets of $G'_j$'s,
$\Leaf(G_{f}(d))=\bigcup_{j=1}^d \Leaf(G'_j)$.

Next, define the labels $\LAB_f(z_i)$ for each $z_i \in \Leaf(G_{f}(d))$.
For every $j \in \{1, \ldots, d-1\}$ and any leaf $z_j \in \Leaf(G'_j)$,
let $\LAB_f(z_j, G_{f}(d))=(u^f_j,u^f_{j+1}) \circ \LAB_{f-1}(z_j, G'_j)$.

Denote the size (number of nodes) of $G_f(d)$ by $\NodesIn(f,d)$,
its depth (maximal distance between two nodes) by  $\depth(f,d)$,
and its number of leaves by  $\NLeaf(f,d) = |\Leaf(G_f(d))|$.
Note that for $f=1$,
$\NodesIn(1,d) = 2d+\sum_{i=1}^d 4+2 \cdot (d-i) \leq 7d^2$,
$\depth(1,d)=6+2(d-1)$ (corresponding to the length of the path $Q^1_1$),
and $\NLeaf(1,d)=d$.
We now observe that the following inductive relations hold.
\begin{observation}
\label{obs:rel}
\begin{description}
\item{(a)}
$\depth(f,d)=O(d^f)$.
\item{(b)}
$\NLeaf(f,d)=d^f$.
\item{(c)}
$\NodesIn(f,d)=c \cdot d^{f+1}$ for some constant $c$.
\end{description}
\end{observation}
\Proof
(a) follows by the length of $Q^f_i$, which implies that
$\depth(f,d)=d \cdot \depth(f-1,d)$.
(b) follows by the fact that the terminals of the paths starting with
$u_1^f, \ldots, u_d^f$ are the terminals of the graphs $G'_1, \ldots, G'_d$
which are disjoint copies of $G(f-1,d)$, so $\NLeaf(f,d)=d \cdot \NLeaf(f-1,d)$.
(c) follows by summing the nodes in the $d$ copies of $G'_i$
(yielding $d \cdot \NodesIn(f,d)$) and the nodes in $d$ vertex disjoint paths,
namely $Q^f_1, \ldots, Q^f_d$ of total $\sum_{i=1}^d (d-i)\depth(f-1,d)$ nodes,
yielding $\NodesIn(f,d)=d \cdot \NodesIn(f-1,d)+\sum_{i=1}^d (d-i)\depth(f-1,d)$.
\QED
Consider the set of $\lambda=\NLeaf(f,d)$ leaves in $G(f,d)$,
$\Leaf(G(f,d)) = \bigcup_{i=1}^d \Leaf(G'_i) = \{z_1, \ldots, z_\lambda\}$,
ordered from left to right according to their appearance in $G(f,d)$.
\begin{lemma}
\label{lem:prop_induc_path}
For every $z_j$ it holds that: \\
(1) The path  $P(z_j, G(f,d))$ is the only $u^f_1-z_j$ path in $G(f,d)$.\\
(2) $P(z_j, G(f,d)) \subseteq G \setminus \LAB_f(z_j, G(f,d))$.\\
(3) $P(z_i, G(f,d)) \not\subseteq G \setminus \LAB_f(z_j, G(f,d))$
for every $i>j$.\\
(4) $|P(z_i, G(f,d))| > |P(z_j, G(f,d))|$ for every $i < j$.
\end{lemma}
\Proof
We prove the claims by induction on $f$.
For $f=1$, the lemma holds by construction.
Assume this holds for every $f' \leq f-1$ and consider $G(f,d)$.
Let $P^*=[u^f_1, \ldots, u^f_d]$, and let $G'_1, \ldots, G'_d$ be $d$ copies
of the graph $G(f-1,d)$, viewed as ordered from left to right,
where $G'_j$ is connected to $u^f_j$.
That is, there are disjoint paths $Q^f_j$ of monotonely increasing length
connecting $u^f_j$ and $\Root(G'_j)$, for $j\in[1..d]$.

By the inductive assumption, there exists a single path $P(z_j, G'_j)$
between the root $\Root(G'_j)$ and the leaf $z_j$, for every $j\in[1..d]$.
We now show that there is a single path between $\Root(G(f,d)) = u^f_1$
and $z_j$ for every $j\in[1..d]$.
Since there is a single path $P'$ connecting $\Root(G(f,d))$ and $\Root(G'_j)$,
where $P'=P^*[u^f_1, u^f_j]\circ Q^f_j$, it follows that
$P(z_j, G(f,d))=P' \circ P(z_j, G'_j)$ is a unique path in $G(f,d)$.

We now show (2). By the inductive assumption,
$P(z_j, G'_j) \in G \setminus \LAB_{f-1}(z_j, G'_j)$.
Since $\LAB_f(z_j, G(f,d))=(u^f_{j}, u^f_{j+1}) \circ \LAB_{f-1}(z_j, G'_j)$,
it remains to show that $e_j=(u^f_{j}, u^f_{j+1})  \notin P'$ for $j \leq d-1$.
Since $P'$ diverges from $P^*=[u^f_1, \ldots, u^f_d]$ at point $u^f_j$,
it holds that $e_j \notin P'$.

Next we consider (3). Let $Z_1=\{z_i \in  \Leaf(G'_j) \mid i>j\}$ be the set
of leaves to the right of $z_j$ that belong to $G'_j$, and let
$Z_2=\{z_i \in \Leaf(G_f(d)) \setminus \Leaf(G'_j) \mid i>j\}$ be
the complementary set of leaves. By the inductive assumption,
$P(z_i, G'_j) \notin G \setminus \LAB_{f-1}(z_j, G'_j)$ for every $z_i \in Z_1$.
Since the order of the leaves in $G'_j$ agrees with their order in $G(f,d)$
and as $P(z_i, G'_j) \subseteq P(z_i, G(f,d))$ and also
$\LAB_{f-1}(z_j, G'_j) \subseteq \LAB_f(z_j, G(f,d))$, the claim holds
for the set $Z_1$. Next, consider the complementary leaf set $Z_2=\{z_k\}$.
Since the divergence point of $P(z_k, G(f,d))$ and $P^{*}$ is at $u^f_k$
for $k >j$, it follows that $e_j=(u^f_j, u^f_{j+1}) \in P(z_k, G(f,d))$,
and thus $P(z_k, G(f,d)) \nsubseteq G \setminus \LAB_f(z_j, G(f,d))$
for every $k >j$.

Finally, consider (4).
Let $Z^\ell_1=\{z_i \in  \Leaf(G'_j) \mid i<j\}$ be the set of leaves
to the left of $z_j$ that belong to $G'_j$ and let
$Z^\ell_2=\{z_i \in \Leaf(G_f(d)) \setminus \Leaf(G'_j) \mid i<j\}$ be
the complementary set of leaves.
First consider $z_i \in Z^\ell_1$. Then, by the inductive assumption,
$P(z_i, G'_j)> P(z_j, G'_j)$. Since $P(z_i, G(f,d))=P' \circ P(z_i, G'_j)$
and $P(z_j, G(f,d))=P' \circ P(z_j, G'_j)$
for $P'=[u^f_1, \ldots, u^f_j] \circ Q^f_j$, the claim holds for the set
$Z^\ell_1$. Consider next the complementary set $Z^\ell_2=\{z_k\}$
which are in $G'_{k}$ for $k>j$.
Since for every such $Q^f_k=[u^f_k, \Root(G'_k)]$ it holds that
$|Q^f_k|>|Q^f_{k+1}|+\depth(G'_{k+1})\geq P(z_j, G(f,d))$, the claim follows.
\QED

\begin{figure}[htb!]
\begin{center}
\includegraphics[scale=0.5]{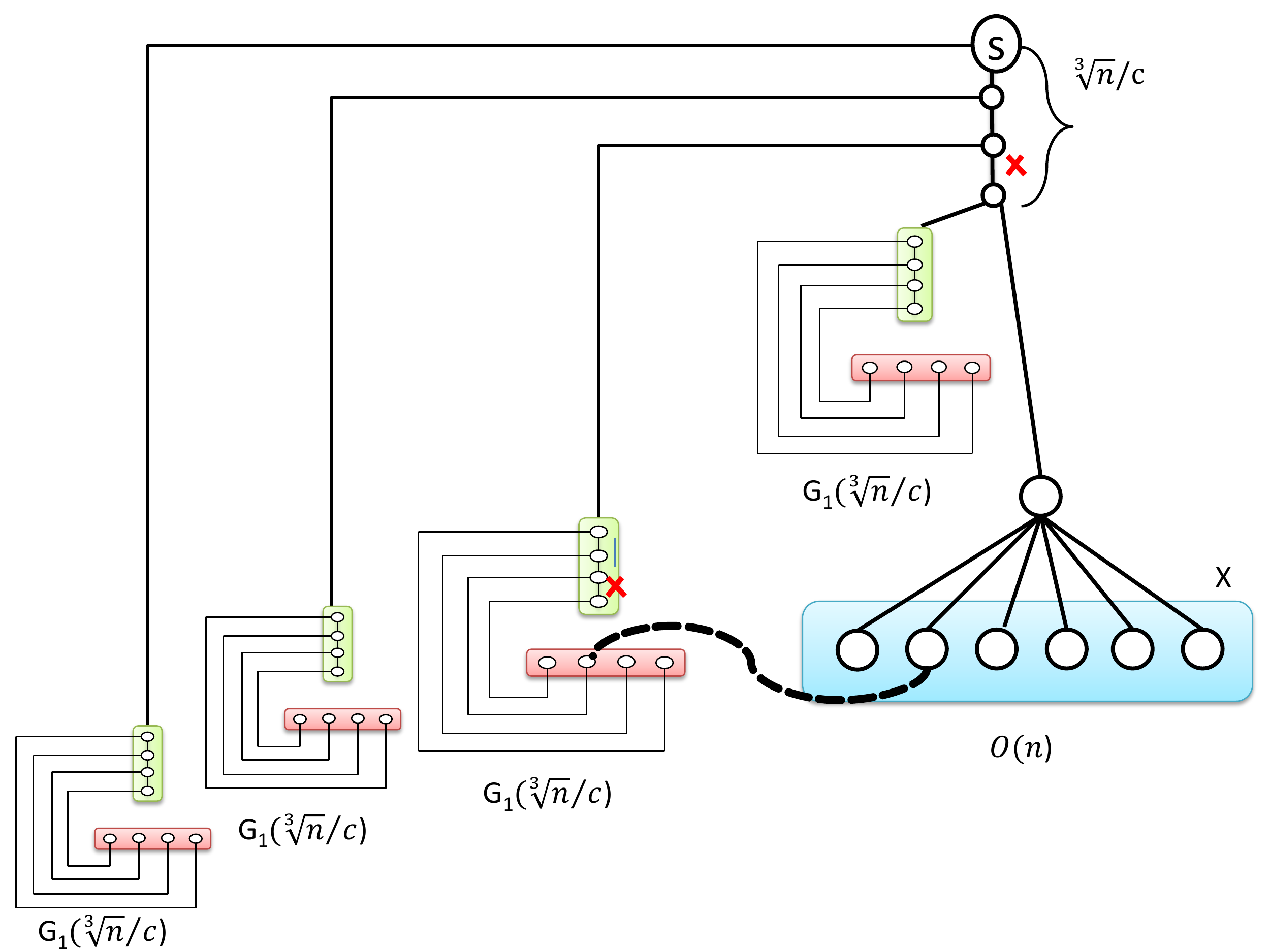}
\caption{\label{fig:lowerboundg2}
Lower bound construction for dual failure \FTBFS\ structure.
The set of X vertices is fully connected to the leaf set of each of the $d$ copies of $G_{1}(d)$. Overall $|E(G^*_2)|=O(n^{5/3})$. The dashed wide edge is required in any dual failure \FTBFS\ structure upon the faults of the edges marked in figure.}
\end{center}
\end{figure}

Finally, we turn to describe the graph $G^*_f(V, E)$ which establishes our
lower bound.
The graph $G^*_f(V, E)$ consists of three components.
The first is the graph $G_{f}(d)$ for $d=\lceil(n/2c)^{1/(f+1)}\rceil$,
where $c$ is some constant to be determined later.
By Obs. \ref{obs:rel}, $n/2 \le |V(G_{f}(d))|$.
Note that $d \le (5/4)^{1/(f+1)} \cdot (n/2c)^{1/(f+1)} = (5n/8c)^{1/(f+1)}$
for sufficiently large $n$,
hence $\NodesIn(f,d)=c \cdot d^{f+1} \le 5n/8$.
The second component of $G^*_f(V, E)$ is a set of nodes
$X=\{x_1, \ldots, x_\chi\}$ and an additional vertex $v^{*}$ that is
connected to $u^f_{d}$ and to all the vertices of $X$.
The cardinality of $X$ is $\chi=n-\NodesIn(f,d)-1$.
The third component of $G^*_f(V, E)$ is a complete bipartite graph $B$
connecting the nodes of $X$ with the leaf set $\Leaf(G_f(d))$, i.e.,
the disjoint leaf sets $\Leaf(G'_1), \ldots, \Leaf(G'_d)$.
The vertex set of the resulting graph is thus
$V=V(G_{f}(d))\cup \{v^{*}\} \cup X$ and hence $|V|=n$.
See Figures \ref{fig:lowerboundg2} and \ref{fig:lowerboundgf}
for illustration of $G^*_2$ and $G^*_f$.

By Prop. (b) of Obs. \ref{obs:rel},
$\NLeaf(G'_i)=d^{f}=\lceil(n/2c)^{1/(f+1)}\rceil^f \ge (n/2c)^{f/(f+1)},$
hence
$|E(B)| \ge (3n/8-1)\cdot (n/2c)^{f/(f+1)} = O(n^{2-1/(f+1)})$.

We now complete the proof of Thm. \ref{thm:lowerbound_f} for the single source case.
\Proof[Thm. \ref{thm:lowerbound_f} for $\NSource=1$]
We show that every $f$-edge  \FTBFS\ structure $H$ with respect to $s=u^f_1$
of $G^*_f(V, E)$ must contain all the edges of $E(B)$.
Let $G'_1, \ldots, G'_d$ be the $d$ copies of $G(f-1,d)$.
Let $z^{*}$ be the rightmost leaf in $G^*_f(V, E)$ (i.e., in $\Leaf(G'_d)$).
We first show that $e'_i=(x_i, z^{*})$ must be included in any
\FTBFS\ structure, for every $i \in \{1, \ldots, |X|\}$.
Assume, towards contradiction, that there exists a \FTBFS\ structure not using
$e'_i$,  i.e., $H \subseteq G^*_f(V, E) \setminus \{e'_i\}$.
Consider the failure of the edge $(u^{f}_{d},v^{*})$.
By Lemma \ref{lem:prop_induc_path}, $P(z^{*}, G(f,d))$ is the unique
shortest-path between $\Root(G(f,d))$ and $z^{*}$, and any other $s-z'$ path
is strictly longer. Hence, we get that
$\dist(s, x_i, H \setminus \{e'_i\})> \dist(s, x_i, G^*_f(V, E) \setminus \{e'_i\})$,
in contradiction to the fact that $H$ is a \FTBFS\ structure.
Next, consider any specific edge $e_{i,j}=(x_i, z_j)$ where
$z_j \in \Leaf(G'_j)$ is not the rightmost leaf, and let
$F=\LAB_f(z_j, G(f,d))$ be the set of edge faults.
Note that by construction, $0<|\LAB_f(z_j, G(f,d))| \leq f$.
It then follows by Lemma \ref{lem:prop_induc_path} that
$P(z_k, G(f,d)) \notin G(f,d) \setminus F$ for every $k >j$.
Thus, $P(z_k, G(f,d)) \notin G^*_f \setminus F$ as well.
In addition, $P(z_k, G(f,d))> P(z_j, G(f,d))$ for every $k>j$,
which implies that
$\dist(s, x_i, G'' \setminus F)>\dist(s, x_i, G^*_f\setminus F)$
for every graph $G'' \subseteq G \setminus \{e_{i,j}\}$.
The theorem follows.
\QED
\def\FIGG1{
\begin{figure}[htb!]
\begin{center}
\includegraphics[scale=0.4]{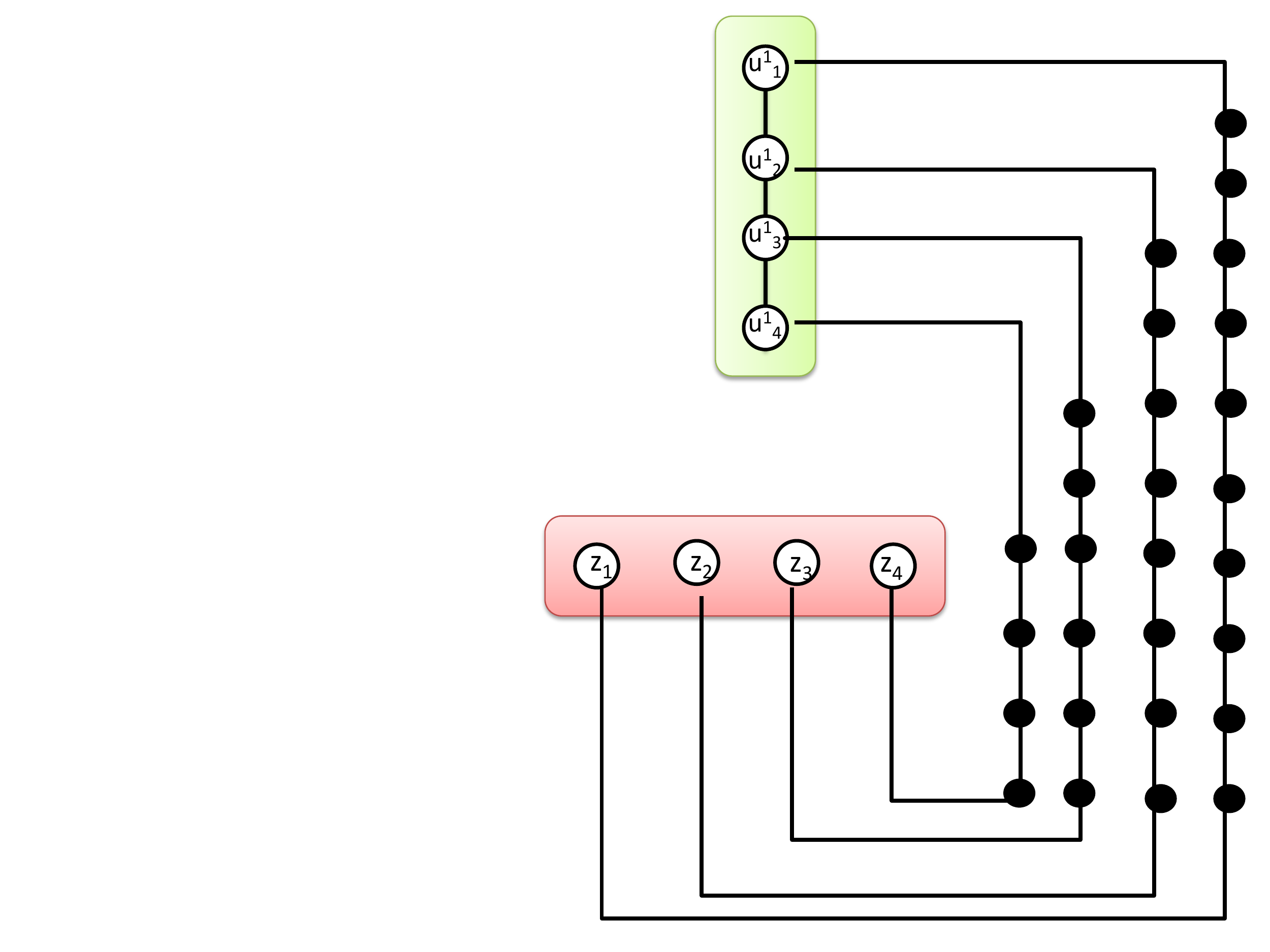}
\caption{\label{fig:lowerboundg1} The graph $G_1(d)$.}
\end{center}
\end{figure}
}%\FIGG1
\FIGG1

\def\FIGGF{
\begin{figure}[h!]
\begin{center}
\includegraphics[scale=0.4]{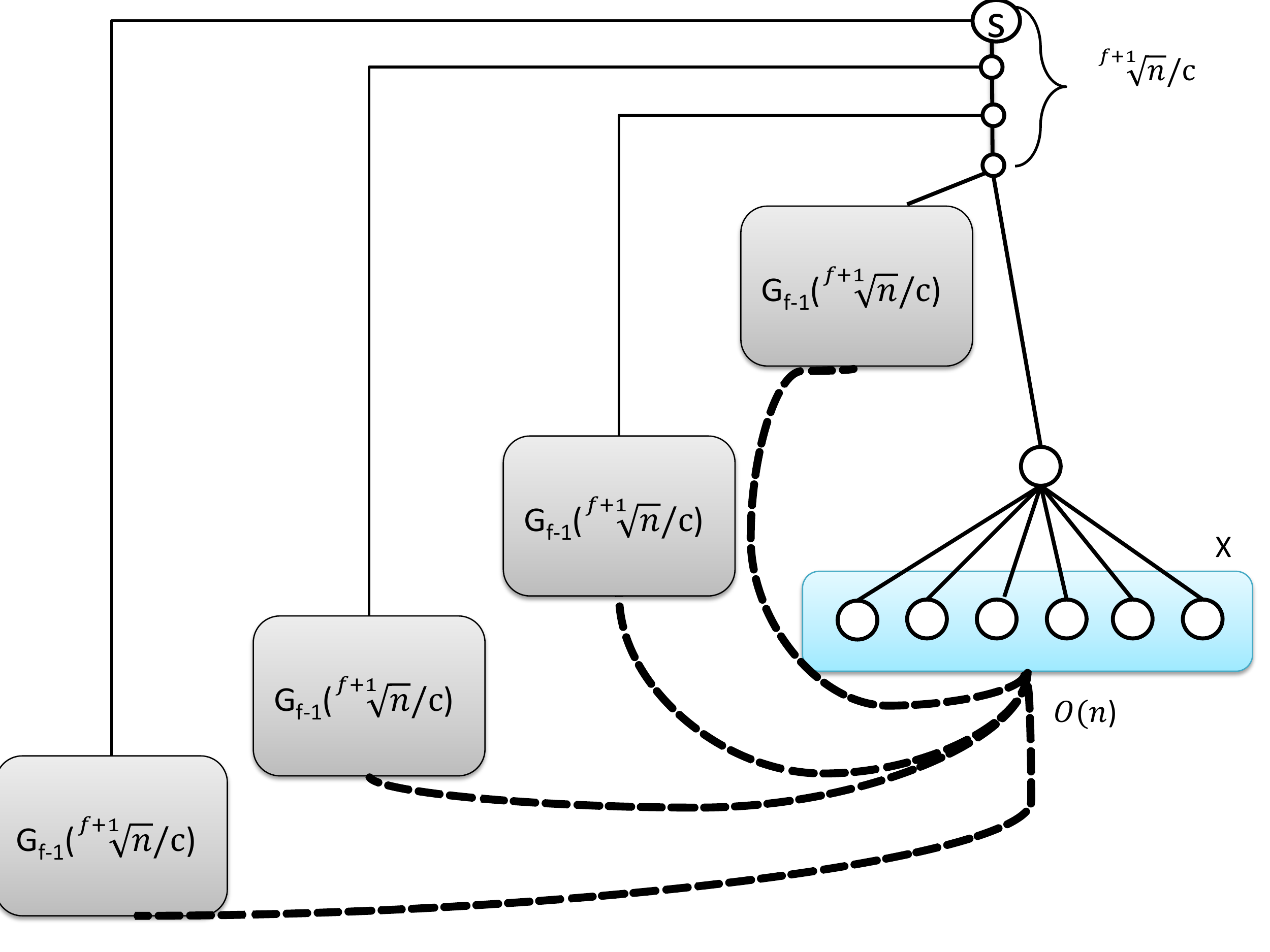}
\caption{\label{fig:lowerboundgf} Schematic illustration of $G^*_f$. The set of $X$ nodes are fully connected to the leaf set of $G_{f-1}(d)$ graphs. Each such edge is necessary for a certain fault set.}
\end{center}
\end{figure}
}%\FIGGF
\FIGGF

\dnsparagraph{\textbf{The multi-source case}}
%In addition, for the case of $\NSource \geq 1$ sources, where it is required to construct a $f$-failure \FTBFS\ with respect to each source $s \in S$ where $|S|=\NSource$, we establish the following lower bound.
%\begin{theorem}
%\label{thm:lowerbound_f_multisource}
%There exists an $n$-vertex graph $G(V, E)$ and a source set $S \subseteq V$
%of cardinality $\NSource$, such that any $f$-failure  \FTBFS\ structure from
%the source set $S$ has $\Omega(n\cdot(\NSource\cdot n)^{f/(f+1)})$ edges.
%\end{theorem}
%\Proof
We now extend the lower bound construction to support the case of multiple sources $S \subseteq V$ for any cardinality of sources. \Proof [Thm. \ref{thm:lowerbound_f} for any $1\leq \NSource \leq n$]
Given a parameter $\NSource$ representing the number of sources,
$\NSource$ copies, $G'_1, \ldots, G'_\NSource$, of $G_f(d)$,
where $d=O((n / \NSource)^{1/(f+1)})$.
By Obs. \ref{obs:rel}, each copy consists of $O(n/\NSource)$ nodes.
Let $y_i$ be the node $u^f_d$ and $s_i=\Root(G'_i)$ in the $i$th copy $G'_i$.
Add a node $v^*$ connected to a set $X$ of $O(n)$ nodes and connect $v^*$
to each of the nodes $y_i$, for $i \in \{1, \ldots, d\}$.
Finally, connect the set $X$ to the $\NSource$ leaf sets
$\Leaf(G'_1), \ldots, \Leaf(G'_\NSource)$ by a complete bipartite graph,
adjusting the size of the set $X$ in the construction so that $|V(G)|=n$.
Since $\NLeaf(G'_i)=\Omega((n / \NSource)^{f/(f+1)})$ (see Obs. \ref{obs:rel}),
overall $|E(G)| = \Omega(n \cdot \NSource \cdot \NLeaf(G_f(d))) =
\Omega(\NSource^{1/(f+1)}\cdot n^{1+f/(f+1)})$.
Since the path from each source $s_i$ to $X$ cannot aid the nodes of $G'_j$
for $j \neq i$, the analysis of the single-source case can be applied
to show that each of the bipartite graph edges in necessary
upon a certain sequence of at most $f$-edge faults.
%See Fig. \ref{fig:lowerboundmulti} for an illustration of
%the multi-source $f=1$ case.
\QED

%
%\begin{figure}[h!]
%\begin{center}
%\includegraphics[scale=0.5]{lowerboundmulti_new}
%\caption{\label{fig:lowerboundmulti}
%Illustration of the lower bound for the multi-source case.}
%\end{center}
%\end{figure}
%%%%%%%%%%%%%%%%%%%%%%%%%%%%%%%%%%%%%%%%%%%%%
\section{$O(\log n)$-Approximation for constructing the minimum $f$-failure FT-BFS structure}
\label{sec:opt}
%%%%%%%%%%%%%%%%%%%%%%%%%%%%%%%%%%%%%%%%%%%%%
In Sec. \ref{sec:analysis}, we presented an algorithm that for every graph $G$ and source $s$ constructs a dual failure \FTBFS\  $H$ with $O(n^{5/3})$ edges.
In Sec. \ref{sec:lb}, we showed that there exist graphs $G$ and source set $S \subseteq V(G)$ for with every \FTBFS\ $H \subseteq G$ with respect to $S$ overcoming up to $f$-faults has $\Omega(|S|^{1-1/(f+1)} \cdot n^{2-1/(f+1)})$ edges, establishing tightness of our algorithm for the case of $|S|=1$ and $f=2$ in the worst-case.

In this section consider the \emph{Minimum \FTMBFS} that aims at finding the minimum size structures that tolerant against $f$-faults for any given set of sources $S$. The \emph{Minimum \FTMBFS} has been defined and studied by \cite{PPFTBFS13} for the single failure case (i.e., $f=1$). We extend the result of \cite{PPFTBFS13} to the general case of constant $f\geq 1$ and provide a $O(\log n)$ approximation algorithm for this problem.

The importance of this result is of twofold.
First, for $f=2$, although our universal upper bound matches the existential lower bound, there are also inputs $(G',s')$ for which the algorithm of Sec. \ref{sec:analysis}, might still produce an \FTBFS\ $H$ which is denser by a factor of
$\Omega(n^{2/3})$ than the size of the optimal \FTBFS\ structure. For the case of $f=1$, an example of such a graph is given by \cite{FTBFSArxiv}, this example can easily be modified to $f=2$ by the lower bound construction of Sec. \ref{sec:lb}.
Second, for general $f\geq 3$, while a tight universal upper bound on the size of $f$-fault \FTBFS\ structures is currently beyond our reach, we can still construct such structures whose size is larger by a factor of $O(\log n)$ than the optimal $f$-fault \FTBFS\ structures.
Although this section is a straightforward extension of  \cite{PPFTBFS13}, for completeness, we provide a full analysis and begin by defining the problem formally.
For a graph $G=(V,E)$, a source set $S \subseteq V$ and number of faults $f \geq 1$, let $\mathcal{H}(S,G,f)$ be the collection of all subgraph $H \subseteq G$ that are \FTMBFS\ with respect to $S$ overcoming up to $f$-faults, that is the subgraphs $H \subseteq G$ satisfying that $\dist(s,v,H\setminus F)=\dist(s,v,G\setminus F)$ for every $(s,v) \in S \times V$ and $F \subseteq E$ where $|F|\leq f$.
Let $\Cost^*(S,G,f)=\min\{|E(H)| ~\mid~ H \in \mathcal{H}(S,G,f)\}.$

In the \emph{Minimum \FTMBFS} problem, we are given a graph $G$, a source set $S \subseteq V$ and number of faults $f\geq 1$ and the goal is to compute an $f$-fault \FTMBFS\ $H \in \mathcal{H}(S,G,f)$
of minimum size, i.e., such that $|E(H)|=\Cost^*(S,G,f)$.

Similarly to \cite{PPFTBFS13}, it can be shown that
the \emph{Minimum \FTMBFS} problem for any constant $f \geq 1$ and $|S|\geq 1$, is NP-hard and moreover, cannot be approximated (under standard complexity assumptions) to within a factor of $\Omega(\log n)$.

We now turn to describe a $O(\log n)$ approximation algorithm given an input $(S,G,f)$.
%This approximation algorithm can easily extended to case of multi source $S \subseteq V$ instead of single source, with the same approximation guarantee. For simplicity, we consider the single source case.
%
To prove theorem  \ref{thm:approx}, we first describe the algorithm and then bound the number of edges.
Let $\ApproxSetCover(\Set,U)$ be an $O(\log n)$
approximation algorithm for the Set-Cover problem,
which given a collection of sets $\Set=\{S_1, \ldots, S_{M}\}$ that covers a universe $U=\{u_1, \ldots, u_N\}$ of size $N$, returns a cover $\Set' \subseteq \Set$ that is larger by at most $O(\log N)$ than any other $\Set'' \subseteq \Set$ that covers $U$ (cf. \cite{Vazirani97}).

\dnsparagraph{\textbf{The Algorithm}}
Starting with $H=\emptyset$, the algorithm adds edges to $H$ until it becomes an $f$-fault \FTMBFS\ structure.
\par Set an arbitrary order on the vertices
$V(G)=\{v_1, \ldots, v_{n}\}$ and define
$U_F=\{F\subseteq (E(G) \cup \{\emptyset\})\}$ be the collection of all possible $k\leq f$ edge failures in $G$. Note that $U_F$ contains also the empty set, corresponding to the fault free case. In addition, note that $|U_F|=O(n^f)$, hence of polynomial size for constant number of faults $f$.
Let
$$U=\{ \langle s_k,F \rangle ~\mid~ s_k \in S,  F \in U_F\}.$$
%be the collection of all possible $k\leq f$ failures in $G$ (note that $U$ contains also the empty set, corresponding to the case where non of the edges fails).
The algorithm consists of $n$ rounds, where in round $i$
it considers $v_i$. Let $\Gamma(v_i, G)=\{u_1, \ldots, u_{d_i}\}$ be the set of neighbors of $v_i$ in some arbitrary order, where $d_i=\deg(v_i,G)$.
For every neighbor $u_j$, define a set $S_{i,j} \subseteq U$ containing elements of $U$.
Informally, a set $S_{i,j}$ contains the pair $\langle s_k, F\rangle \in U$ if there exists an $s_k-v_i$ shortest path in $G \setminus F$ that goes through the neighbor $u_j$ of $v_i$.
Note that $S_{i,j}$ contains the pair $\langle s_k,\emptyset \rangle$ for every $s_k \in S$
iff there exists an $s_k-v_i$ shortest-path in $G$ that goes through $u_j$. Formally, the pair $\langle s_k, F\rangle$ is included
in every set $S_{i,j}$ satisfying that
\begin{equation}
\label{eq:intset}
\dist(s_k, u_j, G \setminus F)=\dist(s_k, v_i, G \setminus F)-1.
\end{equation}
Let $\Set_i=\{S_{i,1}, \ldots, S_{i,d_i}\}$.
The edges incident to $v_i$ that are added to $H$ in round $i$ are now selected by using algorithm $\ApproxSetCover$ to generate an approximate solution for the set cover problem on the collection $\Set=\{S_{i,j}~\mid~ u_j \in \Gamma(v_i, G)\}$.
Let $\Set'_i=\ApproxSetCover(\Set_i,U)$.
For every $S_{i,j}\in \Set'_i$, add the edge
$(u_j,v_i)$ to $H$.
\dnsparagraph{\textbf{Analysis}}
We first show that algorithm constructs an $f$-\FTMBFS\
$H \in \mathcal{H}(S,G,f)$ and then bound its size.
\begin{lemma}
\label{lem:correct}
$H \in \mathcal{H}(S,G,f)$.
\end{lemma}
\Proof
Assume, towards contradiction, that $H \notin \mathcal{H}(S,G,f)$. Let $s \in S$ be some source vertex such that $H \notin \mathcal{H}(\{s\},G,f)$ is not an $f$-\FTBFS\ structure with respect to $s$. By the assumption, such $s$ exits. Let
$$BP=\{(i,F) \mid v_i \in V, F  \in U_F \mbox{~and~}
\dist(s,v_i, H \setminus F) >  \dist(s,v_i,G \setminus F)\}$$
be the set of ``bad pairs," namely, vertex, faulty-set pairs $(i,F)$ for which the $s-v_i$ shortest path distance in $H \setminus F$ is greater than that in $G\setminus F$.
(By the assumption that $H \notin \mathcal{H}(\{s\},G,f)$, it holds that $BP\ne \emptyset$.)

For every pair $(i,F)$, where $v_i \in V \setminus \{s\}$ and $F \in U_F$, define an $s-v_i$ shortest-path $P_{i,F}$ in $G \setminus F$ in the following manner.
Let $u_j \in \Gamma(v_i, G)$ be such that the pair $\langle s, F \rangle \in S_{i,j}$
is covered by the set $S_{i,j}$ of $u_j$ and $S_{i,j} \in \Set'_i$ is included in the cover returned by the algorithm $\ApproxSetCover$ in round $i$. Thus, $(u_j,v_i)\in H$ and
$\dist(s, u_j, G \setminus F)=\dist(s, v_i, G \setminus F)-1$.
Let $P' \in SP(s, u_j, G \setminus F)$ and define
$$P_{i,F}=P' \circ (u_j,v_i).$$
By definition, $|P_{i,F}|=\dist(s, v_i, G \setminus F)$ and by construction, $\LastE(P_{i,F}) \in H$.
Define
$BE(i,F)=P_{i,F} \setminus E(H)$ to be the set of ``bad edges,''
namely, the set of $P_{i,F}$ edges that are missing in $H$.
By definition, $BE(i,F) \neq \emptyset$ for every bad pair $(i,F) \in BP$.
Let $d(i,F)=\max_{e \in BE(i,F)}\{\dist(s,e,P_{i,F})\}$ be the maximal depth
of a missing edge in $BE(i,F)$, and let $DM(i,F)$ denote that ``deepest
missing edge'' for $(i,F)$, i.e., the edge $e$ on $P_{i,F}$ satisfying
$d(i,F) = \dist(s,e,P_{i,F})$.
Finally, let $(i',F') \in BP$ be the pair that minimizes $d(i,F)$,
and let $e_1=(v_{\ell_1}, v_{i_1}) \in BE(i',F')$ be
the deepest missing edge on $P_{i',F'}$, namely, $e_1=DM(i',F')$.
Note that $e_1$ is the {\em shallowest} ``deepest missing edge''
over all bad pairs $(i,F) \in BP$.
Let $P_1=P_{i_1,F'}$, $P_2=P^{*}_{i',F'}[s,v_{i_1}]$ and
$P_3=P^{*}_{i',F'}[v_{i_1}, v_{i'}]$;
Note that since $(i',F') \in BP$, it follows that also $(i_1, F') \in BP$.
(Otherwise, if $(i_1, F') \notin BP$, then any $s-v_{i_1}$ shortest-path
$P' \in SP(s, v_{i_1}, H \setminus F')$
, where $|P'|=|P_{i_1,F'}|$, can be appended to $P_3$ resulting in
$P''=P' \circ P_3$
such that (1) $P'' \subseteq H\setminus F'$ and (2)
$|P''|=|P'|+|P_3|=|P_2|+|P_3|=|P_{i',F'}|$, contradicting the fact that
$(i',F') \in BP$.)
Thus we conclude that $(i_1, F') \in BP$.
Finally, note that $\LastE(P_1) \in H$ by definition, and therefore
the deepest missing edge of $(i,F)$ must be shallower, i.e.,
$d(i_1,F')<d(i',F')$. However, this is in contradiction to our choice
of the pair $(i',F')$. The lemma follows.
\QED
\dnsparagraph{\textbf{Size analysis}}
Let $W:E(G) \to \mathbb{R}_{>0}$ be the weight assignment that guarantees the uniqueness of shortest-paths (i.e., breaks ties between of shortest-paths of the same lengths, in a consistent manner). Note that the algorithm did not use $W$ in the computation of the shortest-paths. For every node $v_i$, let
$\Gamma(v_i,G)=\{u_{1}, \ldots, u_{d_i}\}$ be its ordered neighbor set
as considered by the algorithm.
For every \FTMBFS\ tree
$H\in \mathcal{H}(S,G,f)$, $v_i \in V, F \in U_F$ and source $s_k \in S$, let $\widetilde{P}_i(s_k,F) \in SP(s_k, v_i, H \setminus F,W)$ be an $s_k-v_i$ shortest-path in $H \setminus F$.
Let
$$A_i(H)=\{\LastE(\widetilde{P}_i(s_k,F)) ~\mid~ \langle s_k,F \rangle \in U \}$$
be the edges incident $v_i$ that appear as last edges in the shortest-paths and replacement paths from $S$ to $v_i$ in $H$.
Define
$$\Set_i(H)=\{S_{i,j} ~\mid~ (u_j, v_i) \in A_i(H)\}.$$
We then have that
\begin{equation}
\label{eq:set_size}
|\Set_i(H)|=|A_i(H)|~.
\end{equation}
The correctness of the algorithm (see Lemma \ref{lem:correct})
established that if a subgraph $H \subseteq G$ satisfies that
$\Set_i(H)$ is a cover of
$U$ for every $v_i \in V$, then $H \in \mathcal{H}(S,G,f)$.
We now turn to show the reverse direction.
\begin{lemma}
\label{lem:cover_tree}
For every $H \in \mathcal{H}(S,G,f)$, the collection $\Set_i(H)$ is a cover of $U$, namely,
$\bigcup_{S_{i,j} \in \Set_i(H)}S_{i,j}=U, \mbox{~~for every~~} v_i \in V$.
\end{lemma}
\Proof
Assume, towards contradiction, that there exists an $f$-fault \FTBFS\
$H \in \mathcal{H}(S,G,f)$ and a vertex $v_i \in V$
whose corresponding collection of sets $\Set_i(H)$
does not cover $U$.
Hence there exists at least one uncovered pair $\langle s_k,F \rangle \in U$, i.e.,
\begin{equation}
\label{eq:notcovered}
\langle s_k,F \rangle  \in U \setminus\bigcup_{S_{i,j} \in \Set_i(H)}S_{i,j}~.
\end{equation}
We next claim that $H$ does not contain an optimal
$s_k-v_i$ path when the edges of $F$ fail,
contradicting the fact that $H \in \mathcal{H}(S,G,f)$.
That is, we show that
$$\dist(s_k, v_i, H \setminus F)>\dist(s_k, v_i, G \setminus F).$$
Towards contradiction, assume otherwise, and let
$(u_j, v_i)=\LastE(P_{i,F})$
where $P_{i,F} \in SP(s_k, v_i, H \setminus F,W)$,
hence $(u_j, v_i) \in A_i(H)$
and $S_{i,j} \in \Set_i(H)$.
By the contradictory assumption, $|P_{i,F}|=\dist(s_k, v_i, G \setminus F)$
and hence $\dist(s_k, u_j, G \setminus F)=\dist(s_k, v_i, G \setminus F)-1$.
This implies that $F \in S_{i,j} \in \Set_i(H)$, in contradiction to Eq. (\ref{eq:notcovered}),
stating that $\langle s_k,F \rangle$ is not covered by $\Set_i(H)$.
The lemma follows.
\QED
We now turn to bound that number of edges in $H$.
\begin{lemma}
\label{lem:numbersize}
$|E(H)| \leq O(\log n)\cdot \Cost^*(S, G,f)$.
\end{lemma}
\Proof
Let $H^* \in \mathcal{H}(S,G,f)$ be an optimal $f$-fault \FTMBFS\ satisfying that $|E(H^*)|=\Cost^*(S, G,f)$.
Let $\delta=c \cdot \log n$ be the approximation
ratio guarantee of Algorithm $\ApproxSetCover$.
For ease of notation, let
$O_i=A_i(H^*)$ for every $v_i \in V$.
Let $\Set_i=\{S_{i,1}, \ldots, S_{i,d_i}\}$ be the collection of
$v_i$ sets considered at round $i$ where $S_{i,j}\subseteq U$
is the set of the neighbor $u_j \in \Gamma(v_i,G)$ computed according to
Eq. (\ref{eq:intset}).
\par Let $\Set'_i=\ApproxSetCover(\mathcal{S}_i,U)$
be the cover returned by the algorithm and define
$A_i=\{(u_j, v_i) ~\mid~ S_{i,j} \in \Set'_i\}$
as the collection of edges whose corresponding sets are included in
$\mathcal{S}'_i$.
%and $A_i=A_i(\widehat{T})$
%$\Set'_i=\Set(\widehat{T})$ and
%$\Set^*_i=\Set_i(T^*)$
Thus, by Eq. (\ref{eq:set_size}),
$|O_i|=|\Set_i(H^*)|$ and $|A_i|=|\Set'_i|$ for every $v_i \in V$.
\begin{observation}
\label{cl:a_i}
$|A_i| \leq \delta |O_i|$ for every $v_i \in V$.
\end{observation}
\Proof
Assume, towards contradiction, that there exists some $i$ such that
$|A_i|>\delta |O_i|$. Then by Eq. (\ref{eq:set_size})
and by the approximation guarantee of \ApproxSetCover\, where in
particular $|\Set_i(H)| \leq \delta |\Set''_i|$
for every $\Set''_i \subseteq  \Set_i$ that covers $U$, it follows that $\Set_i(H^*)$ is not a cover of $U$.
Consequently, it follows by Lemma \ref{lem:cover_tree}
that $H^* \notin \mathcal{H}(S,G,f)$, contradiction.
The observation follows.
\QED
%$|E(\widehat{T})|\leq \sum_i |A_i|+ n-1$ and
%$$|E(T^*)| \geq \max\{\sum_i |O_i|/2, n-1\}.$$
%Hence, combining with Cl. \ref{cl:a_i}, we have that
Since $\bigcup A_i$ contains precisely the edges that are added by the algorithm to the constructed $f$-faults \FTMBFS\ structure $H$, we have that
\begin{eqnarray*}
|E(H)|&\leq& \sum_i |A_i| \leq
\delta \sum_i |O_i| \leq 2\delta \cdot \Cost^*(S, G,f)~,
\end{eqnarray*}
where the second inequality follows by Obs. \ref{cl:a_i} and
the third by the fact that $|E(H^*)| \geq \sum_i |O_i|/2$ (as every edge in $\bigcup_{v_i \in V} O_i$ can be counted at most twice, by both its endpoints). The lemma follows.
\QED
Thm. \ref{thm:approx} is established.

\dnsparagraph{\textbf{Acknowledgment}}
I am very grateful to my advisor,
Prof. David Peleg, for many helpful discussions and for
reviewing this paper.
%%%%%%%%%%%%%%%%%%%%%%%%%
\clearpage

%%%\def\thepage{}
%%{\small
%%\bibliographystyle{plain}
%%\bibliography{SINR_v7}
%%}
%%%%%%%%%%%%%%%%%%%%%%%%%

%\small

\end{document}